\documentclass{article}
\usepackage{multirow}
\usepackage{enumitem}
\usepackage{setspace}
\usepackage{arxiv}
\usepackage{xcolor}
\usepackage{floatpag}
\definecolor{darkgreen}{RGB}{0, 150, 0}
\usepackage[utf8]{inputenc} % 
\usepackage[T1]{fontenc}    % 
\usepackage{hyperref}       % 
\usepackage{url}            % 
\usepackage[format=plain, 
justification=raggedright,singlelinecheck=false]{caption}\usepackage[format=plain, 
justification=raggedright,singlelinecheck=false]{caption}
\newcommand{\Bell}{\bm \ell}
\usepackage{booktabs}       % 
\usepackage{amsfonts}       % 
\usepackage{nicefrac}       % 
\usepackage{microtype}      % 
\usepackage{lipsum}		% 
\usepackage{graphicx}
\usepackage[round]{natbib}
\usepackage{doi}

\newcommand{\Tau}{\mathrm{T}}
\usepackage{float}
\usepackage[raggedrightboxes]{ragged2e}
\usepackage{array}
\usepackage{doi}

\usepackage{amsmath,amsfonts,amssymb,amsthm, bm}

\newtheorem{lemma}{Lemma}
\newtheorem{theorem}{Theorem}
\newtheorem{corollary}{Corollary}
\newtheorem{definition}{Definition}

\usepackage{placeins}

\usepackage{array,ragged2e}
\newcolumntype{L}[1]{>{\RaggedRight\arraybackslash}p{#1}}
\newtheorem{Remark}{Remark}

\newtheorem{example}{Example}
\usepackage{bm}
\usepackage{xcolor}
\usepackage{mathtools}
\DeclarePairedDelimiter\ceil{\lceil}{\rceil}
\DeclarePairedDelimiter\floor{\lfloor}{\rfloor}
\usepackage{accents}
\usepackage{placeins}% 
\def\mydefb#1{\expandafter\def\csname b#1\endcsname{\bm{#1}}}
\def\mydefallb#1{\ifx#1\mydefallb\else\mydefb#1\expandafter\mydefallb\fi}
\mydefallb ABCDEFGHIJKLMNOPQRSTUVWXYZabcdeghijklnopqrstuvwxyz\mydefallb

\def\mydefgreek#1{\expandafter\def\csname b#1\endcsname{\text{\boldmath$\mathbf{\csname #1\endcsname}$}}}
\def\mydefallgreek#1{\ifx\mydefallgreek#1\else\mydefgreek{#1}% 
	\lowercase{\mydefgreek{#1}}\expandafter\mydefallgreek\fi}
\mydefallgreek {beta}{Gamma}{alpha}{Delta}{epsilon}{Omega}{Theta}{Iota}{Lambda}{kappa}{mu}{nu}{Xi}{Pi}{chi}{rho}{Sigma}{Psi}{tau}\mydefallgreek

\usepackage{hyperref}   
\hypersetup{% 
	bookmarksnumbered,
	pdfstartview={FitH},
	linkcolor=blue,
	citecolor=blue,
	colorlinks=true,
	pdfpagemode={UseNone},
	plainpages=false
}% 

\newcommand{\C}{\text{\normalfont C}}
\newcommand{\D}{\text{\normalfont D}}

\newcommand{\Tend}{\bar{t}}
\newcommand{\Iend}{\bar{i}}
\newcommand{\calI}{\mathcal{I}}
\newcommand{\calT}{\mathcal{T}}
\newcommand\blfootnote[1]{% 
	\begingroup
	\renewcommand\thefootnote{}\footnote{#1}% 
	\addtocounter{footnote}{-1}% 
	\endgroup
}
\usepackage{algorithm,algpseudocode}
\algnewcommand{\Inputs}[1]{% 
	\State \textbf{Inputs:}
	\Statex \hspace*{\algorithmicindent}\parbox[t]{.8\linewidth}{\raggedright #1}
}
\algnewcommand{\Initialize}[1]{% 
	\State \textbf{Initialize:}
	\Statex \hspace*{\algorithmicindent}\parbox[t]{.8\linewidth}{\raggedright #1}
}
\usepackage{mathtools}

\usepackage{caption}
\usepackage{subcaption}
\usepackage{etoolbox}
\title{Exact statistical analysis for response-adaptive clinical
trials: a general and computationally tractable approach}

\usepackage{authblk}

\usepackage{accents}
\newcommand{\ubar}[1]{\underaccent{\bar}{#1}}

\usepackage{orcidlink}
\usepackage{titlesec}
\titlelabel{\thetitle.\quad}

\usepackage{hyperref}     
\hypersetup{% 
bookmarksnumbered,
pdfstartview={FitH},
linkcolor=blue,
citecolor=blue,
colorlinks=true,
pdfpagemode={UseNone},
plainpages=false
}% 

\usepackage[english]{babel}

\title{Exact statistical analysis for response-adaptive clinical
trials: A general and computationally tractable approach}

\author{Stef Baas\thanks{Corresponding author} \\
	\vspace{-3mm}Stochastic Operations Research Group,
	University~of~Twente, Enschede, The~Netherlands\\
 \phantom{.}\newline 
	Peter Jacko \\
	{Department of Management Science}, {Lancaster University},   {Bailrigg}, {United Kingdom},\\
 and\\
  {Berry Consultants}, { {Abingdon}, {United Kingdom}}\\
 \phantom{.}\newline 
	Sof\'ia S. Villar\\
	{MRC Biostatistics Unit}, {University of Cambridge}, {Cambridge}, {United Kingdom}
}

\date{}

\renewcommand{\shorttitle}{Exact statistical analysis for RA clinical
trials}

\hypersetup{
pdftitle={Exact statistical analysis for response-adaptive clinical
trials: A general and computationally tractable approach},
pdfauthor={Stef Baas, Peter Jacko, Sof\'ia S. Villar}
}

\begin{document}

\maketitle
\begin{abstract}
	Response-adaptive clinical trial designs allow targeting a given objective by skewing the allocation of participants to treatments based on observed outcomes. Response-adaptive designs face greater regulatory scrutiny due to potential type I error rate inflation, which limits their uptake in practice. Existing approaches for type I error control either only work for specific designs, have a risk of Monte Carlo/approximation error, are conservative, or computationally intractable. To this end, a general and computationally tractable approach is developed for exact analysis in two-arm response-adaptive designs with binary outcomes.  This approach can construct exact tests for designs using either a randomized or deterministic response-adaptive procedure. The constructed conditional and unconditional exact tests generalize Fisher's and Barnard's exact tests, respectively. Furthermore, the approach allows for complexities such as delayed outcomes, early stopping, or allocation of participants in blocks. The efficient implementation of forward recursion allows for testing of two-arm trials with 1,000 participants on a standard computer. Through an illustrative computational study of trials using randomized dynamic programming it is shown that, contrary to what is known for equal allocation,  
	the conditional exact Wald test based on total successes has, almost uniformly, higher power than the unconditional exact Wald test. Two real-world trials with the above-mentioned complexities are re-analyzed to demonstrate the value of the new approach in controlling type I errors and/or improving the statistical power.
	\blfootnote{\scriptsize\newline {\bf Abbreviations}: ARREST: Advanced R$^2$Eperfusion STrategies for Refractory
		Cardiac Arrest,  CX: conditional exact, CX-SA test: conditional exact test based on total successes and allocations, CX-S test: conditional exact test based on total successes, DRA: deterministic response-adaptive, FET: Fisher's exact test, M-PTW: modified play-the-winner, OST: optional stopping threshold, PTW: play-the-winner, RA: response-adaptive, RAR: response-adaptive randomization, RDP: randomized dynamic programming,   UX: unconditional exact}
	\hspace{-1.5mm}
\end{abstract}

\keywords{Conditional test\and Design and analysis of experiments\and Exact test\and  Markov chains\and Sequential experiments\and Unconditional test}

\section{Introduction}
Product development and innovation in various industries has become dominated by the use of
randomized experiments as a reliable method of analysis.
Introduction and widespread adoption of 
randomized experiments revolutionized the field of medical research several decades ago in the form of randomized controlled trials~\citep{bhatt2010evolution} and the field of digital marketing as A/B tests in the past decade~\citep{kohavi2020trustworthy}. 

In more recent years, a \emph{sequential} approach to randomized experiments has gained increased popularity, in particular for the design of clinical trials in medical research. Such experiments include \emph{response-adaptive}~(RA) designs, which date back to~\citet{thompson1933likelihood}, and which sequentially adjust the allocation of participants to treatments based on the history of participant outcomes and treatment allocations (\emph{trial history}). A wide variety of RA designs has been developed, able to achieve different objectives, such as increasing the expected participant outcomes, statistical power, or a combination of both~\citep{villar2015multi,robertson2023response}.

In this paper, we refer to a \emph{design} as the complete mathematical description of a clinical trial, which includes the 
number of treatments, endpoint type~(assumed binary in this paper),  allocation procedure, interim decision points (if any beyond final), and statistical test(s). What distinguishes an RA design from a non-RA design is primarily the nature of the allocation procedure used to allocate participants to treatments. An allocation procedure is referred to as an \emph{RA~procedure} if at least one participant allocation depends on the outcomes history, and a design as an RA design when its allocation procedure is RA; otherwise, we call it a non-RA procedure and correspondingly a non-RA design.
Further, we distinguish RA~procedures where the probability of allocating a participant to either treatment is strictly between~0 and~1 for every participant and all possible (finite) trial histories, which we call \emph{response-adaptive randomization}~(RAR) procedures, from RA~procedures where this probability is zero or one for at least one participant, one treatment and one~(finite) trial history, which we call \emph{deterministic response-adaptive}~(DRA) procedures. Note that any procedure that can pause or stop the allocation to a particular treatment (e.g., arm-dropping in a multi-arm trial) is thus considered a DRA procedure as the allocation probability for one of the treatments is zero in these cases.

There is a growing number of applications of RA~designs across both exploratory and confirmatory clinical trials. The uptake of DRA~procedures is growing, predominantly its use in exploratory trials, e.g., in dose-finding designs~\citep{OQuigley1990CRM}, arm-dropping designs~\citep{FDA2019guidance} or arm-adding designs~\citep{FDA2023guidance}. For an example of a DRA procedure applied to a dose-finding trial, see, e.g.,~\citet{PIPAHtrial}.
RAR procedures have been implemented in exploratory or seamless multi-arm trials, where the majority of procedures are tuned Bayesian RAR procedures~\citep[see, e.g.,][]{viele_berry_comment, Pin2025}, while their implementation in confirmatory trials is rare due to ongoing discussions of their risks and benefits~\citep[see, e.g.,][for an overview]{robertson2023response}. A regulatory guidance on adaptive designs~\citet{FDA2019guidance}, encouraged consideration of multiple design options including RA~designs, while mentioning the main arguments and controversies surrounding the use of RA~designs.

A proposal for a clinical trial using a~(response-)adaptive design tends to be reviewed with greater scrutiny than a conventional design, partially due to concerns around type~I~error rate inflation~\citep{bhatt2016adaptive}. As the~(unconditional) distribution of the outcomes changes by adopting an RA design instead of a non-RA design, ignoring the design in the analysis can lead to type~I~error rate inflation~(e.g., in~\autoref{sect:results_numerical} we show that Fisher's exact test induces type I error rate inflation for a specific RA design). Several likelihood-based tests~(e.g., the Wald, score, or likelihood ratio tests) have been shown to asymptotically control type I errors for a special class of RA designs, see, e.g.,~\citet{baldi2018wald,baldi2022simple}.
The current paper focuses on type I error control in finite samples, which is not generally attained by such tests, not even under non-RA designs~(see, e.g.,~\autoref{results_extra_RDP}).
Additional pitfalls occur for asymptotic tests when using an RA~procedure that induces relatively large imbalances, such as a DRA~procedure targeting higher expected participant outcomes, see, e.g.,~\citet{baldi2022simple}.  

The current practice for 
controlling type I errors (in finite samples) under an RA~design consists of two approaches: simulation-based~tests~\citep[see, e.g.,][Section 3.1.1 and Appendix B]{smith2018}
or randomization~tests~\citep[see, e.g., ][]{SIMON2011767}. % 
In a simulation-based test, to target type~I~error control, the critical values are estimated by Monte Carlo simulation of the test statistic under one or several parameter configurations. Under the population model for clinical trials, a randomization test is a nonparametric exact test for the null hypothesis that each treatment outcome is independent of the corresponding treatment allocation conditional on the trial history up to that allocation~\citep{SIMON2011767}.
Randomization tests are robust against unmeasured confounders and time trends~\citep{villar2018} and test the sharp null hypothesis under the randomization model for clinical trials~\citep{rosenberger_randomization}.
In this paper, we define an exact test as a test that results in a type I error rate that is bounded above by the target significance level. 

Our focus is on conditional and unconditional exact tests for the parametric setting where potential outcomes are binary and have a Bernoulli distribution with unknown success rates.
Examples of conditional and unconditional tests for non-RA designs in this setting are,  respectively, Fisher's exact test~\citep{Fisher1934method} and Barnard's test~\citep{barnard1945new}.
Such tests~(henceforth referred to as \emph{exact tests}) have been less studied in the recent literature for RA~designs.
Reasons for this might be that exact tests are less straightforward to apply to RA~designs and are less computationally tractable than simulation-based tests. Furthermore, in comparison to randomization~tests,  exact tests have additional assumptions on the outcomes distribution and are less readily applicable to other outcome types and covariate-adaptive randomization. However, such tests also have important advantages.
The advantage of an exact test is the guarantee of controlling type I errors, whereas a simulation-based test, which
relies on an estimated critical value % 
is at risk of type I error rate inflation due to Monte Carlo error~\citep{robertson2023response}.
The advantage of exact tests over randomization tests is that exact tests can be unconditional or can condition only on certain summary statistics, while, in the population model for a clinical trial, a randomization test is a conditional test that conditions on the entire sequence of outcomes. In response-adaptive designs, the number of histories of allocations leading to the sequence of outcomes may be small, inducing a very discrete test with lower power, especially when using DRA procedures~\citep{villar2018}.

This paper focuses on {(finite-sample)} type~I~error control for tests on binary outcomes (treatment success/failure)  collected from a clinical trial with a control and treatment group (arms) using any RA~procedure.  The focus on binary outcomes is justified by the fact that most response-adaptive designs both in the literature and in practice consider binary primary outcomes. The contributions of the paper are as follows:
(1)~We generalize the Markov chain of summary statistics first introduced in~\citet{wei1990statistical} to allow for, e.g., known fixed delays, batched designs, and optional stopping at interim analyses. (2)~We generalize~\citet[eq. (1)]{yi2013exact} to such Markov chains and to outcomes coming from a finite exchangeable sequence, showing the generality of this formula. (3)~Drawing on pioneering statistical developments such as~\citet{Fisher1934method} and~\citet{ barnard1945new},
we use the generalization of~\citet[eq. (1)]{yi2013exact} to develop an algorithm and provide code to construct and evaluate conditional and unconditional exact tests for trials with binary outcomes that use a, possibly deterministic, RA~procedure.
While~\citet{wei1990statistical} first proposed the idea of conditional and unconditional exact tests for RA~designs, 
the trial size considered 
in that paper was at most 
$20$ participants, which is an unrealistically small size for two-armed confirmatory clinical trials.

In contrast, the current paper uses the computational developments presented in~\citet{jacko2019binarybandit} to calculate the policy-dependent coefficients in the generalized version of~\citet[eq. (1)]{yi2013exact}. While~\citet{yi2013exact} considered a maximum of 85 participants, our method allows us to generate results for % 
trials with up to 960~participants computed on a standard computer. (4)~We present a computational study where a conditional exact test is shown to have higher power for rejecting the null of no treatment effect in comparison to the unconditional exact approach.
(5)~We illustrate the applicability of the proposed methodology for constructing exact tests through two real examples, one of which is an RA design using a DRA procedure and the other one involves optional stopping at interim analyses.

The paper is structured as follows:~\autoref{sect:lit_exact} summarizes relevant literature on exact tests for binary outcomes in two-arm trials,~\autoref{sect:model_methods} introduces the model and methods to construct exact tests based on binary outcomes collected under an RA~procedure,~\autoref{sect:results_numerical} provides results of an illustrative computational study comparing the rejection rates (i.e., type~I~error rate and power) for different exact tests and statistics under the randomized dynamic programming RA procedure,~\autoref{sect:applications} provides the results of two real-world applications of the test procedures in two case studies, and~\autoref{discussion} concludes the paper and indicates topics of future research. 

\section{Conditional and unconditional exact tests }\label{sect:lit_exact}

In this section, we summarize relevant literature on \emph{conditional exact}~(CX) and \emph{unconditional exact}~(UX) tests for the null of no treatment effect based on two-arm trials with binary outcomes.

CX tests were introduced for non-RA designs in~\citet{Fisher1934method} and an exact test for binary outcomes conditioning on the total amount of successes and allocations to both treatment groups is often referred to as~\emph{Fisher's exact test}~(FET).
In a CX test,  the critical value is determined using the conditional null distribution of the sufficient statistic for the parameter of interest (e.g., the treatment effect) where conditioning is done on the sufficient statistics for the nuisance parameters (e.g., the probability of success for the control treatment)~\citep[Section 3.5.7]{agresti_book_cda}. The UX test for binary outcomes, which uses a critical value that bounds the maximum type~I~error rate for all parameter values under the null hypothesis from above by the significance level, was introduced for non-RA designs in~\citet{barnard1945new}.

A discussion on CX versus UX tests for binary outcomes under a non-RA design was given in~\citet[Section~3.5.6]{agresti_book_cda}.
The arguments for a CX or UX test are both practical and philosophical. 
A philosophical argument mentioned in~\citet{agresti_book_cda} considers the sampling model under the null. If the total sum of treatment successes is assumed to be a random variable, then a UX test is reasonable. 
This assumption might be justified when the trial participants can be considered as a random sample from the set of people eligible for the trial.
In rare disease trials, the trial contains a large proportion % 
of the people eligible for inclusion in the trial.
In the latter case, it could be more natural to assume that, under the null, the total sum of treatment successes in the trial is fixed. 
Conditional tests might be more appropriate in such settings, as they control type~I~errors for models with both a random and fixed sum of successes~(see, e.g.,~\autoref{cor:cond_exact}).
Another philosophical argument in favour of a CX test over a UX test is the notion 
that to construct a reference set for the candidate test one should not include situations with severely less (or more) informative allocations than those observed~\citep{Fisher1945new}.

To the best of our knowledge, the first source discussing CX 
and UX tests for RA~designs was~\citet{wei1990statistical}, proposing conditioning on both the total number of successes and the numbers of allocations to each arm at the end of the trial.
The 
philosophical 
argument mentioned above (on the use of less informative allocations to construct p-values) was also made for RA~designs in~\citet{begg1990inferences}
with a recommendation to condition on the number of allocations to each arm in a CX test. 
In~\citet{fleming_discussion}, the issue was raised that such conditioning might lead to a loss of information on the treatment effect, and the suggestion was made to only use the total number of successes. 

Two practical arguments given in~\citet{agresti_book_cda} are that UX tests are less discrete than CX tests for non-RA designs and have more power as a result, while CX tests are less computationally intensive. However, these arguments do not apply in general for binary outcomes in RA~designs. The power comparison heavily depends on the RA procedure and scenario considered. 
We will also show in~\autoref{sect:results_numerical} that there are situations where a CX test has higher power than a UX test. 

The computational complexity of both CX and UX tests and randomization tests remains to be perceived as prohibitive in realistic trial sizes (i.e., exact computation for these approaches may quickly become intractable).~\citet{SIMON2011767} and~\citet{villar2018}  approximate the exact critical values of the randomization test, which can be viewed as a CX test, due to computational intractability for a trial size of~100~participants while~\citet{begg1990inferences} considered CX tests for a trial with a size of 12~participants.
However, computational cost is not a practical issue today and in many cases, randomization tests can be relatively well approximated by simulation. We illustrate the computational tractability of exact tests in \autoref{sect:results_numerical} where we provide and evaluate critical values for UX tests for trial sizes of 960 participants.

\section{Exact analysis for response-adaptive procedures} \label{sect:model_methods}

This section presents the theoretical results of this paper.~\autoref{sect:model} introduces the model and notation for a two-arm trial with binary outcomes using an RA~procedure.
In~\autoref{Sect:MarkovChain}, the evolution of the trial history, describing the accrual of information in the trial, is summarized by a lower-dimensional Markov chain generalizing the Markov chain in~\citet{wei1990statistical} to allow for complex RA designs, which is illustrated by three examples. In~\autoref{sect:likelihood} we show that this Markov chain formulation leads to a simple expression of the data likelihood, extending, e.g.,~\citet[eq. (1)]{yi2013exact}, and we show that the expression also holds when it is only guaranteed that the outcomes are exchangeable. In~\autoref{subsect:exact_tests}, we use this expression of the data likelihood to provide methods for constructing conditional and unconditional exact tests for binary outcomes collected using an RA~procedure.

Throughout the paper, calligraphic font is used for sets, Greek and lowercase letters for deterministic variables, uppercase letters for random variables, bold weight for vectors and matrices, and blackboard bold font for probability and expectation operators, as well as indicators.
For indices, we will use the calligraphic capitalized index variable to refer to the index set and the overlined index to refer to the highest element, e.g.,~$i\leq\Iend$ for all~$i\in\calI$.
Furthermore, tuples are denoted by round brackets, closed intervals by square brackets, half-closed intervals by round and square brackets, and sets by curly brackets.
Notation~\mbox{$x:y:z$} is used to denote a range of values with step-size~$y$ starting at~$x$ and ending at~$z$. We let~$\wedge$~($\vee$) denote logical \emph{and}~(\emph{or}) and define~$\min(\emptyset)=\infty$ with~$\emptyset$ the empty set. \autoref{symbol_table} contains a notation table for the paper.

\subsection{Two-arm response-adaptive design with binary data}\label{sect:model}
In this section, we define the parametric population model, which is the model in which we evaluate the considered exact tests.
Let~\mbox{$\btheta=(\theta_\C,\theta_\D)$} 
contain the (unknown) success probabilities, where~$\C$ denotes the control treatment and~$\D$ 
denotes the developmental treatment. In the following, the same convention (i.e., first~$\C$ then~$\D$) will be used to construct tuples from variables for the control and developmental treatment.
\hbox{Let~$\bY_\C = (Y_{\C,i})_{i=1}^{\Iend}$}, \hbox{$\bY_\D = (Y_{\D,i})_{i=1}^{\Iend}$} 
be two sequences of independent Bernoulli random variables, where~$\mathbb{P}_\btheta(Y_{a,i}=1)=\theta_a$ for~$a\in\{\C,\D\}$ and~$\Iend$ is a natural number denoting the 
(predetermined) maximum trial size. The random variable~$Y_{a,i}$ denotes a potential outcome for trial participant~$i$ under treatment~$a$, which comes from one of two~(i.i.d. Bernoulli) populations.

The adoption of an RA design brings forth the following additions to the population model described above. For an RA design, 
clinical trial participants~$i\in\calI=\{1,2,\dots,\Iend\}$ arrive sequentially
and each participant~$i~$ is allocated to a treatment arm~$A_i$, after which the binary response~$Y_{A_i,i}~\in~\{0,1\}$ is collected~(where~$A_i$ and~$Y_{A_i,i}$ are observed given that the trial has not stopped before the arrival of participant~$i$).
\hbox{Let~$\bH_{i} = (A_1,Y_{A_1,1}, A_2,\dots,A_i, Y_{A_i,i})$} be the  
trial history up to participant~$i\in\calI$, and let~$\bH_0=()$.
Let~$\calI_{\leq i} =\{1,\dots, i\}$ for all~$i\leq \Iend$ and~$\mathcal{H}=\bigcup_{i=0}^{\Iend}\mathcal{H}_i$ for~\mbox{$\mathcal{H}_0=\{()\}$} and $\mathcal{H}_i$ the support of~$\bH_i$, i.e.,
\begin{equation*}
\mathcal{H}_i=\{(a_1,y_{1},a_2, y_{2},\dots, a_i,y_{i}): y_{w}\in\{0,1\},\,a_w\in\{\C,\D\},\;\forall w\in\calI_{\leq i} \}.
\end{equation*}
An RA procedure is a function~$\pi:\mathcal{H}\mapsto [0,1]$, 
where it is assumed that the distribution of~$A_{i+1}$ is non-anticipating and thus only depends on~$\bH_i$, i.e., 
$\pi$ is
such that
\begin{align*}
\pi(\bH_{i}) := \mathbb{P}(A_{i+1}=\C\mid \bH_i)
= 1 - \mathbb{P}(A_{i+1}=\D\mid \bH_i) .
\end{align*}
Henceforth, the probability measure for the outcomes and allocations will be denoted by~$\mathbb{P}_\btheta^\pi$. 
We can now make the distinction between an RAR and a DRA procedure: the procedure~$\pi$ is called an RAR procedure if~\mbox{$0<\pi(\bH_{i})<1$} (almost surely) for all~$i\in\calI$, otherwise~$\pi$ is called a DRA~procedure.

An analysis of outcomes data is modelled by introducing 
a real-valued test statistic~$\Tau_\mathcal{H}$  and rejection region~$\mathcal{R}_\mathcal{H}$ taking as input the trial history~$\bH_i$ up to a participant~$i$. The null hypothesis is rejected and the trial is stopped after observing the outcome of participant~$i$ when~\mbox{$\Tau_\mathcal{H}(\bH_i)\in\mathcal{R}_\mathcal{H}(\bH_i)$.} 
The definition of the rejection region must reflect the particular null hypothesis to be tested~(e.g., two-sided or one-sided), as we will specify in~\autoref{subsect:exact_tests}. In practice, an analysis of trial data and stopping the trial early might only be possible at certain participant indices, i.e., at certain \emph{interim analysis points}. This is covered in the above model by defining the rejection region to be empty for every~$i$ that does not correspond to an interim analysis point.

\subsection{Markov chain model formulation % 
}\label{Sect:MarkovChain}
In this section, we define a Markov chain which we will use to determine the operating characteristics for a given exact test in a specific RA design.
We consider a set of \emph{update times}~$t\in \calT=\{0,1,\dots,{\Tend}\}$ for a natural number~${\Tend}$. At update time~$t\in\calT$ we have access to information of the trial up to trial participant~$i_t\in\calI_0$ where~$\calI_0=\{0,1,\dots,\Iend\}$, $i_t=0$ refers to the information before participant~$i=1$, \mbox{$ i_{t+1}\geq i_t$} for all~$t\in\calT\setminus\{{\Tend}\}$, and we \hbox{have~$i_{\Tend}=\Iend$.} We now define a stochastic process~$(\bX_t)_{t\in\calT}$, denoted in short by~$(\bX_t)_t$, which only contains information in~$(\bH_i)_{i\in\calI}$ needed for the analysis, and hence would typically have lower dimension than~$(\bH_i)_{i\in\calI}$. For all update times~$t$, 
we have~$\bX_t=x_t(\bH_{i_t})$ where~$x_t$ is a function of the history up to trial participant~$i_t$, i.e.,~$x_t:\mathcal{H}_{i_t}\mapsto \mathcal{X}_t$ for a (countable, multi-dimensional) set~$\mathcal{X}_t$. 

We assume that process~$(\bX_t)_t$ has some particular properties.  First,
we require that~$\bX_t$ contains the total number of successes and allocations to each arm~$a$ up to 
\mbox{trial participant~$i_t$}% 
, i.e., letting~$$S_{a,i} = \sum_{i'=1}^i Y_{A_{i'},i'}\mathbb{I}(A_{i'}=a)\;\;\text{ and }\;\; N_{a,i} =  \sum_{i'=1}^i \mathbb{I}(A_{i'}=a),$$ for all~$a\in\{\C,\D\}$, \mbox{$i\in\calI_0$}
be the total sum of successes and allocations, respectively, to treatment arms up to trial participant~$i$;
we \hbox{have~$S_{a,i_t}=s_a(\bX_t)$} and~$N_{a,i_t}=n_a(\bX_t)$ for functions~\mbox{$\bs:\mathcal{X}\mapsto \calI_0^2$}, \mbox{$\bn:\mathcal{X}\mapsto \calI_0^2$}, where~$\mathcal{X}=\cup_t\mathcal{X}_t$. 
Second, we require the existence of a test statistic~$\Tau$ and rejection region~$\mathcal{R}$ such that (early) rejection of the null can be determined from~$\bX_{\Tend}$, i.e., $$\Tau(\bX_{\Tend})\in\mathcal{R}(\bX_{\Tend})\iff \exists i:\Tau_\mathcal{H}(\bH_{i})\in\mathcal{R}_\mathcal{H}(\bH_{i}).$$
Third, we focus on RA~designs where~$(\bX_t)_t$ is a Markov chain 
with, for all~$\bx_{t}\in\mathcal{X}_{t},\,\bx_{t+1}\in\mathcal{X}_{t+1}$, transition structure
\begin{equation}
\hspace{-1mm}\mathbb{P}^\pi_{\btheta}(\bX_{t+1} = \bx_{t+1}|\bX_t=\bx_t) = \
q^{\pi}(\bx_t,\bx_{t+1})\cdot p_{\btheta}(\partial\bs(\bx_t,\bx_{t+1})|\partial\bn(\bx_t,\bx_{t+1})),
\label{Transdyn_state_reduced}
\end{equation}
where~$q^\pi$ is a function depending solely on the RA 
procedure and, for all states~$\bx_t\in\mathcal{X}_t$ and~$\bx_{t+1}\in\mathcal{X}_{t+1}$, we define~$$\partial \bs(\bx_{t},\bx_{t+1})=\bs(\bx_{t+1})-\bs(\bx_t)\;\;\text{   and}   \;\; \partial \bn(\bx_{t},\bx_{t+1})=\bn(\bx_{t+1})-\bn(\bx_t),$$ and, letting~$\mathcal{D}=\{-\Iend,\dots,-1,0,1,\dots,\Iend\}$  denote the range of~$\partial s_a,\partial n_a$ for all~$a\in\{\C,\D\}$ and~$\bY'_\C,\bY'_\D$ be distributed as~$\bY_\C,\bY_\D$ under~$\mathbb{P}_\btheta$ as defined above, we define
\begin{equation}
p_{\btheta}(\partial\bs'\mid\partial\bn') = \prod_{a\in\{\C,\D\}}\mathbb{P}_{\btheta}\left(\sum_{i=1}^{\partial n_a'} Y'_{a,i}=\partial s_a'\right)\;\;\;\;\forall \partial\bs',\partial\bn'\in\mathcal{D}^2\label{prob_transition_extended}.
\end{equation} 

A few things are of note considering the definitions above. The right-hand side of~\eqref{Transdyn_state_reduced} decomposes the distribution of the outcomes collected using an RA~procedure in the distribution over allocations to each arm given the RA~procedure~(represented by~$q^\pi$) times the distribution over outcomes given the allocations to each arm and the parameters~(represented by~$p_\btheta$). 
We require~$\sum_{a\in\{\C,\D\}}\partial n_a(\bx_t,\bx_{t+1})=i_{t+1}-i_t$ and hence the number of additional participants between update times is deterministic.

Section~2 of~\citet{wei1990statistical} introduced the Markov chain~$(\tilde{\bX}_t)_{t\in\calT} = (( N_{\C,t}, S_{t}, S_{\C,t})_t)_{t\in\calT}$, where~$S_t=\sum_{a\in\{\C,\D\}}S_{a,i_t}$,  describing the evolution of the allocations to the control arm, total successes, and successes in the control arm after allocating each participant~$t$ (i.e.,~$t=i$) using the randomized play-the-winner procedure.
Assuming the tests performed in the trial only use the successes and allocations per treatment group up to the point of testing, the Markov chain~$(\tilde{\bX}_t)_{t\in\calT}$  satisfies the three conditions above, hence the above setup generalizes the Markov chain introduced in~\citet{wei1990statistical},  and hence allows the analysis of more complex trials. 

In the following examples, we first present a generic Markov chain and then show how our model can be used for
known fixed delays and blocked RA procedures with early stopping. 
These examples are further extended in~\autoref{sect:applications}.

\begin{example}[Summary statistics Markov chain]~\label{Ex:suffstatMarkov}
For several RA~procedures known from the literature,  the
allocation probability is a function of the summary statistics, i.e., we have~$\pi(\bH_i)=\pi_\C((\bS_i,\bN_i))$ 
for all~\hbox{$i\in\calI$}~\citep{yi2013exact}.
In this case, the summary statistics follow a Markov chain, i.e., 
letting~$\bX_t = (\bS_t, \bN_t)$ and~$i_t=t$ for all~$t$ (i.e., we update after every participant),~$(\bX_t)_t$ is a Markov chain with~$\bX_0=((0,0),(0,0))$, state \hbox{space~$\mathcal{X}=\cup_t\mathcal{X}_t$,} where ~$$\mathcal{X}_t=\mathcal{X}^\text{SS}_t=\{((s'_\C,s'_\D),(n'_\C,n'_\D)):\bs',\bn'\in\calI_0^2, \;\bs'\leq \bn',\;\textstyle\sum_a n'_a=t\},$$  and  transition structure~\eqref{Transdyn_state_reduced} with 
$$q^{\pi}(\bx_t,\bx_{t+1})=\begin{cases}
	\pi_\C(\bx_t),\;\;\;\; &\text{if }\partial n_\C(\bx_t,\bx_{t+1})=1\text{ and }\partial n_\D(\bx_t,\bx_{t+1})=0,\\
	1-\pi_\C(\bx_t),\;\;\;\; &\text{if }\partial n_\C(\bx_t,\bx_{t+1})=0\text{ and }\partial n_\D(\bx_t,\bx_{t+1})=1,\\
	0,&\text{else,}
\end{cases}$$
for all~$\bx_t\in\mathcal{X}_{t},\;\bx_{t+1}\in\mathcal{X}_{t+1}$. 
For instance, this property holds for Thompson sampling~\citep{thompson1933likelihood} where, assuming a prior~$\mathbb{Q}$ for~$\btheta$, by Bayes'~rule for all~$\bx_t\in\mathcal{X}_t$
\begin{align}
	\pi_\C(\bx_t) &=  \mathbb{Q}(\theta_\C\geq\theta_\D\mid\bX_t=\bx_t)\nonumber\\&=\frac{\int_{\theta_\C\geq \theta_\D}\prod_{a\in\{\C,\D\}}\theta_a^{s_a(\bx_t)}(1-\theta_a)^{n_a(\bx_t)-s_a(\bx_t)}\mathbb{Q}(d\btheta)}{\int\prod_{a\in\{\C,\D\}}\theta_a^{s_a(\bx_t)}(1-\theta_a)^{n_a(\bx_t)-s_a(\bx_t)}\mathbb{Q}(d\btheta)}.\label{defTS}
\end{align}
Other examples in which~$(\bX_t)_t$ is a Markov chain are  
randomized play-the-winner~\citep[described, e.g., in][]{wei1990statistical},
the randomized dynamic programming-based procedure introduced in~\citet{cheng2007optimal}, and index-based procedures provided in~\citet{villar2015multi}.
\end{example}

\begin{example}[Responses with known fixed delays]

\autoref{Ex:suffstatMarkov} can be extended to settings where outcomes have a known fixed delay, which is captured by a delayed response in the RA procedure, i.e. for a known fixed delay \hbox{length~$d\in\{1,2,\dots,\Iend\}$} 
and initial allocation \mbox{probability~$p_{\C,0}\in[0,1]$} such that~$\pi(())=p_{\C,0}$ we have
\begin{align*}
	\pi(\bH_i) &= p_{\C,0},\;\;\;\; &&\forall i\in\{1,2,\dots, d\},\\
	\pi(\bH_i) &= \pi_\C((\bS_{i-d},\bN_{i-d})),\;\;\;\; &&\forall i \in\{d+1,d+2\dots,\Iend\}.
\end{align*}
Let~$i_t=t$ for all~$t$ and~$\bX_t = (\bS_{t}, \bN_{t}, \bA_{t})$ with~$\bA_t = (A_{t+1}, \dots, A_{t+d})$.
The process~$(\bX_t)_t$ is a Markov chain with state space~$\mathcal{X}=
\cup_t\mathcal{X}_t$, where~$$\mathcal{X}_t= \{((s'_\C,s'_\D),(n'_\C,n'_\D), \ba): ((s'_\C,s'_\D),(n'_\C,n'_\D))\in\mathcal{X}_t^\text{SS}, \, \ba\in\{\C,\D\}^d\},$$ initial state~$\bX_0=((0,0),(0,0), \bA_{0})$ (where, independently,~$\mathbb{P}(A_{0,d'}=\C)=1-\mathbb{P}(A_{0,d'}=\D) = p_{\C,0}$ for all~$d'\leq d$) and,
letting the function
~$\ba$ be such that~$\ba(\bX_t)=\bA_t$ and~$\iota(\bx_{t+1})=\mathbb{I}(a_{d}(\bx_{t+1})=\C)$,
transition structure~\eqref{Transdyn_state_reduced} where, in case~$\partial n_{a}(\bx_t,\bx_{t+1})=\mathbb{I}(a = a_{1}(\bx_t))$
for \mbox{$a\in\{\C, \D\}$} and~$a_{d'}(\bx_{t+1})=a_{d'+1}(\bx_{t})$
for \mbox{all~$d'\in\{1,\dots, d-1\}$}, we have
$$q^{\pi}(\bx_t,\bx_{t+1})= \pi_\C((\bs(\bx_t),\bn(\bx_t)))^{\iota(\bx_{t+1})}(1-\pi_\C((\bs(\bx_t),\bn(\bx_t))))^{1-\iota(\bx_{t+1})},$$  and~$q^{\pi}(\bx_t,\bx_{t+1})=0$ otherwise. Hence, we add arm~$\C$ to the top of the stack~$\bA_{t+1}$ with probability equal to~$\pi_\C((\bS_t,\bN_t))$, while the next participant is allocated by the treatment at the start of the stack~(based on~$\bH_{\max(0,t-d)}$).
\end{example}

\begin{example}[Bayesian blocked RA design with early stopping]~\label{example:groupedRAR}
Allocation is sometimes performed in groups of participants, where groups of more than one participant are allocated treatment to target an allocation probability. There are different ways to allocate participants to treatments inside groups, for instance by a biased coin, mass-weighted urn, or, most often, a (modified) permuted block design,
worked out below. Furthermore, interim analyses, in addition to an analysis at the end, are often performed to reject equality of the success probabilities when at one of the interim analyses the posterior probability of control superiority,~$\pi_\C((\bS_t,\bN_t))$ determined by~\eqref{defTS},
is smaller than or equal to~$1-\pi^*_\C((\bS_t, \bN_t))$ or larger than or equal to~$\pi^*_\C((\bS_t, \bN_t))$ 
where~$\pi^*_\C((\bS_t, \bN_t))$ is a critical value, determining the rejection region~$\mathcal{R}_\mathcal{H}$ while~$\pi_\C$ determines the test statistic~$\Tau_\mathcal{H}$~\citep[see, e.g.,][pg.~6, where~$\pi^*_\C\equiv 0.986$]{yannopoulos2020advanced}.  

Letting~$\calT_t=\{0,\dots, t\},$ \hbox{$i_0=0$} and~$i_t = \sum_{t'=1}^tb_{t'}$ where~$(b_t)_t$ is the sequence of group sizes,~$U_t=\min(t,U'_t)$ where
\begin{align*}
	U'_t =\min(\{t'\in\calT:\, &\pi_\C((\bS_{t'},\bN_{t'}))\leq 1-\pi^*_\C((\bS_{t'},\bN_{t'}))\\\text{ or }&\pi_\C((\bS_{t'},\bN_{t'}))\geq \pi^*_\C((\bS_{t'},\bN_{t'}))\}),
\end{align*}
we define~$\bX_t = (\bS_{t}, \bN_{t}, U_t)$.

Assume that allocation within blocks is done using a permuted block design. 
Then, letting~$u$ be a function such that~$u(\bX_t)=U_t$, the process~$(\bX_t)_t$ is a Markov chain with state space~$\mathcal{X}=\cup_t \mathcal{X}_t$ where~$$\mathcal{X}_t=\{((s'_\C,s'_\D),(n'_\C,n'_\D), u'):((s'_\C,s'_\D),(n'_\C,n'_\D))\in\mathcal{X}_{i_t}^\text{SS}, u'\in\calT_t\},$$  initial state equal \mbox{to~$\bX_0=((0,0),(0,0), 0)$} and  transition structure~\eqref{Transdyn_state_reduced} with the RA-dependent part~$q^\pi(\bx_{t},\bx_{t+1})~$ targeting an allocation of~$\pi_\C(\bx_t)b_{t}$ participants to the control group, in particular we have that~$q^\pi(\bx_t,\bx_{t+1})$ equals:
\begin{equation*}
	\begin{cases}
		\pi_\C(\bx_t)b_{t} - \floor{\pi_\C(\bx_t)b_{t}}, \; &\text{if }\partial \bn(\bx_{t},\bx_{t+1})=( \ceil{\pi_\C(\bx_t)b_{t}},b_t - \ceil{\pi_\C(\bx_t)b_{t}}),\\&1-\pi^*_\C(\bx_t)<\pi_\C(\bx_{t}) <\pi^*_\C(\bx_t),\\&\text{and } u(\bx_{t+1})=u(\bx_t)+1,\\[2pt]
		\ceil{\pi_\C(\bx_t)b_{t}}-\pi_\C(\bx_t)b_{t},  &\text{if }\partial \bn(\bx_{t},\bx_{t+1})=( \floor{\pi_\C(\bx_t)b_{t}},b_t - \floor{\pi_\C(\bx_t)b_{t}}),\\&1-\pi^*_\C(\bx_t)<\pi_\C(\bx_{t}) <\pi^*_\C(\bx_t),\\&\text{and } u(\bx_{t+1})=u(\bx_t)+1,\\[2pt]
		1,&\text{if }1-\pi^*_\C(\bx_t)\geq \pi_\C(\bx_{t})\text{ and }\bx_{t+1}=\bx_t,\\&\text{or }\pi^*_\C(\bx_t)\leq \pi_\C(\bx_{t})\text{ and }\bx_{t+1}=\bx_t,\\[2pt]
		0,&\text{else,}
	\end{cases} 
\end{equation*} for all~$\bx_t\in\mathcal{X}_{t},\bx_{t+1}\in\mathcal{X}_{t+1}$,   where~$\floor{\delta}=\ceil{\delta}-1$ for all~$\delta\in\mathcal{D}$, $\pi_\C(\bx_t)=\pi_\C((\bs(\bx_t), \bn(\bx_t)))$, and~$\pi^*_\C(\bx_t)$ is defined similarly for all~$\bx_t\in\mathcal{X}_t$. 
The~(random) number of participants included in the trial is captured by~$i_{u(\bX_{\Tend})}$.
Lastly,
we have~$\Tau(\bX_{\Tend})\in\mathcal{R}\iff\exists i:$\mbox{$\Tau_\mathcal{H}(\bH_i)\in\mathcal{R}_\mathcal{H}(\bH_i)$} where~$\mathcal{R}=\{1\}$ and
~$$\Tau(\bX_{\Tend})=\mathbb{I}(u(\bX_{\Tend})<{\Tend},\;\pi_\C(\bX_{\Tend})\leq 1-\pi^*_\C(\bX_{\Tend})\text{, or } \pi_\C(\bX_{\Tend})\geq \pi^*_\C(\bX_{\Tend}))),$$ i.e., the null hypothesis is rejected when the trial stops early or either treatment has a high enough posterior probability of being superior at the end.

\end{example}

\subsection{Data likelihood for a two-arm response-adaptive design}\label{sect:likelihood}

In order to construct exact tests after treatment allocation using an RA~procedure and to compute associated operating characteristics, we build upon~\autoref{sect:model} and we need the following result~(the proof can be found in~\autoref{sect:proofs}), where~$\mathcal{B}([0,1]^2)$ denotes the Borel sigma algebra on~$[0,1]^2$.

\begin{theorem}~\label{theorem:datalikelihood_Bernoulli_RAR}
Assume there is a signed measure~$\mu$ on~$([0,1]^2,\mathcal{B}([0,1]^2))$  such that for all~$\by_\C,\by_\D\in\{0,1\}^{\Iend}$ 
\begin{equation}\mathbb{P}(\bY_{a}=\by_{a}\;\forall a\in\{\C,\D\})=
	\int_{[0,1]^2}\prod_{a\in\{\C,\D\}} \theta_a^{\sum_{i=1}^{\Iend} y_{a,i}}(1-\theta_a)^{\Iend-\sum_{i=1}^{\Iend}y_{a,i}}\mu(d\btheta), \label{prob_seq_exch}\end{equation}
and that the stochastic process~$(\bX_t)_t$ has transition structure~\eqref{Transdyn_state_reduced} with the term with~$p_\btheta$ replaced by the term~$p(\partial\bs(\bx_t,\bx_{t+1})\mid\partial\bn(\bx_t,\bx_{t+1}), \bn(\bx_t), \bs(\bx_t))$  where for all~$\bs',\bn'\in\calI_0^2,\partial\bs',\partial\bn'\in\mathcal{D}^2$~$p(\partial\bs'\mid\partial\bn',\bn',\bs')$ equals
\begin{equation} \mathbb{P}\left(\sum_{i=n_a'+1}^{n_a'+\partial n_a'} Y_{a,i}=\partial s_a'\;\;\forall a\in\{\C,\D\} \;\bigg|\; \sum_{i=1}^{n_a'} Y_{a,i}= s_a'\;\;\forall a\in\{\C,\D\}\right).\label{prob_transition_extended2}
\end{equation}
Then, for all~$\bx_t\in\mathcal{X}_t$, we have 
\begin{equation}\mathbb{P}^\pi(\bX_t=\bx_t)= g_t^\pi(\bx_t)\int_{[0,1]^2}\prod_{a\in\{\C,\D\}}\theta_a^{s_{a}(\bx_t)}(1-\theta_a)^{n_{a}(\bx_t)-s_a(\bx_t)}\mu(d\btheta),\label{expression_likelihood}\end{equation}
where, letting~$\binom{n}{k}$ denote the binomial coefficient for natural numbers~$n,k$, for all~$t,$~$\bx_t\in\mathcal{X}_t$
\begin{align}g_0^\pi(\bx_0)&=1,\nonumber\\ g_t^\pi(\bx_t) &= \sum_{\bx_{t-1}\in\mathcal{X}_{t-1}}\left(\prod_{a\in\{\C,\D\}}\binom{\partial n_a(\bx_{t-1},\bx_t)}{\partial s_a(\bx_{t-1}, \bx_t)}\right)g_{t-1}^\pi(\bx_{t-1} )q^{\pi}(\bx_{t-1},\bx_t).\label{eqn:gdef}\end{align}
\end{theorem}

Following~\citet{jaynes1986_some}, the data likelihood~\eqref{prob_seq_exch} implies that the potential outcomes are participant-exchangeable, i.e.,~$(Y_{a,i})_{i=1}^{\Iend}\stackrel{d}{=}(Y_{a,\rho_{\calI}(i)})_{i=1}^{\Iend}$
for all permutations~$ \rho_\calI$ over~$\mathcal{I }$ respectively, where~$\stackrel{d}{=}$ denotes equality in distribution. 
Note that this is a nonparametric extension of the setting of~\autoref{sect:model} as the outcomes are also participant-exchangeable in each treatment group for the two-population Bernoulli outcomes model.
Furthermore,~$\mu$ is absolutely continuous with respect to the Dirac measure on~$\{\btheta:\theta_\C=\theta_\D\}$ if and only if the distribution of the total sequence of outcomes~$(Y_{a,i})_{ a\in\{\C,\D\}, \;i\in\calI}$ 
is independent of the allocations, i.e.,~$(Y_{\C,i})_{ i=1}^{\Iend}\stackrel{d}{=}(Y_{\D,i})_{i=1}^{\Iend}$, which is a nonparametric version of the null hypothesis~$\theta_\C=\theta_\D$ in the model of~\autoref{sect:model}. 
Hence,~\autoref{theorem:datalikelihood_Bernoulli_RAR} can be applied in these settings, i.e., in a finite population model assuming outcomes come from a participant-exchangeable sequence.
Furthermore, the theorem can also be applied when the outcome sequence can be extended to an infinite exchangeable sequence, in which case the measure~$\mu$ must be a probability measure by de~Finetti's~theorem. Lastly, if the outcomes are i.i.d.~Bernoulli distributed,~$\mu$ is a positive point mass on a parameter value~$\btheta\in[0,1]^2$ and we obtain Equation~(1) from~\citet{yi2013exact}
\begin{equation}
\mathbb{P}^\pi_{\btheta}(\bX_t=\bx_t)=  g_t^\pi(\bx_t)\prod_{a\in\{\C,\D\}}\theta_a^{s_a(\bx_t)}(1-\theta_a)^{n_{a}(\bx_t)-s_a(\bx_t)}\;\;\;\;\forall \bx_t\in\mathcal{X}.\label{eqn:likelihood_bernoulli}
\end{equation}

All settings described above assume that the outcomes collected in a clinical trial come from a population model, and hence not a randomization model~\citep{rosenberger_randomization}. See~\autoref{diff_trial_models} for an overview of the differences between the randomization model, parametric population model, and nonparametric population model for the potential outcomes.

We now give an application of~\autoref{theorem:datalikelihood_Bernoulli_RAR}.
\begin{Remark} \label{gcoef_ER}
Assuming i.i.d. Bernoulli distributed outcomes and the complete randomization non-RA design where~$\pi_\C(\bx)=1/2$ for all~$\bx\in\mathcal{X}$, it can be verified that~$g_t^\pi(\bx_t) = \binom{t}{n_\C(\bx_t)}\binom{n_\C(\bx_t)}{s_\C(\bx_t)}\binom{n_\D(\bx_t)}{s_\D(\bx_t)}/2^{t}$, hence, for all~$\bx_t\in\mathcal{X}$
\begin{equation}
	\mathbb{P}^\pi_{\btheta}(\bX_t=\bx_t) = \frac{\binom{t}{n_\C(\bx_t)}}{2^{t}}\prod_{a\in\{\C,\D\}}\binom{n_a(\bx_t)}{s_a(\bx_t)}\theta_a^{s_{a}(\bx_t)}(1-\theta_a)^{n_{a}(\bx_t)-s_a(\bx_t)}.\label{example_g_ER}
\end{equation}
\end{Remark}

\FloatBarrier
\begin{table}[tbp]
\caption{Differences between the parametric population model, nonparametric population model, and randomization model for the potential outcomes in a clinical trial.}\label{diff_trial_models}
\small

\hspace{5mm}
\renewcommand{\arraystretch}{2}
\begin{tabular}{L{19mm}L{31mm}L{45mm}L{41mm}}
	\hline
	& \textbf{Parametric\newline population model\newline (\autoref{sect:model})}                                                                       & \textbf{Nonparametric\newline population model}                                                                                                                                  & \textbf{Randomization model}                                                                                            \\ \hline
	\textbf{Potential outcomes}                                                                  &$Y_{\C,i}\sim\text{Bern}(\theta_\C)$ \newline $Y_{\D,i}\sim \text{Bern}(\theta_\D)$\newline i.i.d. for all~$i$&$(\bY_{\C}, \bY_{\D})\sim \mathbb{P}$
	\newline$(Y_{a,i})_{i=1}^{\Iend} \stackrel{d}{=}(Y_{a, \rho_\calI(i)})_{i=1}^{\Iend}$ \newline for all permutations~$\rho_{\calI}$ on~$\calI$ & Unknown, fixed \newline$y_{\C,1},\dots, y_{\C,\Iend}$ \newline$y_{\D,1},\dots, y_{\D,\Iend}$                          \\
	\textbf{Null\newline hypothesis}                                                             & Parametric null~$H_0$:\newline$\theta_{\C}=\theta_{\D}$                                                                                                        & Nonparametric null~$H_0^\text{NP}$:\newline$(Y_{\C,i})_{i=1}^{\Iend} \stackrel{d}{=}(Y_{\D, i})_{i=1}^{\Iend}$                                                                                     & Sharp null~$H_0^\text{sharp}:$\newline$y_{\C,i}=y_{\D,i}\;\forall i$                                                      \\
	\textbf{Reference set}                                                                       & One reference set, equal to state space~$\mathcal{X}$ containing treatment group sizes and successes per arm                                                        & Several conditional reference sets~$\mathcal{X}_{\zeta}(z)$ for each data summary~$z\in\mathcal{Z}$
	& One reference set: all possible allocations given RA procedure and fixed outcomes,  e.g., for DRA procedures such as PTW contains two possible paths \\
	{\bf Probability\newline over states} & Distribution over states follows from joint probability measure~$\mathbb{P}_{\btheta}^\pi$                                              & Conditional probability $$g_{{\Tend}}^\pi(\bx_{\Tend})/\sum_{\bx_{\Tend}\in\mathcal{X}_\zeta(z)} g_{{\Tend}}^\pi(\bx_{\Tend})$$ of state~$\bx_{\Tend}$ given~$\zeta=z$   & Sum of probabilities of allocation sequences leading to the state,  given outcomes                                                    \\
	\textbf{Critical values}                                                               & One (unconditional exact) minimax critical value, ensuring type I error control for all \newline parameters under null hypothesis                                                 & Several (conditional exact) critical values, one for each conditional reference set, determined by conditional probability distribution                         & One (unconditional exact) critical value, determined by probability distribution over allocations                       \\
	\textbf{Test~
		under fixed equal treatment groups}                                           & Barnard's test                                                                                                                                                & Fisher's exact test (for test statistic depending on~$S_{\C,\Iend}$)                                                                                              & Randomization test~(in general)                                                                                 \\
	\textbf{Two-sided \newline alternative\newline hypothesis}                                    &$H_1:\theta_{\C}\neq\theta_{\D}$                                                                                                                                 &$H^\text{NP}_1:(Y_{\C,i})_{i=1}^{\Iend} \stackrel{d}{\neq}(Y_{\D,i})_{i=1}^{\Iend}$\newline$\stackrel{d}{\neq}$ denotes inequality in distribution                                                          &$H'_1:(y_{\C,i})_{i=1}^{\Iend}\neq (y_{\D,i})_{i=1}^{\Iend}$                                                                                    \\ \hline

\end{tabular}\\

\end{table}

\FloatBarrier

\noindent 
The recursive definition~\eqref{eqn:gdef} leads to an efficient implementation of a forward recursion algorithm to compute~$g_t^\pi(\bx_t)$ for all states~$\bx_t\in\mathcal{X}$. This algorithm is based on the same principles for efficient calculation and storage of value functions and policies as those outlined in~\citet{jacko2019binarybandit}, i.e., uses the conservation law for the states, a storage mapping function, and overwrites values that are not used further in the algorithm.
Equation~\eqref{eqn:likelihood_bernoulli} states that for the setting described above, the probability of reaching a state~$\bx_t$ can be decomposed as the product of a coefficient~$g_t^\pi(\bx_t)$, corresponding to the uncertainty in reaching~$\bx_t$ due to the allocation~procedure, and a term proportional to the likelihood of the outcomes data conditioned on the allocations, independent of the allocation~procedure. In particular, it means that if one has access to~$g_t^\pi$, computation of expectations with respect to~$\bX_t$ under different values of~$\btheta$ comes down to taking the inner product of~$g_t^\pi$ with the (scaled) likelihood under these values of~$\btheta$. This procedure can be used to calculate the rejection rate
~$\mathbb{P}^\pi_{\btheta}(\Tau(\bX_{\Tend})\in\mathcal{R}(\bX_{\Tend}))$,
making it possible to efficiently compute the exact tests in the following sections.

\subsection{Exact tests}\label{subsect:exact_tests}

In the following, we consider a test for the null hypothesis of % 
no % 
treatment effect% 
,
i.e., we test~$$H_0:\theta_\C= \theta_\D \text{ v.s. } H_1:\theta_\C\neq \theta_\D.$$ For a chosen test statistic~$\Tau$ 
and upper/lower critical values~$c_u,c_{\ell}$,
this test rejects~$H_0$ whenever~$\Tau(\bX_{\Tend})\geq c_u(\bX_{\Tend})$ or~$\Tau(\bX_{\Tend})\leq c_{\ell}(\bX_{\Tend})$.
This test can be made one-sided by setting~$c_u \equiv\infty$ or~$c_{\ell} \equiv-\infty$. 

\subsubsection{Conditional exact tests}\label{sect:RT}

In this subsection, we discuss a conditional test for RA designs extending, e.g., Fisher's exact test~\citep{Fisher1934method}.

\begin{definition}[Conditional Test] \label{def:CX_SA}
Let~$\zeta:\mathcal{X}_{\Tend}\mapsto \mathcal{Z}$ 
be a conditioning (summary) function for~$\bx_{\Tend}\in\mathcal{X}_{\Tend}$ and~$\mathcal{X}_\zeta(z)$ be the pre-image of~$z\in\mathcal{Z}$ under~$\zeta$, denoted the conditional reference set of states.
A conditional test based on~$\zeta$ for a test statistic function~$\Tau$, RA~procedure~$\pi$, and significance level~$0<\alpha<1$ rejects \mbox{when~$\Tau(\bX_{\Tend})\geq c_u(\zeta(\bX_{\Tend}))$ or~$\Tau(\bX_{\Tend})\leq c_{\ell}(\zeta(\bX_{\Tend}))$} 
where, for~$0<\alpha_u,\,\alpha_{\ell}<1$ such that \mbox{$\alpha_u+\alpha_{\ell}=\alpha$}, we have for all~$z\in\mathcal{Z}$
\begin{align}
	c_u(z)=\min&\left\{c\in \Tau^+(\mathcal{X}_\zeta(z)):\frac{\sum_{\bx_{\Tend}\in\mathcal{X}_\zeta(z): \;\Tau(\bx_{\Tend})\geq c} g_{{\Tend}}^\pi(\bx_{\Tend})}{\sum_{\bx_{\Tend}\in\mathcal{X}_\zeta(z)} g_{{\Tend}}^\pi(\bx_{\Tend})}\leq \alpha_u\right\},\label{critval_cond_upper}\\
	c_{\ell}(z)=\max&\left\{c\in \Tau^+(\mathcal{X}_\zeta(z)):\frac{\sum_{\bx_{\Tend}\in\mathcal{X}_\zeta(z):\; \Tau(\bx_{\Tend})\leq c} g_{{\Tend}}^\pi(\bx_{\Tend})}{\sum_{\bx_{\Tend}\in\mathcal{X}_\zeta(z)} g_{{\Tend}}^\pi(\bx_{\Tend})}\leq \alpha_{\ell}\right\}.\label{critval_cond_lower}
\end{align}
In the above,~$\Tau(E)$ denotes the image of~$E\subseteq\mathcal{X}_{\Tend}$ under~$\Tau$, while we define~$\Tau^+(E)=$\hbox{$\Tau(E)\cup\{-\infty,\infty\}$.}
\end{definition}

Let~$s(\bx)=\sum_{a\in\{\C,\D\}}s_a(\bx)$ for all~$\bx\in\mathcal{X}$. The next result, for which the proof can be found in~\autoref{sect:proofs}, states that if~$\zeta(\bX_{\Tend})$ contains the total number of successes, a conditional test based on~$\zeta$ is exact under the assumptions of~\autoref{theorem:datalikelihood_Bernoulli_RAR} and will be denoted the \text{CX-$\zeta$} test.

\begin{corollary}\label{cor:cond_exact}
If~$\zeta:\mathcal{X}_{\Tend}\mapsto\mathcal{Z}$,~$s(\bX_{\Tend})=\tilde{s}(\zeta(\bX_{\Tend}))$ for a 
function~$\tilde{s}:\mathcal{Z}\mapsto \calI_0$ and the assumptions of~\autoref{theorem:datalikelihood_Bernoulli_RAR} hold 
then under the nonparametric null hypothesis~$H_0^\text{NP}$ in~\autoref{diff_trial_models} we have~$$\mathbb{P}^\pi_{\btheta}\Big(\Tau(\bX_{\Tend})\geq c_u(\zeta(\bX_{\Tend}))\text{ or  }\Tau(\bX_{\Tend})\leq c_{\ell}(\zeta(\bX_{\Tend}))\Big)\leq \alpha.$$
\end{corollary}

Let~$\text{SA}(\bx_t) = (s(\bx_t),n_\C(\bx_t))$ and~$\text{S}(\bx_t)=s(\bx_t)$, according to~\autoref{cor:cond_exact} the \emph{conditional exact test based on total successes and allocations}~(CX-SA test) and \emph{conditional exact test based on total successes}~(CX-S test) are conditional exact tests under the assumptions of~\autoref{theorem:datalikelihood_Bernoulli_RAR}, hence the nonparametric null hypothesis in~\autoref{diff_trial_models}.
Note that the CX-S test is the conditional exact test only conditioning on total successes, i.e., a single margin of the 2$\times$2 contingency table,  agreeing with the test suggested by a reviewer of~\citet{begg1990inferences}. The CX-SA test conditions on both margins of the 2$\times$2 contingency table, and was argued for from a philosophical standpoint in~\citet{begg1990inferences}~(see also~\autoref{sect:lit_exact}).

The next remark shows that~\autoref{cor:cond_exact} recovers the result that FET is exact under the nonparametric null hypothesis and the complete randomization non-RA design~\citep{berger2021roadmap}.

\begin{Remark} \label{example_FET}
In the setting of~\autoref{gcoef_ER}, it follows from~\eqref{example_g_ER} that for all pairs~$(s',n'_\C)\in\calI_0^2$ of successes and allocations, we have that~$c_{\ell}((s',n'_\C))$ equals
\begin{align*}
	&\max\left\{c\in 
	\Tau^+(\mathcal{X}_\text{SA}
	((s',n'_\C))):\sum_{\bx_{\Tend}\in\mathcal{X}_\text{SA}((s',n'_\C)): \;\Tau(\bx_{\Tend})\leq c} 
	p_{\text{FET}}(\bx_{\Tend})\leq \alpha_{\ell}\right\},
\end{align*}
where the conditional probability of state~$\bx_{\Tend}$ is given by\begin{equation}
	p_\text{FET}(\bx_{\Tend})=\frac{\binom{n_\C(\bx_{\Tend})}{s_\C(\bx_{\Tend})}\binom{n_\D(\bx_{\Tend})}{s_\D(\bx_{\Tend})}}{\sum_{\bx'_{\Tend}\in\mathcal{X}_{\text{SA}}((s(\bx_{\Tend}), n_\C(\bx_{\Tend})))} \binom{n_\C(\bx'_{\Tend})}{s_\C(\bx'_{\Tend})}\binom{n_\D(\bx'_{\Tend})}{s_\D(\bx'_{\Tend})}}.
\end{equation} The upper critical value~$c_u(\bX_{\Tend})$ is defined similarly.
Hence, \hbox{letting~$\alpha_u=0$}, \hbox{$\alpha_{\ell}=\alpha$} and choosing the test statistic equal to the conditional probability of seeing a state more extreme than~$\bx_{\Tend}$ (in terms of~$p_{\text{FET}}$) \begin{equation}
	T_\text{FET}(\bx_{\Tend}) = \sum_{\substack{\bx_{\Tend}'\in\mathcal{X}_\text{SA}((s(\bx_{\Tend}), n_\C(\bx_{\Tend})))\\\;p_\text{FET}(\bx'_{\Tend})\leq p_\text{FET}(\bx_{\Tend})}}p_\text{FET}(\bx'_{\Tend})\label{eq:Stat_FET},
\end{equation}
we have that~$c_u((s(\bx_{\Tend}), n_\C(\bx_{\Tend})))=\infty,\; c_{\ell}((s(\bx_{\Tend}), n_\C(\bx_{\Tend})))\in[0,\alpha]$, and the CX-SA test  corresponds to 
the two-sided FET given in~\citet[Section 3.5.3]{agresti_book_cda}.
\end{Remark}

\subsubsection{Unconditional exact test}

In this subsection we discuss an unconditional test for RA designs, generalizing Barnard's test~\citep{barnard1945new}.
\begin{definition}[Unconditional Test]~\label{def:uncond_test}
An unconditional test for test statistic function~$\Tau$,
RA~procedure~$\pi$, and significance level~$0<\alpha<1$ 
rejects the null hypothesis when~$\Tau(\bX_{\Tend})\geq c_u$ or
when~$\Tau(\bX_{\Tend})\leq c_{\ell}$ where, for significance levels~$0<\alpha_u,\,\alpha_{\ell}<1$ such 
that~$\alpha_u+\alpha_{\ell}=\alpha$, we have
\begin{align}
	c_u&=\min\left\{c\in \Tau^+(\mathcal{X}_{\Tend}):\max_{\substack{\btheta\in[0,1]^2,\\\theta_\C=\theta_\D}}\sum_{\bx_{\Tend}\in\mathcal{X}_{\Tend}:\,\Tau(\bx_{\Tend})\geq c}\mathbb{P}^\pi_{\btheta}(\bX_{\Tend}=\bx_{\Tend})\leq \alpha_u\right\},\nonumber\\
	c_{\ell}&=\max\left\{c\in \Tau^+(\mathcal{X}_{\Tend}):\max_{\substack{\btheta\in[0,1]^2,\\\theta_\C=\theta_\D}}\sum_{\bx_{\Tend}\in\mathcal{X}_{\Tend}:\,\Tau(\bx_{\Tend})\leq c}\mathbb{P}^\pi_{\btheta}(\bX_{\Tend}=\bx_{\Tend})\leq \alpha_{\ell}\right\},\nonumber
\end{align}
with~$\mathbb{P}^\pi_{\btheta}(\bX_{\Tend}=\bx_{\Tend})$ given as in~\eqref{eqn:likelihood_bernoulli}.
\end{definition}

The next result, for which the proof can be found in~\autoref{sect:proofs}, states that the unconditional test is exact, i.e., is a UX test, when the assumptions of~\autoref{theorem:datalikelihood_Bernoulli_RAR} are met and~$\mu$ is a probability measure. We note that this is a specific result for the null hypothesis~$H_0^\text{NP}$, while in general, for different null hypotheses, the UX test is exact only under the parametric population model in~\autoref{diff_trial_models}.

\begin{corollary}\label{cor:uncond_exact}
If the assumptions of 
~\autoref{theorem:datalikelihood_Bernoulli_RAR} hold with~$\mu$ a probability measure, then under~$H_0^\text{NP}$ it holds \mbox{that 
	~$\mathbb{P}^\pi_{\btheta}\Big(\Tau(\bX_{\Tend})\geq c_u\text{ or }\Tau(\bX_{\Tend})\leq c_{\ell}\Big)\leq \alpha,$}
where~$c_u,\,c_{\ell}$ are as given in~\autoref{def:uncond_test}.
\end{corollary}

No simple formula such as~\eqref{critval_cond_upper} exists for exact calculation of the critical values for a UX test, hence one needs to resort to computational techniques and bound the critical values from above or below respectively. We propose using~\autoref{alg_unconditional} for this, which uses the Lipschitz property of the rejection rate function. 

\section{Analysis of randomized dynamic programming}\label{sect:results_numerical}

\noindent This section evaluates the rejection rate for several tests under the \emph{randomized dynamic programming}~(RDP) RA procedure introduced in~\citet{cheng2007optimal} as a way to optimally  % 
balance statistical 
and participant benefit considerations. 
\citet{williamson2017bayesian} suggested that a good balance is achieved by using the degree of randomization~$0.9$, % 
which however may still result in a large imbalance in the final allocations.
The decision to declare a treatment as superior was implemented as a Bayesian decision rule in~\citet{cheng2007optimal} 
while~\citet{williamson2017bayesian} used the Fisher's exact test.
The Markov chain modelling the trial information necessary for statistical testing is the summary statistics Markov chain introduced in~\autoref{Ex:suffstatMarkov}.

In this section, we focus on the Wald statistic for the simple difference of the unknown parameters, 
as the Wald test was shown to be first-order efficient for specific RA designs and was furthermore shown to outperform other likelihood-based tests in RA designs with small samples~\citep{baldi2022simple}.
For all~$\bx_{\Tend}\in\mathcal{X}_{\Tend}$ we define the (adjusted) Wald statistic as 
\begin{equation}\Tau_\text{WS}(\bx_{\Tend}) = \frac{\hat{\theta}_\D(\bx_{\Tend})-\hat{\theta}_\C(\bx_{\Tend})}{\sqrt{\frac{\hat{\theta}_\C(\bx_{\Tend})(1-\hat{\theta}_\C(\bx_{\Tend}))}{\tilde{n}_{\C}(\bx_{\Tend})} + \frac{\hat{\theta}_\D(\bx_{\Tend})(1-\hat{\theta}_\D(\bx_{\Tend}))}{\tilde{n}_{\D}(\bx_{\Tend})}}},\label{defn:WS}\end{equation}
where~$\hat{\theta}_a(\bx_{\Tend}) = (s_{a}(\bx_{\Tend})+1)/\tilde{n}_{a}(\bx_{\Tend})$ and~$\tilde{n}_a(\bx_{\Tend})=n_a(\bx_{\Tend})+2$ for all~$a$. The adjustment of adding two observations~(one success and one failure) to each treatment group to compute the Wald statistic induces that the statistic can be computed for all possible values of~$S_{a,{\Tend}}, N_{a,{\Tend}}$ and was, e.g., suggested in~\hbox{\cite[Note 3.2]{agresti_book_cda}}. We note that for RA designs where~\hbox{$\frac{n_\C(
	\bX_t
	)}{i_t}\rightarrow \rho$} for $\rho\in(0,1)$ almost surely, the effect of this adjustment vanishes in the limit and testing with~$T_{\text{WS}}$ is asymptotically equivalent to testing under the uncorrected Wald statistic as, e.g., given in~\citet{Hu01092003}.
	We consider RA designs where, after allocation with the RDP RA procedure, either~% 
	a naive 
	FET (i.e., ignoring the RA design) or the\mbox{ CX-S, CX-SA, UX,} or asymptotic Wald tests are performed, where the asymptotic test assumes an asymptotic standard normal distribution for~$\Tau_\text{WS}(\bX_{\Tend})$.

	The maximum randomized allocation rate
	for RDP was set to~$0.9.$
	We illustrate the results for~\hbox{$\Iend\in\{60,240,960\}$} computed on a standard laptop~(our memory-efficient code in Julia programming language allows one to consider trial sizes up to around~$1,000$ on a computer with 64 GB of RAM). 
	For trial sizes~$\Iend\in\{60,240\}$~\autoref{alg_unconditional} was used to calculate the critical value for the UX test; for larger trial sizes, due to the increased computational complexity, the critical value is approximated by the critical value ensuring type I error control for null success rates~\hbox{$\theta_\C=\theta_\D\in\{0.00,0.01,\dots,0.99,1.00\}$.}  For the Wald tests, we use the upper and lower significance levels~\hbox{$\alpha_u=\alpha_{\ell}=0.025$} and for FET, we set~$\alpha_{\ell}=0.050$ and~$\alpha_u=0.0$ i.e., we reject if the outcomes data has low likelihood under the~(naive) hypergeometric null distribution.
	
\begin{figure}[tbp]
\centering
\includegraphics[ width = \textwidth]{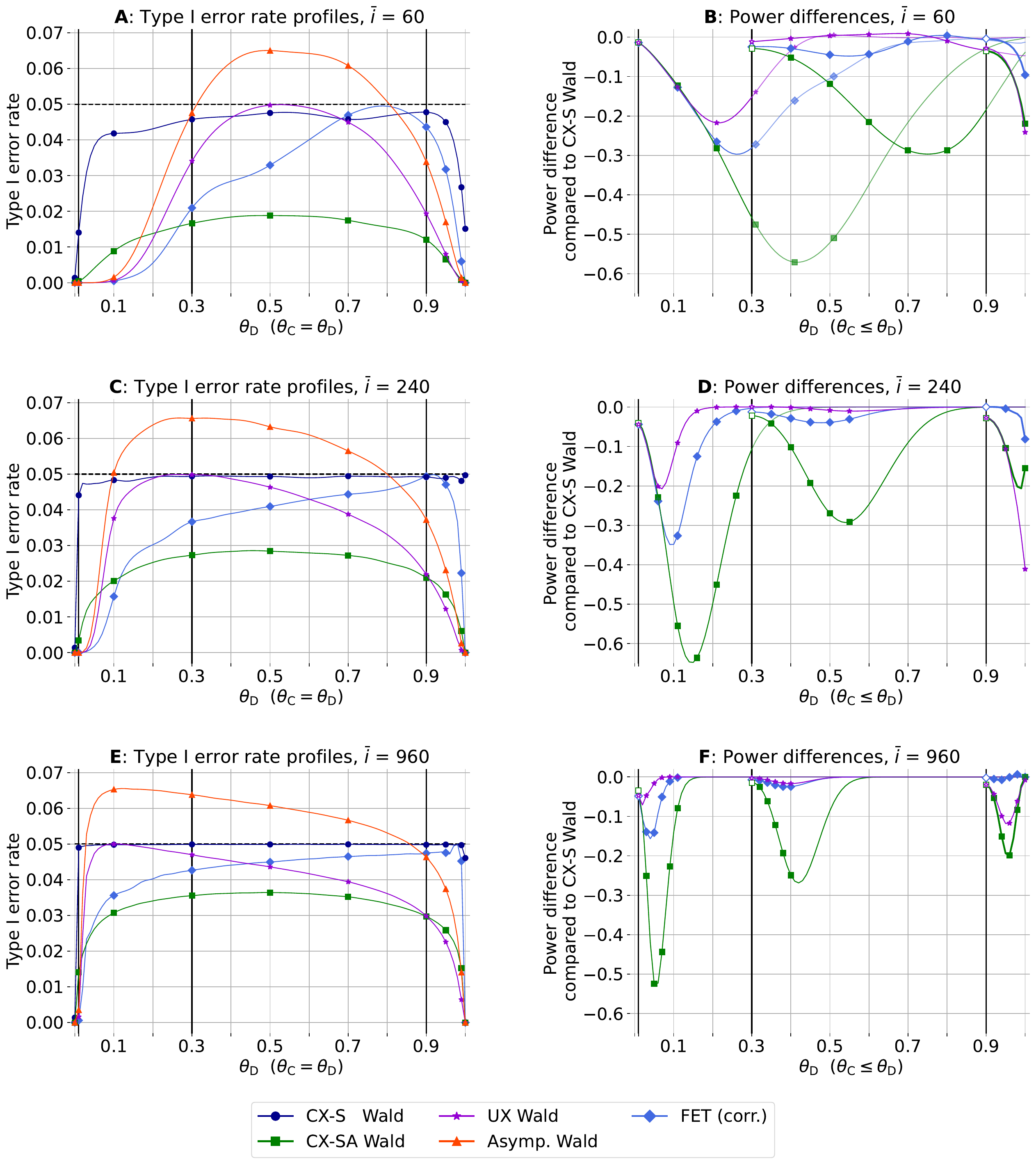}

\caption[]{{\bf Type I error rate and power difference plots for RDP RA procedure.}\par \small Subfigures A, C, E: type~I~error rate under the RDP RA procedure for the two-sided \mbox{CX-S}, CX-SA, UX, asymptotic Wald tests, and FET (corrected for type~I~error rate inflation) and different trial sizes~$\Iend$~(corresponding to rows of the figure).\par
Subfigures~B, D, F: Power difference under the RDP RA procedure for the two-sided CX-SA and UX Wald tests and FET (corr.) compared to the CX-S Wald test, \mbox{for~$\theta_\C\in\{0.01,0.3, 0.9\}$} and~$\theta_\D\geq \theta_\C$. The asymptotic Wald test~(orange line with triangle markers) is~omitted in Subfigures~B,D,F  as this test does not control type I errors.}\label{fig:comparison_tests}
\end{figure}

	Subfigures~A, C, E of~\autoref{fig:comparison_tests} show the type~I~error rates for the considered tests. The type~I~error rates for the asymptotic Wald test lie above~$0.05$ for several values of~$\theta_\D=\theta_\C$ for all values of~$\Iend$,  even for large values of~$\Iend.$ 
	This behaviour is also indicated
	by the UX critical values, which are approximately~$2.095,\;2.103,$ and~$2.100$ for~$\Iend=60,\,240,\,960$~(in comparison to the usual asymptotic critical value around 1.960) for the Wald statistic respectively. 
	While widely believed to be conservative, FET did not control type~I~errors under RDP for large success rates under the null, and hence the critical value of FET was corrected to the minimum of~$0.05$ and the UX critical value. 
	The corrected critical values for FET
	are approximately~0.050, 0.044, and 0.043 
	for~$\Iend=60,\,240,\,960$ respectively. Note that the corrected critical values for FET being below 5\% indicates that FET (at the usual 5\% level) would inflate the type I error rate.

	Subfigures A, C, and E show that the type~I~error rate of the CX-SA test lies well below~$0.05$ for each value of~$\theta_\D=\theta_\C$, reflecting the discreteness of this test. In principle, the rejection rate of this test can be increased in several ways to increase the power of the test, but in this case, the test would no longer be conditionally exact at level~$\alpha$.
	The type~I~error rate for the corrected FET is highest for high success rates for all considered trial sizes. Comparing the type~I~error rate of the CX and UX tests, the errors for the UX tests are less homogeneous over different values of the success rate, whereas the type~I~error rate for the CX-S test is almost constant and equal to~$0.05$ for~$\Iend=960.$

	Subfigures B, D, F in~\autoref{fig:comparison_tests} show the difference in power (rejection rate when~$\theta_\D\neq\theta_\C$) for the CX-SA test, UX Wald test, and corrected FET compared to the CX-S Wald test, where each curve corresponds to a control success rate~$\theta_\C~\in~\{0.01,0.3,0.9\}$~(indicated by white markers and black vertical lines) and~$\theta_\D\geq \theta_\C$, these three values of~$\theta_\C$ were chosen as they correspond to three different types of behaviour seen when considering all possible power difference curves.
	The power differences are shown instead of absolute power as we are mainly interested in which exact test has the highest power~(while absolute power values are shown in~\autoref{tab:comparison_tests_RDP_60}, \autoref{tab:comparison_tests_RDP_240}, and~\autoref{tab:comparison_tests_RDP_960}). We chose to present the differences in power with respect to the CX-S test as this test often shows highest absolute power, hence the power differences shown often have the same sign (negative) and can be more easily distinguished. The power difference for the asymptotic Wald test is omitted from Subfigures B, D, and F as this test did not control type~I~errors in Subfigures A, C, and E.

	The CX-SA  Wald test often has lower power than the other tests, and only has higher power than the UX Wald test and FET (corr.) \hbox{for~$\theta_\C = 0.01$} and small values of~$\theta_\D$. Both the UX and CX-S Wald test outperform (in terms of power) FET (corr.) 		
	\noindent when~$\theta_\C= 0.01$ and when~$\theta_\C=0.3$ for~$\theta_\D\leq 0.7$, while FET (corr.) substantially outperforms the UX Wald test but not the CX-S Wald test \hbox{when~$\theta_\C,\theta_\D\geq 0.9$.}
	Comparing the UX and CX-S Wald tests, the CX-S Wald test often shows the highest power, where the difference is most notable for smaller trial sizes and~$\theta_\C\in\{0.01, 0.9\}$.
	\autoref{fig:comparison_tests} shows that CX-S is slightly outperformed by FET (corr.) for certain values of~$\theta_\D$  \hbox{when~$\theta_\C=0.9$} and~$\Iend = 960$. 
	In~\autoref{results_extra_RDP}, we present more results for the RDP procedure, as well as an analogous analysis for and comparison to a non-RA design with equal allocation, i.e., when both treatment group sizes are (deterministically) equal to~$\Iend/2$.
\FloatBarrier
\noindent

\section{Application to real-world trials}\label{sect:applications}
\subsection{Application I: Modified play-the-winner trial}
We investigate the results in~\citet{reiertsen1993}, who analyzed the safety of administering Enoxaparin (developmental)  versus Dextran-70 (control)  during digestive surgery. 
We illustrate the advantages of our approach using this trial as it considered a moderate trial size, used a \emph{modified play-the-winner} (M-PTW) DRA procedure, and used a non-standard testing approach based on a log-rank test. 

\subsubsection{Design}
In~\citet{reiertsen1993}, a success represented the absence of any of a set of adverse events in the first week after surgery.
The trial considered~\hbox{$\Iend = 327$} participants and was designed based on a modification of the 
{play-the-winner}~(PTW) procedure.  
The PTW procedure allocates the next participant to current ``winning" \mbox{treatment~$W_i\in\{\C,\D\}$,} i.e.,~$A_{i+1}=W_i$, where~$W_0$ is chosen uniformly at random prior to assigning the first participant~(corresponding to clinical equipoise) and $W_{i}\neq W_{i-1}$ when
the allocation~$A_i$ results in a failure. 
In M-PTW,~$W_i$ furthermore switches when the same treatment is allocated 15 times in a row (denoted a cut-off), i.e., $W_{i+1}\neq W_i$ when~\mbox{$L_i=15$} where we define $L_0=1$, $L_i = L_{i-1}\cdot\mathbb{I}(W_{i}=W_{i-1}) + 1$ for~$i\geq 1$
to be the amount of subsequent allocations up to and including participant~$i+1$~(note that this is a function of~$\bH_i$). The modification of PTW described above results in a more balanced allocation to both treatments than under PTW. 
As the follow-up time was one week, a new M-PTW sequence as described above was started whenever a participant arrived and all current M-PTW sequences were awaiting a new outcome~\mbox{\citep[see, e.g.,][Fig.~2]{reiertsen1993}.} Under the model of~\autoref{sect:model_methods} these sequences can be modelled by concatenating all M-PTW sequences, forcing~$L_{i}=1$ and resampling~$W_i$ uniformly at random for certain participant indices~$i$ in the setting above, corresponding to an M-PTW sequence reaching its end at participant $i-1$.
\\
Let~$L^\prime_{1} = \min\{i\geq 1:L_{i}=1\}$ and~$$L^\prime_{k+1}=\min\{i\geq 1:L_{i+\sum_{k' = 1}^{k}L^\prime_{k'}}=1\text{ or } \textstyle\sum_{k'=1}^kL^\prime_{k'}+i=\Iend\}$$ 
be the treatment allocation sequence 
lengths 
and~$W_k^\prime=W_{\sum_{k'=1}^kL^\prime_{k'}-1 }$ 
be the treatment allocation indicators, i.e., the treatment being administered, 
during treatment allocation sequence~$k$.
After the trial was completed, a test from survival analysis, 
which we assumed 
to be the log-rank test
was performed on~$(\bm\ell^\prime,\bw^\prime,\bm \delta)$, where~$\bm \ell^\prime,\;\bw^\prime$ are the realizations of~$\bL^\prime,\;\bW^\prime$ (resp.) and~$\delta_k$ are the censoring indicators denoting whether treatment allocation sequence~$k$ was right-censored due to cut-off, 
switching the winner by misclassification of a success as a failure,
or reaching the end of an M-PTW sequence.
In~\citet{reiertsen1993} the log-rank test resulted in a p-value of~$0.05$,
while the success rates were estimated at~$0.830$  and~$0.748$ for Enoxaparin and Dextran-70 respectively. 
\subsubsection{Results}
\begin{figure}[h!]	\includegraphics[width = \linewidth]{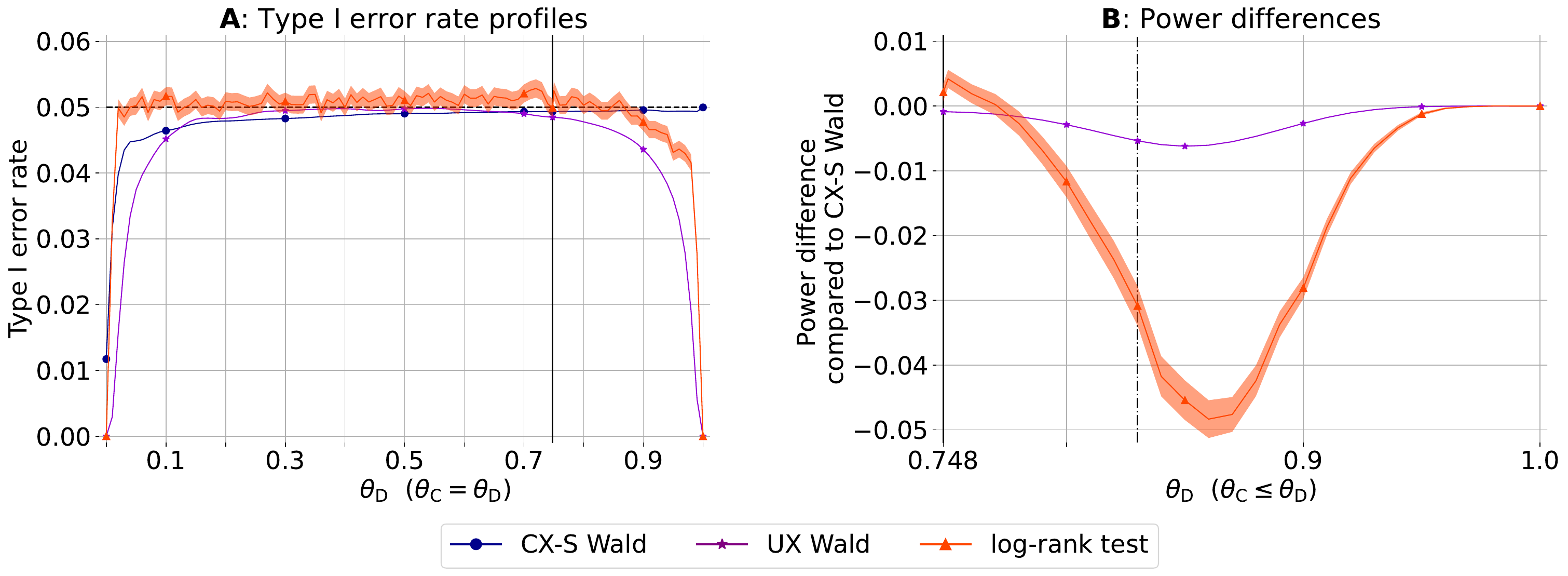}
\caption{{\bf Type I error rate and power difference plots for M-PTW trial}.\newline Subfigure A: type~I~error rate for the CX-S and UX Wald tests, as well as the log-rank test~(with 95\% simulation-based confidence interval). Subfigure B:~Power difference for~$\theta_{\C} = 0.748$ and~$\theta_{\D}\geq \theta_{\C}$ of the UX Wald and log-rank compared to the CX-S Wald test. The vertical dash-dotted line denotes~$\theta_{\D}=0.830.$ The upper and lower significance levels both equal~$2.5\%.$ In both subfigures, the vertical solid line denotes~$\theta_\C=\theta_{\D}=0.748.$ }\label{fig:comparison_M-PTW_CRDP_DB}
\end{figure}

The Markov chain modelling the M-PTW DRA procedure is the Markov chain described in~\autoref{Ex:suffstatMarkov}, augmented with~$W_i$~and~$L_i$.
The lengths of the \mbox{M-PTW} sequences were not reported in~\citet{reiertsen1993} and hence an assumption needs to be made on this part of the data~(see~\autoref{tab:comparison_tests_ex1}). 
The multiple independent \mbox{M-PTW} sequences started in the trial are modelled by restarting the \mbox{M-PTW} procedure (keeping the successes and allocations per arm) whenever the length of the current \mbox{M-PTW} reaches a value in~\autoref{tab:comparison_tests_ex1}. The transition dynamics of the Markov chain~(fully described in~\autoref{MPTW_MCform}) take the update of the winner and the treatment sequence length into account, as well as a potential cut-off of the treatment sequence length. As the frequency of misclassified outcomes was low in the trial, the choice was made not to take this possibility into account in the model.
\\
\autoref{fig:comparison_M-PTW_CRDP_DB} shows 
the type~I~error rate for the log-rank test (based on 100,000 simulations, with~95\% confidence interval based on a normal approximation), UX Wald test and CX-S Wald test, as well as the difference in power for the UX  Wald test and log-rank test over the \mbox{CX-S} Wald test for~$\theta_\C = 0.748$ and~$\theta_\D\geq \theta_\C$ based on the M-PTW procedure.
Results for the CX-S and the UX Wald test are presented, as these tests showed highest power in \autoref{sect:results_numerical}.
We evaluate the log-rank test by simulation, and not direct computation, as the log-rank statistic depends on the joint distribution of the treatment lengths, treatment values, and censoring indicators, hence directly calculating the distribution of the log-rank statistic is outside the scope of this paper.   
\\

The type~I~error rate is bounded by~$\alpha=0.05$ for the CX-S and UX Wald tests, while the estimated type~I~error rate for the log-rank test is often close to~$0.05$ (\autoref{fig:comparison_M-PTW_CRDP_DB}, Subfigure A). Empirically, for~28.4\% of the considered values of~$\theta=\theta_\C=\theta_\D$ the lower bound for the 95\% confidence interval of the rejection rate was higher than~$0.05,$ indicating that the log-rank test, being an asymptotic test, does not control  type~I~errors, although simulations indicate that the type I error rate is close to 5\% for all null success rates.
\FloatBarrier
\FloatBarrier

The power of the log-rank test is lower than the power of both the exact tests for M-PTW (\autoref{fig:comparison_M-PTW_CRDP_DB}, Subfigure B), except for values~$\theta_\D$ close \hbox{to~$\theta_\C=0.748$} (up to around~$0.77$, which might not be considered clinically relevant). Under the success rates~$\theta_\C = 0.748$ and~$\theta_\D = 0.830$ found in~\citet{reiertsen1993}, the {difference} in power for the log-rank test over the CX-S Wald test is about~$-0.03$, while for the UX Wald test it is about~$-0.005$; hence for both exact tests, the power is higher at the estimated value of~$\btheta$.
\\
In~\autoref{sect:reiertsentrial_extra}, we present more information on the model and results for the M-PTW trial, such as a justification of the log-rank test, the absolute power values under all considered designs and expected
proportion of allocations on the superior arm.

\subsection{Application II: ARREST trial}\label{arresttrial}
In this section, 
we  illustrate our approach by analyzing a recent clinical trial using a conventional RA design % 
where the batched allocation and interim analyses (i.e., to allow for early stopping) make the RA design more difficult to analyze than a fully sequential and fixed sample size RA~design.
We present results for a real-world trial that used a Bayesian RA design based on a modification of Thompson sampling~\citep{thompson1933likelihood}, and which required a tailored approach for its analysis, where we compare our proposed tests.
We consider the \emph{Advanced R$^2$Eperfusion STrategies for Refractory
Cardiac Arrest}~(ARREST) trial~\citep{yannopoulos2020advanced}, where extracorporeal membrane oxygenation
facilitated resuscitation~(developmental) was compared to
standard advanced cardiac life support (control) 
in adults who experienced an out-of-hospital
cardiac arrest and refractory ventricular fibrillation. 
\subsubsection{Design}
In~\citet{yannopoulos2020advanced}, a success represented survival to hospital discharge. Participants were allocated to treatment in groups of 30 under a permuted block design, with allocation probability to control equal to the posterior probability (based on independent uniform priors) that the control treatment is superior, restricted between~0.25 and~0.75. 
While not explicitly stated in~\citet{yannopoulos2020advanced}, it is assumed for our calculations that all outcomes are available up to each interim analysis, i.e., there is no delay, possibly through truncation of the time to hospital discharge after a given amount of days. 
If at any of the interim analyses, the posterior probability of superiority for one of the treatments became higher than an \emph{optional stopping threshold}~(OST) of~$\pi^*_\C = 0.986$, the recommendation was made to stop the trial early. This OST bounds the type~I~error rate by 0.05 based on a simulation study of 10,000 samples under the scenario~$\theta_\C=\theta_\D=0.12$. The target effect was a difference in (survival) probabilities~of~$0.25$, corresponding to the parameter configuration~$\theta_\C=0.12,\;\theta_\D = 0.37$ considered under the alternative hypothesis, which led to a power of 90\% under a maximum trial size of 150 participants. 

The ARREST trial ended with a recommendation of the extracorporeal membrane oxygenation treatment after allocating the first group of 30 participants, with a posterior probability of superiority of~0.9861. 
Using the notation of~\autoref{example:groupedRAR}, three different specifications of~$\pi^*_\C$, i.e., RA designs, are considered, the simulation-based OST~\hbox{$\pi^*_\C = 0.986$}~\citep[pg. 6]{yannopoulos2020advanced}, a UX OST calculated as~$\pi^*_\C = 0.992$, and a CX-S OST with thresholds defined by~\eqref{critval_cond_upper} and~\eqref{critval_cond_lower} with~$\alpha_{\ell}=\alpha_u=0.05/10$ (where division by 10 is due to the five two-sided tests)
according to a Bonferroni correction.

\subsubsection{Results}
Given an OST~$\pi^*_\C$, the Markov chain in~\autoref{example:groupedRAR} is used for calculating the operating characteristics for the ARREST trial with~$$\tilde{\pi}_\C(\bx_t)=\mathbb{Q}(\theta_\C>\theta_\D\mid \bX_t=\bx_t),\quad \pi_\C=\min(0.75,\max(0.25,\tilde{\pi}_\C(\bx_t))),$$  trial size~$\Iend=150$, and block size~$b_t=30$ for all~$t = \{ 1, 2, \dots, {\Tend} \}$ with~$ {\Tend} = 5$ update times.  
To calculate the UX OST, a separate Markov chain~(see~\autoref{ARREST_MC}) was used, where the state variable~$U_t$ was replaced by a state variable~$M_t$ denoting the highest value in a finite set~$\mathcal{M}
$ 
that was crossed by~$\tilde{\pi}_\C(\bX_{t'})$ for update time~$t'$ up to and including~$t$, applying~\autoref{alg_unconditional} to the distribution of the Markov chain at update time~${\Tend}$ when using~$ M_t$ as the test statistic results in a UX OST for the ARREST trial.
This approach was performed twice, first for~$$\mathcal{M}_1=\{0.5,0.6,\dots,0.9,0.95\}\cup 0.986:0.013/24:0.999,$$ and then for~$\mathcal{M}_2$ consisting of value~$0.5$ and~29 equidistant points between the UX OST  found in the first run and the highest value in~$\mathcal{M}_1$ strictly below the UX OST, leading to a critical value of approximately 0.992. % 
Note that~$0.986\in\mathcal{M}_1$ hence it is possible that the UX OST equals the simulation-based OST. 
To have deterministic error guarantees and a deterministic critical value for the CX-S tests, values of~$\tilde{\pi}_\C(\bx_t)$ were calculated as two-dimensio- nal integrals using adaptive Gaussian quadrature~(absolute tolerance~$10^{-3}$).

\begin{figure}[h!]
\centering
\includegraphics[width=\linewidth]{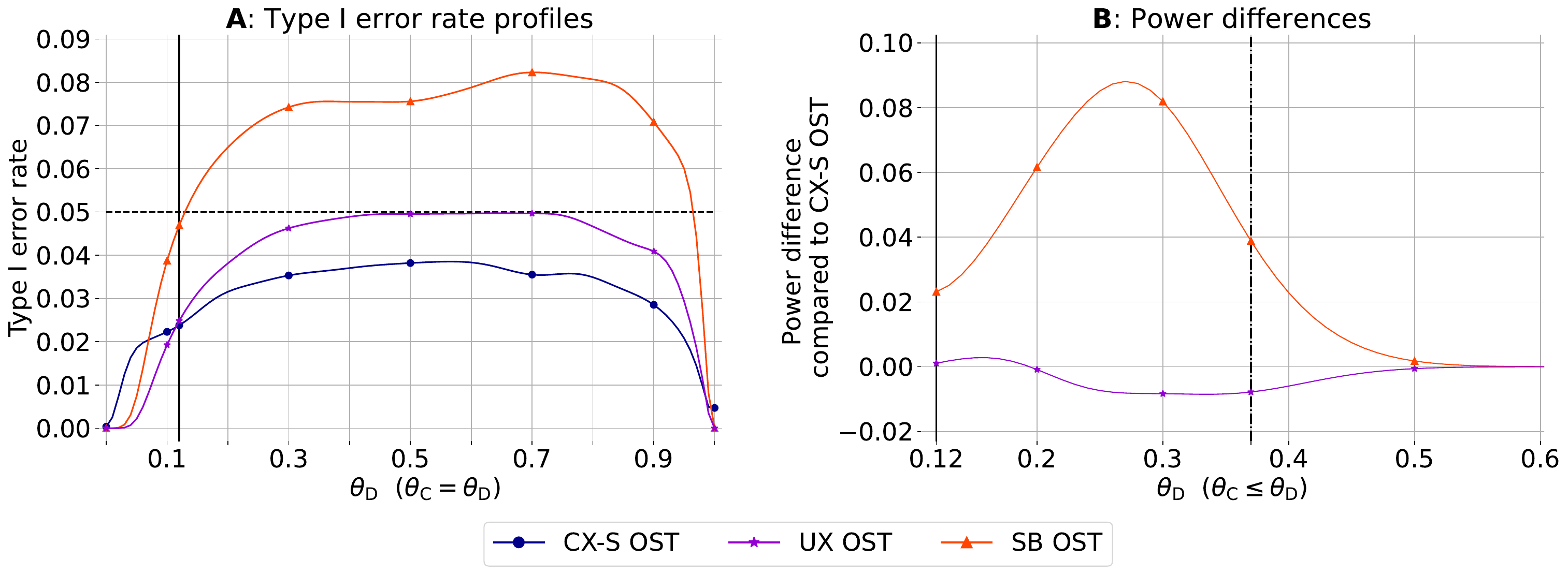}
\caption[Results for ARREST trial]{ {\bf  Type I error rate and power difference plots for ARREST trial.}\par\small Subfigure A: Type~I~error rate under the  CX-S, UX, and simulation-based~(SB) OST for~$\theta_\C=\theta_\D$. Subfigure B: Power difference
	for the UX and simulation-based OST  compared to the CX-S OST,~$\theta_\C = 0.12$ and~$\theta_\D\geq \theta_\C$, where the vertical dash-dotted line denotes~$\theta_\D=0.37.$ The upper and lower significance levels both equal 2.5\%. In both subfigures, the vertical solid line \hbox{denotes~$\theta_\C=\theta_\D=0.12$.}} \label{fig:comparison_ARREST_top}
	\end{figure}

\autoref{fig:comparison_ARREST_top} shows the type~I~error rate for the considered OSTs and difference in rejection rates.
The simulation-based OST controls type~I~errors given that~$\theta=\theta_\C=\theta_\D=0.12$, when~$\theta>0.12$ the type~I~error rate grows to a value higher than~$0.05$, reaching a maximum of about~$0.08$~(\autoref{fig:comparison_ARREST_top}, Subfigure A).
The type~I~error rate of the UX OST is close to 0.05 at its maximum.
The power of all exact tests is below that of the simulation-based OST (\autoref{fig:comparison_ARREST_top}, Subfigure B), where the UX OST performs best at first, while \hbox{CX-S} performs better for larger treatment differences.
Note that while the simulation-based OST has highest power,
it does not control type~I~errors for values \hbox{of~$\theta_\D=\theta_\C$} that deviate even a slight bit from the \hbox{assumed~$\theta_\C=0.12.$} The higher degree of conservativeness for the CX-S OST 
could possibly be explained by the discreteness of the CX-S test for early updates due to the Bonferroni correction.

\FloatBarrier

	\FloatBarrier

	In~\autoref{extra_results_arrest}, we present more results for the ARREST trial, such as the absolute power values and expected
	proportion of allocations on the superior arm under all considered designs. Due to the optional stopping component and definition of allocations on the superior arm in this setting, we see the same trade-off of the latter measure with exactness as for power (where in terms of these measures the UX OST outperforms the CX-S OST) which wouldn't be the case for an RA design without optional stopping.
	\section{Discussion} \label{discussion}
	
	We considered the theory and computation of exact tests for binary outcomes collected in a two-arm clinical trial using a \emph{response-adaptive}~(RA) procedure with possibly deterministic allocations. This approach allows for exact type~I~error control for RA designs in finite samples, even in cases where an asymptotic test does not yield type I error control which, as shown in the paper, can even occur in non-RA designs with equal allocation and moderately large trial sizes. % 
	The Markov chain introduced in~\citet{wei1990statistical} was generalized to allow for more elaborate designs and to construct exact conditional and unconditional tests for such designs. % 
	
	The first key takeaway from our results is that, while the conditional exact test is often outperformed in terms of power by the unconditional exact test in non-RA designs with equal allocation, our results show that the opposite is true for several RA designs.
	This was seen for both the application to the randomized dynamic programming design and the modified play-the-winner design, 
	and to a slightly lesser extent in~\autoref{arresttrial} which is possibly due to the use of a Bonferroni correction to account for possible type I error rate inflation from early stopping.
	This is in line with the findings in~\citet{mehrotra2003cautionary}, where the conditional exact test outperformed the unconditional exact Wald test in terms of power for non-RA designs with unequal treatment group sizes, something likely to occur for more aggressive RA procedures.
	Out of the exact approaches considered, the conditional exact test based on total successes~(CX-S test) often showed the highest power, is the most computationally tractable, and is exact when the outcomes are only assumed to be exchangeable.
	Lastly, the conservativeness of the unconditional and asymptotic tests close to the parameter space boundaries (i.e., for very low or very high success rates), both for RA and non-RA designs, was not found for the CX-S test, which results in a significant advantage in terms of power by the CX-S test in such cases.
	This indicates that CX-S tests are a good candidate test for RA designs, as RA designs are often applied in cases with low or high success rates due to ethical reasons.
	
	A second crucial lesson from our work is that, as seen in, e.g., our real-world application, the type I error rate inflation due to misspecification of the null success rate in simulation-based tests can be substantial. The proposed approach yields a more robust test where type I errors are under control no matter the null success rate, while, e.g., optimization over a subset of~$[0,1]$ for the unconditional threshold can result in an intermediate test with higher power~\citep[see, e.g.,][]{yi2013exact}.
	
	In conclusion, in cases where type I control is of primary importance we recommend the consideration of exact tests (where tractable) instead of today's commonly-used tests for RA designs.
	Our evaluations of scenarios not involving early stopping show that the CX-S tests often had the highest power. 
	The unconditional exact test could provide higher power than the CX-S test when early stopping is allowed or when type I error control is only enforced in a small set of parameters under the null hypothesis. 
	In situations where approximate type I error control is acceptable,  we again recommend considering the use of exact tests, as asymptotic tests may suffer from slow, non-monotonic, convergence of the type I error to the target level or it may not converge at all.
	
	In the paper we have considered two applications to real-life clinical trials. In the application to the ARREST trial, we have assumed that responses were not delayed. 
	In the application to the trial of~\citet{reiertsen1993}, an assumption was made on the M-PTW sequence lengths, and the choice was made not to model the (small) probability of misclassifying an outcome as a failure. We do not expect these assumptions, which facilitate our analysis, to have a substantial effect on the conclusions reported in this paper, while we note that when devising a test for a specific clinical trial (i.e., in a research project more tailored to a specific trial), we advise that the effect and necessity of such assumptions should be carefully assessed.
	We note that the above assumptions would also have to be made for a simulation-based analysis, where our computational~(non-simulation-based) analysis approach has the advantage that it has no Monte Carlo error.
	When analyzing the ARREST trial, we used numerical integration to compute allocation probabilities,  used a Bonferroni correction to construct a conditional exact test, used a two-step approximation procedure to calculate the unconditional critical value. Future research could consider alternatives to these choices. 
	
	Note that our results and conclusions are in large part based on exact Wald tests for the null hypothesis of no treatment effect. In many situations, one could be interested in hypotheses that use a different estimand of interest~\citep[e.g., a log-odds ratio or a relative risk, see, e.g.,][]{Pin2024} and while our approach still applies, results may be different when changing the Wald test definition. 
	
	The assumption of participant-exchangeable outcomes can be viewed as quite a stringent one in certain settings. Indeed, many researchers mention time trends as a reason of not implementing RA designs. We note here that the likelihood-based and simulation-based tests performed on data collected by an RA procedure also have this assumption. We furthermore note that for binary outcomes,~\citet{villar2018} showed that the trend needs to be quite severe to induce substantial type~I~error inflation. Future research could consider how to  loosen this restriction while retaining high power.
	
	Future research is needed to extend the approach to more general settings, such as multiple outcomes per participant,  other hypotheses (e.g., non-inferiority tests), random enrollment of participants, alternative adaptive procedures for clinical trials such as covariate-adjusted response-adaptive procedures, and multi-arm seamless phase II/III trials. As the latter setting mainly involves an extension of the Markov chain considered in the current paper to include more arms, we expect that the main limitation to extend the methodology to multi-arm multi-phase/stage designs is the computational tractability.
	It would furthermore be interesting to consider whether results such as~\autoref{theorem:datalikelihood_Bernoulli_RAR} could be extended to construct exact tests for other outcome types collected using a response-adaptive procedure, where a first next direction would be to consider categorical data. 
	Most tests considered in this paper used the Wald statistic, and future research could focus on a comparison to other statistics, including % 
	randomization tests.  
	Other topics in statistical inference such as estimation and providing reliable confidence intervals in response-adaptive designs are also important topics of future research. 
	
		\bibliographystyle{elsarticle-harv} 
 \bibliography{exact_analysis_RAD_arxiv.bib}
	\begin{appendix}
	
	\begin{table}
		\section{Table of symbols}\label{symbol_table}
		
		{
			\setlength{\tabcolsep}{2mm}
			\renewcommand{\arraystretch}{1.2}\begin{tabular}{m{16mm} m{120mm} r}\hline
				Symbol  & Description & \multicolumn{1}{l}{Defined in} \\
				&   &   \multicolumn{1}{l}{Section} \\
				\midrule
				$x:y:z$&Set of values with step-size~$y$ starting at~$x$ and ending at~$z$&3.1\\ $\wedge,\vee$ & Logical \emph{and} and~\emph{or}&3.1\\ 
				$\emptyset$ & Empty set&3.1\\ 
				$\C,\D$ & Control and developmental treatment indicators (resp.)&3.1\\$\theta_\C,\theta_\D,\btheta$ & Success rates for control and developmental treatment, and parameter vector~(resp.)& 3.1\\
				$a$ &  Treatment arm (control or developmental) & 3.1 \\ 
				$\bY_a$ &Potential outcome sequence for arm~$a$ (mostly assumed i.i.d. Bernoulli, exchangeable in general)&3.1\\
				$\mathbb{P}_\btheta$& Probability measure for outcomes under parameter~$\btheta$ &3.1\\
				$i,\calI,\Iend$& Participant index, set of participant indices, and trial size (resp.) &3.1\\
				$A_i$& Treatment allocation for participant~$i$ &3.1\\
				$(\bH_i)_{i\in\calI}$&Trial history process~(random sequence/process)& 3.1\\
				$\mathcal{H},\mathcal{H}_i$& Support of trial histories, trial histories up to participant~$i$ (resp.) &3.1\\
				$\pi$& Response-adaptive procedure & 3.1\\
				$\mathbb{P}^\pi_\btheta$& Probability measure for allocations and outcomes under parameter~$\btheta$ and RA procedure~$\pi$  &3.1\\
				$\Tau_{\mathcal{H}},\mathcal{R}_{\mathcal{H}}$& Test statistic and rejection region based on the full trial history&3.1\\
				$t, \calT, \Tend$ & Update time, set of 
				update times, and final update time (resp.)  & 3.2\\
				$i_t$&Participant index coupled to update time~$t$& 
				3.2\\
				$\bX_t$ &   State (random) of  Markov chain describing trial information at update time~$t$& 3.2\\
				$\calI_0$&$\calI\cup\{0\}$& 
				3.2\\
				$\bx,x_t$ &  State of Markov chain~(deterministic) and function to obtain~$\bX_t$ from~$\bH_{i_t}$~(resp.) & 3.2 \\
				$\mathcal{X}, 
				\mathcal{X}_t$& State space of Markov chain~$(\bX_t)_t$ and support of state~$\bX_t$~(resp.) &3.2\\
				$S_{a,i},N_{a,i}$& Total successes and treatment group sizes for arm~$a$ up to trial participant~$i$~(resp.)&3.2\\
				$s_a,n_a$&  Functions to obtain successes and treatment group sizes (resp.) for arm~$a$ from the state &3.2\\
				$\Tau, \mathcal{R}$& Test statistic and rejection region~(resp.) &3.2\\
				$q^\pi$& Part of transition kernel of $\bX_t$ dependent on RA procedure~$\pi$ &3.2\\ $p_{\btheta},p$&Part of transition kernel of $\bX_t$ dependent on the outcomes model&3.2, 3.3\\
				$\partial\bs, \partial \bn$& Functions reading difference of successes and treatment group sizes between states &3.2\\
				$\mathcal{X}_t^{\text{SS}}$&State space containing successes and allocations per arm up to participant~$i_t$ &3.2\\
				$\mathbb{Q}$&Prior for~$\btheta$ used in Thompson sampling &3.2\\
				$b_t$&Group sizes for group-sequential RA design &3.2\\
				$U_t,u $&Early stopping point of trial, function reading~$U_t$ from~$\bX_t$~(resp.) &3.2\\
				
				$\mu$& Signed measure defining the outcome distribution for exchangeable outcomes model &3.3\\
				$g^\pi_t$& Part of the distribution of~$\bX_t$ induced by RA procedure &3.3\\
				$H_0$&Null hypothesis of no treatment effect & 3.4\\
				$c_u,c_{\ell}$& Upper and lower critical values for test (resp.)&3.4\\
				$\zeta, \mathcal{X}_{\zeta}$& Summary of state used in conditional exact test, pre-image in $\mathcal{X}_{\Tend}$ under~$\zeta$~(resp.)&3.4\\
				
				$\alpha,\alpha_{\ell},\alpha_u
				$& Significance level, lower and upper significance levels (resp.)  &3.4\\
				$\Tau^+(E)$& Image of $E$ under $\Tau$ extended with $-\infty,\infty$ &3.4\\
				\hline
		\end{tabular}}
	\end{table} 
	\FloatBarrier
	
	\section{Proofs}\label{sect:proofs}
	In this section, we restate the theorem and corollaries of the main paper, and give their proofs.
	\setcounter{theorem}{0}
	\begin{theorem}~\label{theorem:datalikelihood_Bernoulli_RAR_2}
		Assume there is a signed measure~$\mu$ on~$([0,1]^2,\mathcal{B}([0,1]^2))$  such that for all~$\by_\C,\by_\D\in\{0,1\}^{\Iend}$ 
		\begin{equation}\mathbb{P}(\bY_{a}=\by_{a}\;\;\forall a\in\{\C,\D\})=
			\int_{[0,1]^2}\prod_{a\in\{\C,\D\}} \theta_a^{\sum_{i=1}^{\Iend} y_{a,i}}(1-\theta_a)^{\Iend-\sum_{i=1}^{\Iend}y_{a,i}}\mu(d\btheta) \label{prob_seq_exch}\end{equation}
		and the stochastic process~$(\bX_t)_t$ has transition structure~(1) with the term with~$p_\btheta$ replaced by~$p(\partial\bs(\bx_t,\bx_{t+1})\mid\partial\bn(\bx_t,\bx_{t+1}), \bn(\bx_t), \bs(\bx_t))$  where for all~$\bs',\bn'\in\calI_0^2,\partial\bs',\partial\bn'\in\mathcal{D}^2$ the term~$p(\partial\bs'\mid\partial\bn',\bn',\bs')$ equals
		\begin{equation} \mathbb{P}\left(\sum_{i=n_a'+1}^{n_a'+\partial n_a'} Y_{a,i}=\partial s_a'\;\;\forall a\in\{\C,\D\} \;\bigg|\; \sum_{i=1}^{n_a'} Y_{a,i}= s_a'\;\;\forall a\in\{\C,\D\}\right)\label{prob_transition_extended2}
		\end{equation}
		then for all~$\bx_t\in\mathcal{X}_t$ 
		\begin{equation}\mathbb{P}^\pi(\bX_t=\bx_t)= g_t^\pi(\bx_t)\int_{[0,1]^2}\prod_{a\in\{\C,\D\}}\theta_a^{s_{a}(\bx_t)}(1-\theta_a)^{n_{a}(\bx_t)-s_a(\bx_t)}\mu(d\btheta),\label{expression_likelihood}\end{equation}
		where, letting~$\binom{n}{k}$ denote the binomial coefficient for natural numbers~$n,k$, for all~$t,$~$\bx_t\in\mathcal{X}_t$
		\begin{align}g_0^\pi(\bx_0)&=1,\nonumber\\ g_t^\pi(\bx_t) &= \sum_{\bx_{t-1}\in\mathcal{X}_{t-1}}\left(\prod_{a\in\{\C,\D\}}\binom{\partial n_a(\bx_{t-1},\bx_t)}{\partial s_a(\bx_{t-1}, \bx_t)}\right)g_{t-1}^\pi(\bx_{t-1} )q^{\pi}(\bx_{t-1},\bx_t).\label{eqn:gdef}\end{align}
	\end{theorem}
	
	\begin{proof}
		Examining \eqref{prob_transition_extended2} and  summing over paths that contain~$s_a'$ successes in the first~$n_a'$ observations and~$\partial s_a'$ in the next~$\partial n_a'$ gives for all~$\bn',\bs'\in\calI_0^2,\partial\bs',\partial\bn'\in\mathcal{D}^2$
		$$p(\partial\bs'\mid\partial\bn',  \bn',\bs') = \frac{\int_{[0,1]^2}\prod_{a\in\{\C,\D\}}\binom{n_a' }{s_a' }\binom{ \partial n_a}{ \partial s_a}\theta_a^{s_a'+\partial s_a'}(1-\theta_a)^{(n_a' + \partial n_a'-s_a'-\partial s_a')}\mu(d\btheta)}{\int_{[0,1]^2}\prod_{a\in\{\C,\D\}}\binom{ n_a'}{  s_a'}\theta_a^{s_a'}(1-\theta_a)^{(n_a'-s_a')}\mu(d\btheta)}.$$
		The statement of the theorem is shown by induction, with (trivial) base case~$t=0$ \mbox{as~$S_{a,0}~=N_{a,0}=0$.}
		Assume the statement holds up to and including~$t-1$, then by~(1) we have that~$\mathbb{P}^\pi(\bX_{t}=\bx_{t})$ equals
		\begin{align*}
			& \sum_{\bx_{t-1}\in\mathcal{X}_{t-1}} \mathbb{P}^\pi(\bX_{t-1} = \bx_{t-1})
			q^{\pi}(\bx_{t-1},\bx_{t})\cdot p(\partial\bs(\bx_{t-1},\bx_{t})\mid\partial\bn(\bx_{t-1},\bx_{t}),\bn(\bx_{t-1}),\bs(\bx_{t-1}))
			\\
			&= \sum_{\bx_{t-1}\in\mathcal{X}_{t-1}} g_{t-1}^\pi(\bx_{t-1})q^{\pi}(\bx_{t-1},\bx_{t})\int_{[0,1]^2}\prod_a\binom{\partial n_a(\bx_{t-1},\bx_t)}{\partial s_a(\bx_{t-1}, \bx_t)}\theta_a^{s_a(\bx_{t})}(1-\theta_a)^{(n_a(\bx_{t})-s_a(\bx_{t}))}\mu(d\btheta)\\&=\left(\sum_{\bx_{t-1}\in\mathcal{X}_{t-1}} \left(\prod_a\binom{\partial n_a(\bx_{t-1},\bx_{t})}{\partial s_a(\bx_{t-1}, \bx_{t})}\right)g_{t-1}^\pi(\bx_{t-1})q^{\pi}(\bx_{t-1},\bx_{t})\right)\\&\cdot \left(\int_{[0,1]^2} \prod_{a\in\{\C,\D\}}\theta_a^{s_a(\bx_{t})}(1-\theta_a)^{(n_a(\bx_{t}) -s_a(\bx_{t}))}\mu(d\btheta)\right) \\&= g_{t}^\pi(\bx_{t})\int_{[0,1]^2} \prod_{a\in\{\C,\D\}}\theta_a^{s_a(\bx_{t})}(1-\theta_a)^{(n_a(\bx_{t}) -s_a(\bx_{t}))}\mu(d\btheta)\quad \forall \bx_t\in\mathcal{X}_t.
		\end{align*}
		Hence, the statement follows by mathematical induction.
	\end{proof}
	
	\begin{corollary}\label{cor:cond_exact}
		If~$\zeta:\mathcal{X}_{\Tend}\mapsto\mathcal{Z}$, 
		~$s(\bX_{\Tend})=\tilde{s}(\zeta(\bX_{\Tend}))$ for a 
		function~$\tilde{s}:\mathcal{Z}\mapsto \calI_0$ and the assumptions of~\autoref{theorem:datalikelihood_Bernoulli_RAR_2} hold 
		then under the nonparametric null hypothesis~$H_0^\text{NP}$ in~\autoref{diff_trial_models} we have~$\mathbb{P}^\pi_{\btheta}\Big(\Tau(\bX_{\Tend})\geq c_u(\zeta(\bX_{\Tend}))\text{ or  }\Tau(\bX_{\Tend})\leq c_{\ell}(\zeta(\bX_{\Tend}))\Big)\leq \alpha.$
	\end{corollary}
	
	\begin{proof}
		Using the convention~$0/0=1$, following \autoref{theorem:datalikelihood_Bernoulli_RAR_2} we have that for all~$z\in\mathcal{Z}$ for \mbox{which~$\mathbb{P}^\pi(\zeta(\bX_{\Tend})=z)>0$:}
		\begin{align*}&\mathbb{P}^\pi(\Tau(\bX_{\Tend})\geq c_u(\zeta(\bX_{\Tend}))\mid \zeta(\bX_{\Tend})=z)=\mathbb{P}^\pi(\Tau(\bX_{\Tend})\geq c_u(z)\mid \zeta(\bX_{\Tend})=z)\\&\stackrel{\text{(\autoref{theorem:datalikelihood_Bernoulli_RAR_2})}}{=}\frac{\sum_{\bx_{\Tend}\in\mathcal{X}_{\zeta}(z)\,:\, \Tau(\bx_{\Tend})\geq c_u(z)} g_{\Tend}^\pi(\bx_{\Tend})\int_{[0,1]^2}\prod_{a\in\{\C,\D\}}\theta_a^{s_{a}(\bx_{\Tend})}(1-\theta_a)^{n_{a}(\bx_{\Tend})-s_a(\bx_{\Tend})}\mu(d\btheta)}{\sum_{\bx_{\Tend}\in\mathcal{X}_{\zeta}(z)} g_{\Tend}^\pi(\bx_{\Tend})\int_{[0,1]^2}\prod_{a\in\{\C,\D\}}\theta_a^{s_{a}(\bx_{\Tend})}(1-\theta_a)^{n_{a}(\bx_{\Tend})-s_a(\bx_{\Tend})}\mu(d\btheta)}\\
			&=\frac{\sum_{\bx_{\Tend}\in\mathcal{X}_{\zeta}(z)\,:\, \Tau(\bx_{\Tend})\geq c_u(z)} g_{\Tend}^\pi(\bx_{\Tend})\int_{\{\btheta\in[0,1]^2\,:\,\theta_\C=\theta_\D\}}\theta_\C^{s(\bx_{\Tend})}(1-\theta_\C)^{\Iend-s(\bx_{\Tend})}\mu(d\btheta)}{\sum_{\bx_{\Tend}\in\mathcal{X}_{\zeta}(z)} g_{\Tend}^\pi(\bx_{\Tend})\int_{\{\btheta\in[0,1]^2\,:\,\theta_\C=\theta_\D\}}\theta_\C^{s(\bx_{\Tend})}(1-\theta_\C)^{\Iend-s(\bx_{\Tend})}\mu(d\btheta)}\\&=\frac{\int_{\{\btheta\in[0,1]^2\,:\,\theta_\C=\theta_\D\}}\theta_\C^{\tilde{s}(z)}(1-\theta_\C)^{\Iend-\tilde{s}(z)}\mu(d\btheta)\cdot \sum_{\bx_{\Tend}\in\mathcal{X}_{\zeta}(z)\,:\, \Tau(\bx_{\Tend})\geq c_u(z)} g_{\Tend}^\pi(\bx_{\Tend})}{\int_{\{\btheta\in[0,1]^2\,:\,\theta_\C=\theta_\D\}}\theta_\C^{\tilde{s}(z)}(1-\theta_\C)^{\Iend-\tilde{s}(z)}\mu(d\btheta)\cdot \sum_{\bx_{\Tend}\in\mathcal{X}_{\zeta}(z)} g_{\Tend}^\pi(\bx_{\Tend})}\\&=\frac{ \sum_{\bx_{\Tend}\in\mathcal{X}_{\zeta}(z)\,:\, \Tau(\bx_{\Tend})\geq c_u(z)} g_{\Tend}^\pi(\bx_{\Tend})}{\sum_{\bx_{\Tend}\in\mathcal{X}_{\zeta}(z)} g_{\Tend}^\pi(\bx_{\Tend})}\stackrel{(11)}{\leq} \alpha_u.
		\end{align*}
		Similarly~$\mathbb{P}^\pi(\Tau(\bX_{\Tend})\leq c_{\ell}(\zeta(\bX_{\Tend}))\mid \zeta(\bX_{\Tend})=z)\leq \alpha_{\ell}$, and hence the statement of the theorem is proven by the law of total expectation.
	\end{proof}
	
	\begin{corollary}\label{cor:uncond_exact}
		If the assumptions of 
		\autoref{theorem:datalikelihood_Bernoulli_RAR_2} hold with~$\mu$ a probability measure, then under~$H_0^\text{NP}$ it holds \mbox{that 
			~$\mathbb{P}^\pi_{\btheta}\Big(\Tau(\bX_{\Tend})\geq c_u\text{ or }\Tau(\bX_{\Tend})\leq c_{\ell}\Big)\leq \alpha,$}
		where~$c_u,c_{\ell}$ are as given in~Definition~2.
	\end{corollary}
	
	\begin{proof}
		By \autoref{theorem:datalikelihood_Bernoulli_RAR_2}, we have 
		\begin{align*}&\mathbb{P}^\pi(\Tau(\bX_{\Tend})\geq c_u)=\int_{[0,1]^2}\mathbb{P}^\pi_\btheta(\Tau(\bX_{\Tend})\geq c_u)\mu(d\btheta) \\&= \int_{\{\btheta\in[0,1]^2\,:\,\theta_\C=\theta_\D\}}\mathbb{P}^\pi_\btheta(\Tau(\bX_{\Tend})\geq c_u )\mu(d\btheta)\leq \max_{\btheta\in[0,1]^2:\;\theta_\C=\theta_\D}\mathbb{P}^\pi_\btheta(\Tau(\bX_{\Tend})\geq c_u)\leq \alpha_u.
		\end{align*}
		Where the penultimate inequality follows by taking the maximum inside the integral and as~$\mu$ is a probability measure. Similarly~$\mathbb{P}^\pi(\Tau(\bX_{\Tend})\leq c_{\ell})\leq \alpha_{\ell}$, proving the statement.
	\end{proof}

	\section{Algorithms and computation time results}
	\subsection{Algorithm for calculating an upper bound for unconditional critical values}
	
	This section provides an algorithm to compute a (tight) upper bound for a UX upper critical value~$c_u$, based on~\autoref{theorem:datalikelihood_Bernoulli_RAR_2}.
	Often, due to symmetry, it suffices to only compute~$c_u$, otherwise, the computational procedure has to be adjusted in a straightforward manner. Hence, we focus on the computation of a (tight) upper bound for~$c_u$.
	In order to determine a numerical procedure for calculating an upper bound of~$c_u$ the following result, bounding the difference between the rejection rate at different parameter values, is of use.

	\begin{lemma}~\label{Thm:Lipschitz}
		Let~$r_c(\theta) = \mathbb{P}^\pi_{(\theta,\theta)}(\Tau(\bX_{\Tend})\geq c)$ for all~$\theta\in[0,1]$. Then, for~$\theta_1,\,\theta_2\in[0,1]$ where~$\theta_1\leq\theta_2$ we have
		~$$r_c(\theta_3) - r_c(\theta_1)\leq k_c(\theta_1,\theta_2)(\theta_3-\theta_1)\quad\forall \theta_3\in[\theta_1,\theta_2],$$
		where
		\begin{align*} k_c(\theta_1,\theta_2)&=\sum_{\bx_{\Tend}\in\mathcal{X}_{\Tend}\,:\, \Tau(\bx_{\Tend})\geq c} g_{\Tend}^\pi(\bx_{\Tend})  \max(0, h^*(\bx_{\Tend},\theta_1,\theta_2))\\
			h^*(\bx_{\Tend},\theta_1,\theta_2)&=\max(\{h(\bx_{\Tend}, \theta_1),h(\bx_{\Tend}, \theta_2),h(\bx_{\Tend}, Proj_{[\theta_1,\theta_2]}(\theta^*_{-}(\bx_{\Tend}))),  h(\bx_{\Tend},Proj_{[\theta_1,\theta_2]}(\theta^*_{+}(\bx_{\Tend})))\})\\
			h(\bx_{\Tend},\theta) &=(s(\bx_{\Tend}) - {\Iend}\cdot \theta)\theta^{s(\bx_{\Tend})-1}(1-\theta)^{\Iend-s(\bx_{\Tend})-1},
		\end{align*}
		~$Proj_{[\theta_1,\theta_2]}(x)=\min(\theta_1,\max(\theta_2,x))$, and 
		$$\theta^*_{\pm}(\bx_{\Tend})=\frac{2s(\bx_{\Tend})(\Iend-1) \pm \sqrt{4s(\bx_{\Tend})^2(\Iend-1)^2-4\Iend(\Iend-1)s(\bx_{\Tend})(s(\bx_{\Tend})-1)})}{2\Iend(\Iend-1)}.$$
		
	\end{lemma}
	\begin{proof}
		From (8) we have~$r_c(\theta) =  \sum_{\bx_{\Tend}\in\mathcal{X}_{\Tend}\,:\, \Tau(\bx_{\Tend})\geq c} g_{\Tend}^\pi(\bx_{\Tend})\cdot \theta^{s(\bx_{\Tend})}(1-\theta)^{\Iend-s(\bx_{\Tend})}$
		hence~\begin{align*}
			r'_c(\theta)&=\sum_{\bx_{\Tend}\in\mathcal{X}_{\Tend}\,:\, \Tau(\bx_{\Tend})\geq c} g_{\Tend}^\pi(\bx_{\Tend}) h(\bx_{\Tend},\theta) 
			,\quad r''_c(\theta) =\sum_{\bx_{\Tend}\in\mathcal{X}_{\Tend}\,:\, \Tau(\bx_{\Tend})\geq c} g_{\Tend}^\pi(\bx_{\Tend})h' (\bx_{\Tend},\theta)
		\end{align*}
		where~\begin{align*}h'(\bx_{\Tend},\theta) &= ( \Iend(\Iend -1)\theta^2  - 2s(\bx_{\Tend})(\Iend - 1)\theta +  s(\bx_{\Tend})(s(\bx_{\Tend})-1)  )   \theta^{s(\bx_{\Tend})-2}(1-\theta)^{\Iend-s(\bx_{\Tend})-2}.\end{align*}
		Note that~$h'(\bx_{\Tend},\theta) = 0$~(i.e., $h$ has extreme values)  at~$\theta^*_{\pm}(\bx_{\Tend})$ and~$\{0,1\}$.
		By the mean value theorem, there exists~$\theta_3\in[\theta_1,\theta_2]$ such that 
		~$r_c(\theta_1) - r_c(\theta_2)= r'_c(\theta_3)(\theta_1-\theta_2).$
		The bound now follows as we can bound~$r'_c(\theta_3)$ by maximizing each function~$h(\bx_{\Tend},\theta_3)$ over~$\theta_3\in[\theta_1,\theta_2]$, making use of the fact that the maximum occurs either at $\theta_1,\theta_2$ 
		or at the zeros~$\theta^*_{\pm}(\bx_{\Tend})$ of~$h'$ whenever they fall in~$[\theta_1,\theta_2]$.
	\end{proof}

	From~\autoref{Thm:Lipschitz}, for~$0<\epsilon<1$, if~$\theta_1-\theta_2\leq \epsilon/k_c(\theta_1,\theta_2)$ then \hbox{$r_c(\theta_1) - r_c(\theta_3)~\leq~\epsilon$} for all~$\theta_3\in[\theta_1,\theta_2].$
	Hence, if we have a sequence of points~$0=\theta_0<\theta_1< \cdots<\theta_m=1$ such that the bound~\mbox{$\theta_\ell-\theta_{\ell+1}\leq \epsilon/k_c(\theta_\ell,\theta_{\ell+1})$} holds and~$r_c(\theta_\ell) \leq \alpha_u - \epsilon$ for all~$\ell$ then~$r_c(\theta) \leq \alpha_u~$ for all~$\theta\in[0,1]$ hence~$c\geq c_u.$ This is the idea behind~\autoref{alg_unconditional} for calculating an upper bound for~$c_u$, which first determines the constants~$k_c(\theta_1,\theta_2)$ for a coarse grid, and then refines the grid such that the distance between consecutive points is less than~$\epsilon/k_c(\theta_1,\theta_2)$, using the fact that ~$k_c(\theta_1,\theta_2)$ is decreasing in the difference between~$\theta_1$ and~$\theta_2$ and in~$c$.
	Finally, the algorithm calculates~$c$ such that~$r_{c}(\theta)\leq \alpha_u-\epsilon$ for all values of~$\theta$ in the grid, where certain values can be skipped depending on the difference between the current rejection rate and~$\alpha_u-\epsilon.$ A similar approach was introduced in~\citet{suissaExact1985}.

	\FloatBarrier
	\begin{algorithm}[h!]
		\caption{Algorithm for calculating upper critical values of~$\bv$ for significance level~$\alpha$ and probability weight vector 
			$\bw$}\label{right_tail_algo}
		\setstretch{1.3}
		\begin{algorithmic}[1]
			\Inputs{$\bv,\bw,\alpha$;}
			\State Sort~$\bv$ and permute $\bw$ accordingly;
			\State Set~$i=1$;
			\While{$\sum_{i'=1}^iw_{i'}<1-\alpha$}
			\State~$i:=i+1$;
			\EndWhile
			\If{$\sum_{i':v_{i'}\geq v_i}w_{i'}>\alpha$  }\quad \emph{(this happens in case of ties in $\bv$)}
			\State Set $i := \min\{i'\geq i:v_{i'}> v_i\}$;
			\EndIf
			\\{\bf Outputs:}~$v_i$.
		\end{algorithmic}
	\end{algorithm}
	\begin{algorithm}[tbp]
		\caption{Algorithm for calculating an upper bound for UX critical value~$c_u$}\label{alg_unconditional}
		
		\begin{algorithmic}[1]
			\Inputs{$0<\epsilon,\alpha_u,\theta_0<1,\,m\in\{1,2,\dots\}$;}
			\Initialize {Set~$\bv=()$ and~$c_u=\min\{c\in\Tau^+(\mathcal{X}_{\Tend}):r_c(\theta_0)\leq \alpha_u-\epsilon\}$~using~\autoref{right_tail_algo};}
			\For{$\ell\in\{1,\dots, m\}$}
			\If{$\epsilon/(k_{c_u}((\ell-1)/m,\,\ell/m))\in(0, 1/m)$}
			\State Set~$d\theta~=~\epsilon/(k_{c_u}((\ell-1)/m,\,\ell/m));~$
			\Else 
			\State Set~$d\theta = 1/m;$
			\EndIf
			
			\State Update~$\bv:=\bv^\frown(\ell-1)/m:d\theta\,:\, \ell/m$\,;\quad\text{($\frown$ denotes concatenation)}
			\EndFor
			\State Set~$k=1$;
			\While{$k\leq |\bv|$}
			\State Set~$c_u'=\min\{c\in\Tau^+(\mathcal{X}_{\Tend})
			:r_c(v_k)\leq \alpha_u-\epsilon\}$~using~\autoref{right_tail_algo};
			\State Update~$c_u:=\max(c_u',c_u);$
			\State Update~$k := k + \max(1, \floor{(\alpha_u - r_{c_u}(v_k))/\epsilon});$
			\EndWhile
			\\{\bf Outputs:}~$c_u$
		\end{algorithmic}
	\end{algorithm}
	
	\subsection{Considerations regarding computation time}\label{comptime}
	
	\vbox{\autoref{right_tail_algo}, \autoref{alg_unconditional}, 
	\autoref{calc_ocs}, and \autoref{FET_algo}, together with~\autoref{calc_g} analyzed in the next section, form the most important algorithms for generating the results of this paper, and can be used to calculate the operating characteristics, critical values, the p-values under FET, the policy-dependent coefficients~$g_t^\pi$, and the CX-S critical values, respectively. 
	From \autoref{calc_ocs} it can be seen that the complexity of calculating the OCs in the state space size is~$O(|\mathcal{X}_{\Tend}|)$ (coming from the for-loop), while \autoref{right_tail_algo} with~$|\bv|=|\mathcal{X}_{\Tend}|$, due to the sorting step, will have complexity~\mbox{$O(|\mathcal{X}_{\Tend}|\cdot \log(|\mathcal{X}_{\Tend}|))$}. As \autoref{alg_unconditional} also uses \autoref{right_tail_algo} to determine the critical value for values in a grid the complexity in terms of the state space size is also~$O(|\mathcal{X}_{\Tend}|\cdot\log(|\mathcal{X}_{\Tend}|))$.  Lastly, \autoref{FET_algo} has complexity~\mbox{$O(|\mathcal{X}_{\Tend}|\cdot \max_{s',n'_\C}|\mathcal{X}_\text{SA}(s',n'_\C)|)$} due to the for-loop over~$\mathcal{X}_{\text{SA}}(s',n'_\C)$ including the sum over a vector of size~$\mathcal{X}_{\text{SA}}(s',n'_\C)$.}
	
\begin{figure}[h!]
			\centering
			\includegraphics[width =.8\textwidth]{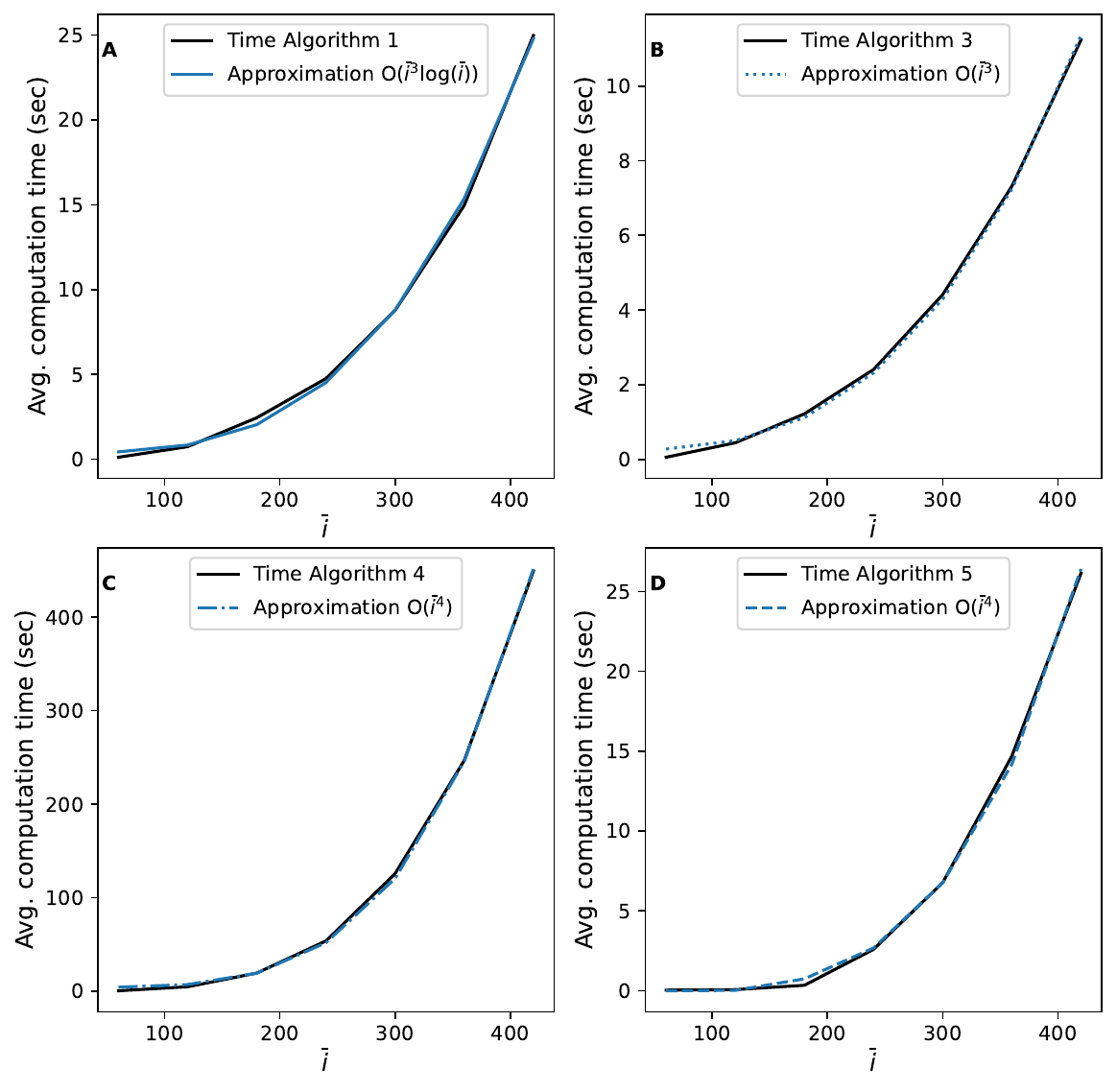}
			\caption{Average computation times (25 runs) vs. linear regressions on the time complexities for~\autoref{right_tail_algo} (Subfigure~A), \autoref{calc_ocs} (Subfigure~B), \autoref{FET_algo} (Subfigure~C), and \autoref{calc_g} (Subfigure~D).}
			\label{figure:regression_time_complexities}
		\end{figure}

	In case of the Markov chain in Example~1, we will have that~\mbox{$|\mathcal{X}_{\Tend}|\in O(\Iend^3)$}~\citep[see, e.g., ][]{jacko2019binarybandit}, hence the complexities in~$\Iend$ become~$O(\Iend^3),\; O(\Iend^3\cdot \log(\Iend)),$ and~$O(\Iend^4)$, as for Example~1 we have~$|\mathcal{X}_\text{SA}(s',n'_\C)|=\min(s',n'_\C)\leq \Iend.$ 
	\autoref{figure:regression_time_complexities} is in agreement with these theoretical results, where it is shown that a linear regression~(found using ordinary least squares) on the above theoretical orders including intercept results in a good fit for~\autoref{right_tail_algo}, 
	\autoref{calc_ocs}, and~\autoref{FET_algo}.

	\FloatBarrier
	{
		\begin{algorithm}[tbp]
			\caption{Algorithm for calculating operating characteristics~$\mathbb{E}_\btheta^\pi[f(\bX_{\Tend})]$ where~$i$ is an index mapping function for~$\mathcal{X}_{\Tend}$.}\label{calc_ocs}
			
			\begin{algorithmic}[1]
				\Inputs{$i,f,g_{\Tend}^\pi,\btheta$;}
				\State Set~$\bp,\bg,\bv=\bm 0_{|\mathcal{X}_{\Tend}|}$;
				\For{$\bx_{\Tend}\in\mathcal{X}_{\Tend}$}
				\State Set~$p_{i(\bx_{\Tend})}=\prod_{a\in\{\C,\D\}}\theta_a^{s_a(\bx_{\Tend})}(1-\theta_a)^{n_a(\bx_{\Tend})-s_a(\bx_{\Tend})}$;
				\State Set~$g_{i(\bx_{\Tend})} = g_{\Tend}^\pi(\bx_{\Tend})$ and~$v_{i(\bx_{\Tend})} = f(\bx_{\Tend})$;
				\EndFor
				\\{\bf Outputs:}~$ \bp^\top(\bg\circ\bv)$ (where~$\top$ means transpose and~$\circ$ the Hadamard product)
			\end{algorithmic}
		\end{algorithm}
		
		\begin{algorithm}[tbp]
			\caption{Algorithm for calculating~$\Tau_\text{FET}$ for every state as a vector~$\bv_\Tau$.}\label{FET_algo}
			\setstretch{1.3}
			\begin{algorithmic}[1]
				\Inputs{ index mapping function~$i$ for~$\mathcal{X}_{\Tend}$;}
				\State Set~$\bp',\bv_\Tau=\bm 0_{|\mathcal{X}_{\Tend}|}$;
				\For{$s',n'_\C\in\calI_0$}
				\State Set~$\mathcal{P}=\emptyset$;
				\For{$\bx_{\Tend}\in\mathcal{X}_\text{SA}((s',n'_\C))$}
				\State Set~$p'_{i(\bx_{\Tend})}=p_\text{FET}(\bx_{\Tend})$ with~$p_\text{FET}$ as defined in~(12);
				\State Update~$\mathcal{P}:=\mathcal{P}\cup\{p'_{i(\bx_{\Tend})}\}$ ;
				\EndFor
				\For{$\bx_{\Tend}\in\mathcal{X}_\text{SA}((s',n'_\C))$}
				\State Set~$v_{\Tau,i(\bx_{\Tend})}=\sum_{p''\in\mathcal{P}\,:\,p''\leq p'_{i(\bx_{\Tend})}}p''$;
				\EndFor
				\EndFor
				\\{\bf Outputs:}~$ \bv_\Tau$ 
			\end{algorithmic}
		\end{algorithm}
	}
	\FloatBarrier
		{	\subsubsection{Time complexity of calculating the conditional exact test}
			
				\autoref{calc_g}, with time complexity order~$\mathcal{O}(|B|\sum_{t=1}^{\Tend}|\mathcal{X}_t|)$, calculates the vector~$\bg^\pi$ of policy-dependent coefficients for all states in~$\mathcal{X}$, as well as the CX-S upper critical values~(lower critical values are found similarly), and hence the CX-S test outcome for all possible datasets. The set~$B$ is the minimal set of backward transformations that for all states~$\bx\in\mathcal{X}$ yields all states with positive probability of leading to~$\bx$.  Given~$\bg^\pi$, the CX-S test is found by 
			calculating upper quantiles of the conditional distribution of~T over~$\mathcal{X}_{\text{S}}(s)$ using \autoref{right_tail_algo} for all~$s\in\mathcal{I}_0$, yielding time complexity order~$\mathcal{O}(|\mathcal{X}_{\Tend}|\cdot\log(|\mathcal{X}_{\Tend}|))$.

			Using Example~1 for illustration, we have~$|B|=4$ as a state~$\bx_t$ can only be reached by a success or a failure on either arm, and the time complexity order in~$\Iend$ is~$\mathcal{O}(\Iend^4)$ (verified in~\autoref{figure:regression_time_complexities}D).
			In comparison, finding the randomization test outcome \emph{for all possible datasets} has
			time complexity order~$\mathcal{O}(\Iend^3 2^{\Iend})$ in the case of Example~1, using the procedure given in~\citet{wei1988exactRT}.
			To compare, for~$\Iend=240$ we have~$\Iend^4 \approx 3.3\cdot10^9$, which is $70$ orders of magnitude smaller than~$\Iend^3 2^{\Iend}\approx 2.4\cdot10^{79}$. We note that, while having a much larger time complexity, the randomization test is exact under more scenarios. 
			\FloatBarrier
		\begin{algorithm}[h!]
			\caption{Calculating~$g_{t}^\pi(\bx_{t})$ for all~$\bx_{t},t$ and critical values for the CX-S test.}\label{calc_g}
			\setstretch{1.3}
			\begin{algorithmic}[1]
				\Inputs{$\Tend,\Iend,q^\pi$, index mapping function~$i$ for~$\mathcal{X}$,~$B$ a set of backward operators for~$(\bX_t)_t$,~$\alpha_u$, test statistic function~T;}
				\State Define~$\bg^\pi$ such that~$g^{\pi}_{i(\bx)}=\mathbb{I}(\bx = \bx_0)$ for all~$\bx\in\mathcal{X}$, let~$\bc_\ell,\bc_u=\bm 0_{\Iend+1}$;
				\For{$t\in\{1,\dots,\Tend\}$, $\bx_{t}\in\mathcal{X}_{t}$,~$\beta\in B$} 
				\State Set $g_t^\pi(\bx_t) = g_t^\pi(\bx_t) +\left(\prod_{a\in\{\C,\D\}}\binom{\partial n_a(\beta(\bx_{t}),\bx_t)}{\partial s_a(\beta(\bx_{t}), \bx_t)}\right)g_{t-1}^\pi(\beta(\bx_{t}) )q^{\pi}(\beta(\bx_{t}),\bx_t)$;
				\EndFor
				\For{$s\in\{0,\dots,\Iend\}$}
				
				\State Find~$c_{u,s}$ using~\autoref{right_tail_algo} with inputs~(T$(\mathcal{X}_\text{S}(s)), g^\pi_{i(\mathcal{X}_\text{S}(s))},\alpha_u)$;
				\EndFor
				\\{\bf Outputs: $\bg^\pi,\bc_u$}
			\end{algorithmic}
	\end{algorithm}}
	
	\FloatBarrier

		\FloatBarrier

		\FloatBarrier
		
		\section{Additional results for randomized dynamic programming and equal allocation design}\label{results_extra_RDP}
		\autoref{tab:comparison_tests_RDP_60}, \autoref{tab:comparison_tests_RDP_240}, and  \autoref{tab:comparison_tests_RDP_960} show the rejection rates for the RDP RAR procedure under several parameter configurations, tests, and trial sizes, where~$
		\alpha_u=\alpha_{\ell}=2.5\%$. The last column of the tables shows the \emph{expected
			proportion of allocations on the superior arm}~(EPASA). The rows of the table where~$\theta_\C\in\{0.01, 0.3,0.9\}$ correspond to the markers in Subfigures~B, D, F of~\autoref{fig:comparison_tests}.
	The asymptotic Wald test is excluded from consideration in the power comparison as it does not control type I errors.
	
	The tables show that the CX-S test often has highest power, and in cases where it is outperformed the power differences are less than~$1\%$. For very small and very high control success rates the increase in power for the CX-S test over other tests is quite substantial, e.g., for~$\theta_\C=0.0$, power differences around~$20-30\%$ are seen for all values of~$\Iend$. Something that also stands out is that the rejection rate at the points~$\theta_\C=\theta_\D\in\{0,1\}$ for the CX-S test is higher than 0, and even around~$5\%$ for~$\Iend=960$ and~$\theta_\C=\theta_\D=1.0.$ This is because, while the treatment outcomes are deterministic in such cases, the allocations are still a source of randomness.  The rejection rate for the UX Wald test and FET (corr.) is zero in these cases because null success rates~$0$ and~$1$ induce significantly less variance in the test statistic in comparison to the parameter configuration where the type I error rate is highest. As also seen in~\autoref{fig:comparison_tests}, the type I error rate inflation under the asymptotic test does not reduce for large trial sizes. As expected, the EPASA grows in~$|\theta_\D-\theta_\C|$, where we note that the range of~$|\theta_\D-\theta_\C|$ decreases from~0.5 to 0.1 in~\autoref{tab:comparison_tests_RDP_60}, \autoref{tab:comparison_tests_RDP_240} , and~\autoref{tab:comparison_tests_RDP_960}, hence the EPASA stays below the maximum value of 90\% for RDP in the tables. 
	
\begin{figure}[tbp]\thisfloatpagestyle{empty}
		\includegraphics[height=0.84\textheight, width = \textwidth]{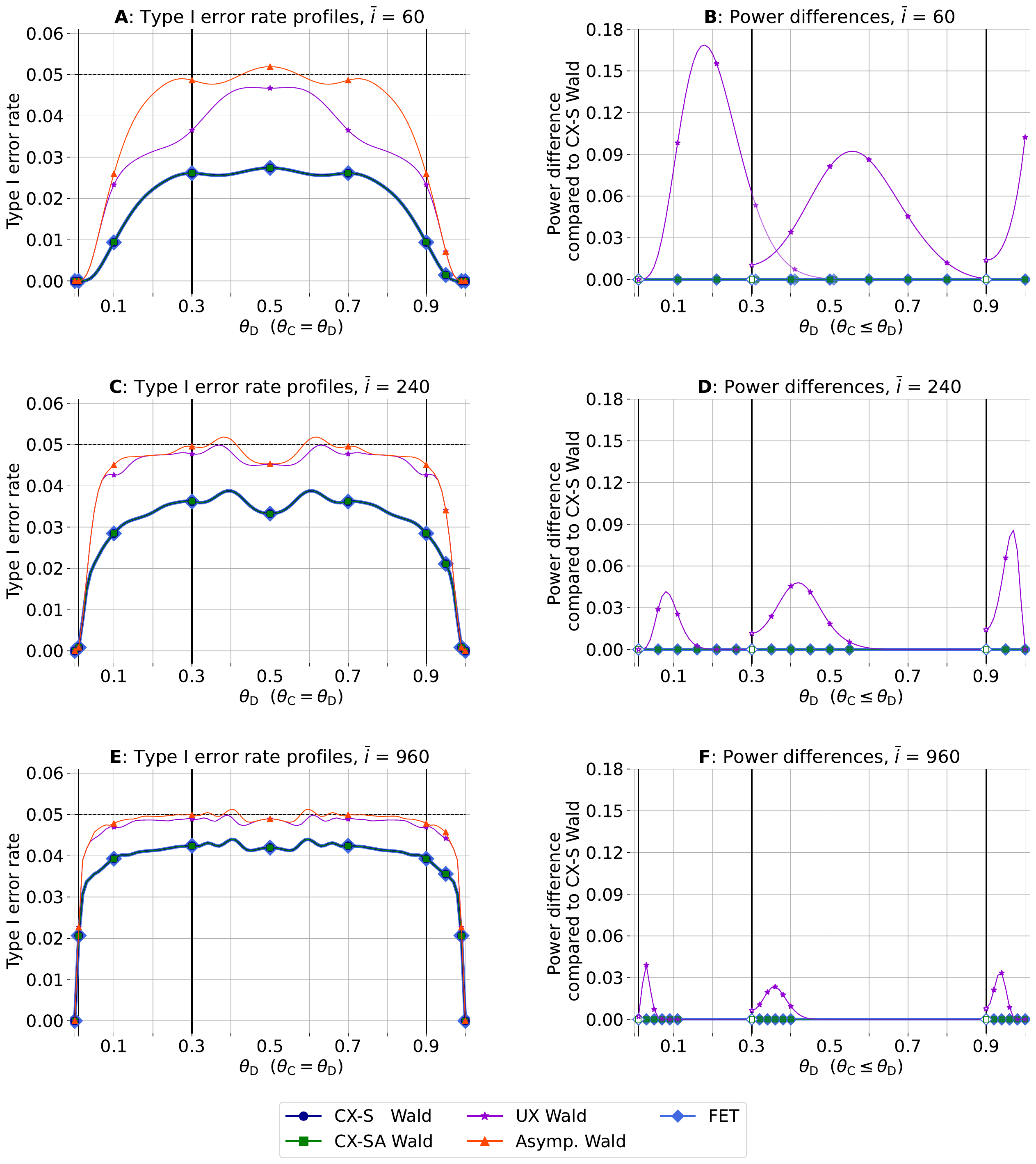}
		\caption{
			Subfigures A, B, C: Type~I~error rate under the equal allocation (EA) non-RA procedure of 
			\mbox{CX-S}, CX-SA, UX, asymptotic Wald tests, and FET (one-sided significance level~$2.5\%$) for different trial sizes~$\Iend$~(corresponding to rows of the figure).} Subfigures B, D, F: Power difference under the EA non-RA procedure of two-sided  CX-SA, UX Wald tests, and FET compared to the \mbox{CX-S} Wald test, \mbox{for~$\theta_\C\in\{0.01,0.3, 0.9\}$} and~\mbox{$\theta_\D\geq \theta_\C$}. The asymptotic Wald test~(orange line with triangle markers) is~omitted in the subfigures on the right as it does not control type I errors.  \label{fig:comparison_tests_TB}
	\end{figure}

	\autoref{fig:comparison_tests_TB} shows the comparison made in~Figure 1 of the paper when both treatment group sizes are (deterministically) equal to~$\Iend/2$, i.e. in case of a non-RA design with \emph{equal allocation}~(EA), such as a design with truncated binomial allocation or 1:1 permuted block allocation. As FET controls type~I~errors under EA, FET does not need to be corrected. As there is a one-to-one relation between the test statistic outcome and the number of successes in an arm when both allocations and total successes are fixed, the rejection rates of the  FET, CX-S, and CX-SA  Wald test coincide. 
	
	\autoref{fig:comparison_tests_TB} shows that the UX Wald test often has the highest power, where the maximum difference in power decreases with~$\Iend$. The asymptotic Wald test does not control type~I~errors for each value of~ the null success rate, even under a non-RA design. However, the maximum type~I~error rate becomes 
	closer to~$\alpha$ when~$\Iend$ increases. 
	The UX critical values~$c_u$ also indicate this behaviour,
	with values 2.066, 1.978, and 1.970 for~$\Iend=60,\,240,\, 960$ respectively, in comparison to the asymptotic critical value around 1.960~(see \autoref{tab:critical}). Similar curves can be found in, e.g.,~\citet[pg. 18]{shan2016exact}. 
	
	\autoref{tab:comparison_tests_TB_60}, 
	\autoref{tab:comparison_tests_TB_240}, and~\autoref{tab:comparison_tests_TB_960} show the (absolute) rejection rates under all tests for EA for~$\Iend=60, 240$ and~$960$. The rows of the table for~$\theta_\C=0.01,0.3$ and~$0.9$ correspond to the markers in \autoref{fig:comparison_tests_TB}, while each vertical line in the figure corresponds to a row in the tables where~$\theta_\C=\theta_\D$.  First, the tables indicate that the type~I~error rate inflation occurring under EA is easily missed, as there is only one case (for~$\Iend=60$) where type~I~error rate inflation occurs. Second, the tables show that the UX Wald test uniformly has highest power out of the considered tests and parameter values, as the asymptotic Wald test is not considered in the power comparison as it does not yield type I error control. Third, the rejection rate of the CX-S test at~$\theta_\C=\theta_\D\in\{0,1\}$ is~$0$, as EA is a fixed design.
	
		\autoref{tab:critical} shows the critical values of the unconditional Wald test and FET for both the RDP and EA designs for several trial sizes. The UX FET under an EA design is a special case of Boschloo's test~\citep{boschlooraised}. Even for~$\Iend=960,$ the critical values for the Wald test and FET are off from their respective asymptotic values~$1.96$ and~$5\%$ for both designs, where the critical value for the Wald test under the EA non-RA design 
	is closest to~$1.96.$ Note that the asymptotic Wald test inflates the type I error rate under the EA design for~$\Iend = 960$ as the UX critical value is above 1.96. As FET is a CX-S (or CX-SA) test for the EA design, the UX critical value is always above $5\%$ so that the UX test corrects for type~I~error rate deflation. In~\autoref{fig:comparison_tests}, the critical value for FET (corr.) under the RDP design was set to the minimum of $5\%$ and the value reported in \autoref{tab:critical} to make sure to only correct for type~I~error rate inflation, while in \autoref{fig:comparison_tests_TB}, the critical value for FET under the EA design was $5\%$ in accordance to the usual FET. 
	\begin{table}[tbp]
		\centering
		\vspace{-10mm}\thisfloatpagestyle{empty}
		\caption{Rejection rates and expected
			proportion of allocations on the superior arm~(EPASA, last column), both in percentage points, for the RDP RAR procedure under several parameter configurations, tests, and trial size~$\Iend = 60$.  Red indicates type I error rate inflation, green indicates highest power (excluding asymptotic test).}
		\label{tab:comparison_tests_RDP_60}
		 \begin{tabular}{lllrrrrrr}
\toprule
$\Iend$ & $\theta_\text{C}$ & $\theta_{\text{D}}$ & CX-S Wald & CX-SA Wald & UX Wald & Asymp. Wald & FET (corr.) & EPASA \\
\midrule
\multirow[t]{51}{*}{60} & \multirow[t]{6}{*}{0.0} & 0.0 & \phantom{00}0.14 & \phantom{00}0.00 & \phantom{00}0.00 & \phantom{00}0.00 & \phantom{00}0.00 & \phantom{0}50.00 \\
 &  & 0.1 & \textcolor{darkgreen}{\bf\phantom{0}13.64} & \phantom{00}0.49 & \phantom{00}0.17 & \phantom{00}0.40 & \phantom{00}0.15 & \phantom{0}73.16 \\
 &  & 0.2 & \textcolor{darkgreen}{\bf\phantom{0}32.39} & \phantom{00}2.69 & \phantom{00}8.40 & \phantom{0}13.64 & \phantom{00}4.09 & \phantom{0}80.94 \\
 &  & 0.3 & \textcolor{darkgreen}{\bf\phantom{0}57.80} & \phantom{00}8.04 & \phantom{0}41.51 & \phantom{0}51.78 & \phantom{0}27.42 & \phantom{0}84.24 \\
 &  & 0.4 & \textcolor{darkgreen}{\bf\phantom{0}79.73} & \phantom{0}19.85 & \phantom{0}76.16 & \phantom{0}82.16 & \phantom{0}62.08 & \phantom{0}86.00 \\
 &  & 0.5 & \phantom{0}91.83 & \phantom{0}38.95 & \textcolor{darkgreen}{\bf\phantom{0}92.29} & \phantom{0}94.37 & \phantom{0}81.57 & \phantom{0}87.08 \\
\cline{2-9}
 & \multirow[t]{6}{*}{0.01} & 0.01 & \phantom{00}1.40 & \phantom{00}0.04 & \phantom{00}0.00 & \phantom{00}0.00 & \phantom{00}0.00 & \phantom{0}50.00 \\
 &  & 0.11 & \textcolor{darkgreen}{\bf\phantom{0}12.95} & \phantom{00}0.69 & \phantom{00}0.26 & \phantom{00}0.58 & \phantom{00}0.21 & \phantom{0}72.34 \\
 &  & 0.21 & \textcolor{darkgreen}{\bf\phantom{0}31.19} & \phantom{00}2.98 & \phantom{00}9.46 & \phantom{0}14.92 & \phantom{00}4.65 & \phantom{0}80.56 \\
 &  & 0.31 & \textcolor{darkgreen}{\bf\phantom{0}55.96} & \phantom{00}8.42 & \phantom{0}42.03 & \phantom{0}51.54 & \phantom{0}28.76 & \phantom{0}84.08 \\
 &  & 0.41 & \textcolor{darkgreen}{\bf\phantom{0}77.33} & \phantom{0}20.26 & \phantom{0}74.72 & \phantom{0}80.43 & \phantom{0}61.22 & \phantom{0}85.93 \\
 &  & 0.51 & \phantom{0}90.10 & \phantom{0}39.11 & \textcolor{darkgreen}{\bf\phantom{0}90.62} & \phantom{0}93.01 & \phantom{0}80.14 & \phantom{0}87.05 \\
\cline{2-9}
 & \multirow[t]{6}{*}{0.05} & 0.05 & \phantom{00}3.83 & \phantom{00}0.41 & \phantom{00}0.00 & \phantom{00}0.01 & \phantom{00}0.00 & \phantom{0}50.00 \\
 &  & 0.15 & \textcolor{darkgreen}{\bf\phantom{0}10.98} & \phantom{00}1.37 & \phantom{00}0.93 & \phantom{00}1.81 & \phantom{00}0.50 & \phantom{0}69.86 \\
 &  & 0.25 & \textcolor{darkgreen}{\bf\phantom{0}27.37} & \phantom{00}3.79 & \phantom{0}13.38 & \phantom{0}19.00 & \phantom{00}7.52 & \phantom{0}79.18 \\
 &  & 0.35 & \textcolor{darkgreen}{\bf\phantom{0}49.49} & \phantom{00}9.66 & \phantom{0}42.59 & \phantom{0}49.83 & \phantom{0}31.86 & \phantom{0}83.47 \\
 &  & 0.45 & \textcolor{darkgreen}{\bf\phantom{0}69.72} & \phantom{0}21.64 & \phantom{0}69.41 & \phantom{0}74.81 & \phantom{0}57.13 & \phantom{0}85.67 \\
 &  & 0.55 & \phantom{0}84.63 & \phantom{0}39.59 & \textcolor{darkgreen}{\bf\phantom{0}85.23} & \phantom{0}88.45 & \phantom{0}76.05 & \phantom{0}86.95 \\
\cline{2-9}
 & \multirow[t]{6}{*}{0.1} & 0.1 & \phantom{00}4.18 & \phantom{00}0.88 & \phantom{00}0.06 & \phantom{00}0.15 & \phantom{00}0.05 & \phantom{0}50.00 \\
 &  & 0.2 & \textcolor{darkgreen}{\bf\phantom{00}9.63} & \phantom{00}1.82 & \phantom{00}2.45 & \phantom{00}4.10 & \phantom{00}1.17 & \phantom{0}67.86 \\
 &  & 0.3 & \textcolor{darkgreen}{\bf\phantom{0}23.95} & \phantom{00}4.49 & \phantom{0}16.79 & \phantom{0}21.65 & \phantom{0}11.18 & \phantom{0}77.82 \\
 &  & 0.4 & \textcolor{darkgreen}{\bf\phantom{0}43.44} & \phantom{0}10.93 & \phantom{0}41.30 & \phantom{0}47.43 & \phantom{0}31.87 & \phantom{0}82.82 \\
 &  & 0.5 & \phantom{0}63.63 & \phantom{0}22.87 & \textcolor{darkgreen}{\bf\phantom{0}64.22} & \phantom{0}69.79 & \phantom{0}53.08 & \phantom{0}85.40 \\
 &  & 0.6 & \phantom{0}80.21 & \phantom{0}40.14 & \textcolor{darkgreen}{\bf\phantom{0}80.60} & \phantom{0}84.31 & \phantom{0}73.81 & \phantom{0}86.86 \\
\cline{2-9}
 & \multirow[t]{6}{*}{0.3} & 0.3 & \phantom{00}4.57 & \phantom{00}1.66 & \phantom{00}3.41 & \phantom{00}4.75 & \phantom{00}2.10 & \phantom{0}50.00 \\
 &  & 0.4 & \textcolor{darkgreen}{\bf\phantom{00}7.97} & \phantom{00}2.78 & \phantom{00}7.62 & \phantom{00}9.85 & \phantom{00}5.09 & \phantom{0}64.61 \\
 &  & 0.5 & \phantom{0}18.36 & \phantom{00}6.49 & \textcolor{darkgreen}{\bf\phantom{0}18.69} & \phantom{0}22.66 & \phantom{0}13.88 & \phantom{0}75.19 \\
 &  & 0.6 & \phantom{0}35.54 & \phantom{0}13.97 & \textcolor{darkgreen}{\bf\phantom{0}36.22} & \phantom{0}41.36 & \phantom{0}31.20 & \phantom{0}81.59 \\
 &  & 0.7 & \phantom{0}55.89 & \phantom{0}27.18 & \textcolor{darkgreen}{\bf\phantom{0}56.75} & \phantom{0}62.17 & \phantom{0}54.79 & \phantom{0}85.12 \\
 &  & 0.8 & \phantom{0}75.88 & \phantom{0}47.21 & \phantom{0}74.90 & \phantom{0}78.96 & \textcolor{darkgreen}{\bf\phantom{0}76.26} & \phantom{0}87.08 \\
\cline{2-9}
 & \multirow[t]{6}{*}{0.5} & 0.5 & \phantom{00}4.75 & \phantom{00}1.88 & \phantom{00}4.97 & \textcolor{red}{\bf \phantom{00}6.50} & \phantom{00}3.29 & \phantom{0}50.00 \\
 &  & 0.6 & \phantom{00}8.03 & \phantom{00}3.15 & \textcolor{darkgreen}{\bf\phantom{00}8.34} & \phantom{0}10.48 & \phantom{00}6.77 & \phantom{0}64.26 \\
 &  & 0.7 & \phantom{0}18.96 & \phantom{00}8.00 & \textcolor{darkgreen}{\bf\phantom{0}19.25} & \phantom{0}23.20 & \phantom{0}18.82 & \phantom{0}75.48 \\
 &  & 0.8 & \phantom{0}39.92 & \phantom{0}19.26 & \phantom{0}37.99 & \phantom{0}43.42 & \textcolor{darkgreen}{\bf\phantom{0}41.04} & \phantom{0}82.57 \\
 &  & 0.9 & \textcolor{darkgreen}{\bf\phantom{0}68.02} & \phantom{0}42.33 & \phantom{0}60.18 & \phantom{0}66.98 & \phantom{0}67.86 & \phantom{0}86.42 \\
 &  & 1.0 & \textcolor{darkgreen}{\bf\phantom{0}95.77} & \phantom{0}83.40 & \phantom{0}80.02 & \phantom{0}85.71 & \phantom{0}91.33 & \phantom{0}88.42 \\
\cline{2-9}
 & \multirow[t]{4}{*}{0.7} & 0.7 & \phantom{00}4.58 & \phantom{00}1.75 & \phantom{00}4.50 & \textcolor{red}{\bf \phantom{00}6.08} & \phantom{00}4.69 & \phantom{0}50.00 \\
 &  & 0.8 & \phantom{00}9.44 & \phantom{00}3.67 & \phantom{00}8.14 & \phantom{0}10.60 & \textcolor{darkgreen}{\bf\phantom{00}9.97} & \phantom{0}66.11 \\
 &  & 0.9 & \textcolor{darkgreen}{\bf\phantom{0}30.16} & \phantom{0}13.73 & \phantom{0}21.85 & \phantom{0}28.10 & \phantom{0}30.12 & \phantom{0}79.07 \\
 &  & 1.0 & \textcolor{darkgreen}{\bf\phantom{0}81.20} & \phantom{0}53.68 & \phantom{0}48.85 & \phantom{0}59.06 & \phantom{0}72.11 & \phantom{0}86.52 \\
\cline{2-9}
 & \multirow[t]{2}{*}{0.9} & 0.9 & \phantom{00}4.78 & \phantom{00}1.21 & \phantom{00}1.93 & \phantom{00}3.38 & \phantom{00}4.36 & \phantom{0}50.00 \\
 &  & 1.0 & \textcolor{darkgreen}{\bf\phantom{0}30.58} & \phantom{00}8.58 & \phantom{00}6.45 & \phantom{0}10.77 & \phantom{0}20.99 & \phantom{0}73.28 \\
\cline{2-9}
 & \multirow[t]{6}{*}{0.95} & 0.95 & \phantom{00}4.50 & \phantom{00}0.66 & \phantom{00}0.80 & \phantom{00}1.70 & \phantom{00}3.17 & \phantom{0}50.00 \\
 &  & 0.96 & \textcolor{darkgreen}{\bf\phantom{00}4.64} & \phantom{00}0.64 & \phantom{00}0.73 & \phantom{00}1.57 & \phantom{00}3.11 & \phantom{0}52.60 \\
 &  & 0.97 & \textcolor{darkgreen}{\bf\phantom{00}5.29} & \phantom{00}0.72 & \phantom{00}0.73 & \phantom{00}1.59 & \phantom{00}3.36 & \phantom{0}55.20 \\
 &  & 0.98 & \textcolor{darkgreen}{\bf\phantom{00}6.61} & \phantom{00}0.92 & \phantom{00}0.81 & \phantom{00}1.75 & \phantom{00}3.96 & \phantom{0}57.75 \\
 &  & 0.99 & \textcolor{darkgreen}{\bf\phantom{00}8.81} & \phantom{00}1.26 & \phantom{00}0.96 & \phantom{00}2.03 & \phantom{00}4.99 & \phantom{0}60.22 \\
 &  & 1.0 & \textcolor{darkgreen}{\bf\phantom{0}12.24} & \phantom{00}1.78 & \phantom{00}1.19 & \phantom{00}2.44 & \phantom{00}6.51 & \phantom{0}62.55 \\
\cline{2-9}
 & \multirow[t]{2}{*}{0.99} & 0.99 & \phantom{00}2.67 & \phantom{00}0.08 & \phantom{00}0.05 & \phantom{00}0.14 & \phantom{00}0.60 & \phantom{0}50.00 \\
 &  & 1.0 & \textcolor{darkgreen}{\bf\phantom{00}2.57} & \phantom{00}0.06 & \phantom{00}0.03 & \phantom{00}0.08 & \phantom{00}0.45 & \phantom{0}52.31 \\
\cline{2-9}
 & 1.0 & 1.0 & \phantom{00}1.51 & \phantom{00}0.00 & \phantom{00}0.00 & \phantom{00}0.00 & \phantom{00}0.00 & \phantom{0}50.00 \\
\cline{1-9} \cline{2-9}

\end{tabular}

	\end{table}
	
	\begin{table}[tbp]
		\centering
		\vspace{-20mm}\thisfloatpagestyle{empty}
		\caption{Rejection rates and expected
			proportion of allocations on the superior arm~(EPASA, last column), both in percentage points, for the RDP RAR procedure  under several parameter configurations, tests, and trial size ~$\Iend = 240$.  Red indicates type I error rate inflation, green indicates highest power (excluding asymptotic test).}
		\label{tab:comparison_tests_RDP_240}
		 \begin{tabular}{lllrrrrrr}
\toprule
$\Iend$ & $\theta_\text{C}$ & $\theta_{\text{D}}$ & CX-S Wald & CX-SA Wald & UX Wald & Asymp. Wald & FET (corr.) & EPASA \\
\midrule
\multirow[t]{54}{*}{240} & \multirow[t]{6}{*}{0.0} & 0.0 & \phantom{00}0.14 & \phantom{00}0.00 & \phantom{00}0.00 & \phantom{00}0.00 & \phantom{00}0.00 & \phantom{0}50.00 \\
 &  & 0.05 & \textcolor{darkgreen}{\bf\phantom{0}31.22} & \phantom{00}2.46 & \phantom{00}3.51 & \phantom{00}6.35 & \phantom{00}1.48 & \phantom{0}79.78 \\
 &  & 0.1 & \textcolor{darkgreen}{\bf\phantom{0}75.72} & \phantom{00}8.44 & \phantom{0}60.05 & \phantom{0}69.84 & \phantom{0}30.52 & \phantom{0}84.76 \\
 &  & 0.15 & \textcolor{darkgreen}{\bf\phantom{0}97.89} & \phantom{0}24.88 & \phantom{0}96.73 & \phantom{0}98.11 & \phantom{0}83.91 & \phantom{0}86.59 \\
 &  & 0.2 & \textcolor{darkgreen}{\bf\phantom{0}99.92} & \phantom{0}55.10 & \phantom{0}99.90 & \phantom{0}99.95 & \phantom{0}98.38 & \phantom{0}87.53 \\
 &  & 0.25 & \textcolor{darkgreen}{\bf100.00} & \phantom{0}82.16 & \textcolor{darkgreen}{\bf100.00} & 100.00 & \phantom{0}99.88 & \phantom{0}88.10 \\
\cline{2-9}
 & \multirow[t]{6}{*}{0.01} & 0.01 & \phantom{00}4.41 & \phantom{00}0.34 & \phantom{00}0.00 & \phantom{00}0.00 & \phantom{00}0.00 & \phantom{0}50.00 \\
 &  & 0.06 & \textcolor{darkgreen}{\bf\phantom{0}25.94} & \phantom{00}3.07 & \phantom{00}5.82 & \phantom{00}9.48 & \phantom{00}2.08 & \phantom{0}77.93 \\
 &  & 0.11 & \textcolor{darkgreen}{\bf\phantom{0}64.62} & \phantom{00}9.12 & \phantom{0}55.51 & \phantom{0}63.18 & \phantom{0}31.98 & \phantom{0}84.17 \\
 &  & 0.16 & \textcolor{darkgreen}{\bf\phantom{0}88.55} & \phantom{0}24.94 & \phantom{0}87.54 & \phantom{0}90.60 & \phantom{0}76.03 & \phantom{0}86.33 \\
 &  & 0.21 & \textcolor{darkgreen}{\bf\phantom{0}97.16} & \phantom{0}52.07 & \phantom{0}97.09 & \phantom{0}98.02 & \phantom{0}93.43 & \phantom{0}87.40 \\
 &  & 0.26 & \phantom{0}99.44 & \phantom{0}76.94 & \textcolor{darkgreen}{\bf\phantom{0}99.45} & \phantom{0}99.66 & \phantom{0}98.43 & \phantom{0}88.02 \\
\cline{2-9}
 & \multirow[t]{6}{*}{0.05} & 0.05 & \phantom{00}4.74 & \phantom{00}1.53 & \phantom{00}0.57 & \phantom{00}1.09 & \phantom{00}0.18 & \phantom{0}50.00 \\
 &  & 0.1 & \textcolor{darkgreen}{\bf\phantom{0}15.71} & \phantom{00}3.92 & \phantom{0}11.88 & \phantom{0}15.20 & \phantom{00}5.58 & \phantom{0}72.90 \\
 &  & 0.15 & \textcolor{darkgreen}{\bf\phantom{0}40.80} & \phantom{0}10.52 & \phantom{0}39.12 & \phantom{0}44.89 & \phantom{0}28.45 & \phantom{0}81.93 \\
 &  & 0.2 & \textcolor{darkgreen}{\bf\phantom{0}67.94} & \phantom{0}23.73 & \phantom{0}67.51 & \phantom{0}72.72 & \phantom{0}56.55 & \phantom{0}85.33 \\
 &  & 0.25 & \phantom{0}86.18 & \phantom{0}43.10 & \textcolor{darkgreen}{\bf\phantom{0}86.23} & \phantom{0}89.32 & \phantom{0}78.67 & \phantom{0}86.88 \\
 &  & 0.3 & \phantom{0}95.04 & \phantom{0}63.84 & \textcolor{darkgreen}{\bf\phantom{0}95.10} & \phantom{0}96.47 & \phantom{0}91.79 & \phantom{0}87.73 \\
\cline{2-9}
 & \multirow[t]{6}{*}{0.1} & 0.1 & \phantom{00}4.83 & \phantom{00}2.01 & \phantom{00}3.76 & \textcolor{red}{\bf \phantom{00}5.04} & \phantom{00}1.57 & \phantom{0}50.00 \\
 &  & 0.15 & \textcolor{darkgreen}{\bf\phantom{0}11.68} & \phantom{00}4.28 & \phantom{0}11.04 & \phantom{0}13.76 & \phantom{00}7.08 & \phantom{0}69.53 \\
 &  & 0.2 & \textcolor{darkgreen}{\bf\phantom{0}30.64} & \phantom{0}10.71 & \phantom{0}30.28 & \phantom{0}35.36 & \phantom{0}22.53 & \phantom{0}79.72 \\
 &  & 0.25 & \phantom{0}54.90 & \phantom{0}21.66 & \textcolor{darkgreen}{\bf\phantom{0}54.95} & \phantom{0}60.71 & \phantom{0}45.49 & \phantom{0}84.20 \\
 &  & 0.3 & \phantom{0}75.51 & \phantom{0}37.24 & \textcolor{darkgreen}{\bf\phantom{0}75.71} & \phantom{0}80.16 & \phantom{0}68.84 & \phantom{0}86.28 \\
 &  & 0.35 & \phantom{0}88.85 & \phantom{0}55.23 & \textcolor{darkgreen}{\bf\phantom{0}88.94} & \phantom{0}91.49 & \phantom{0}84.99 & \phantom{0}87.39 \\
\cline{2-9}
 & \multirow[t]{6}{*}{0.3} & 0.3 & \phantom{00}4.94 & \phantom{00}2.73 & \phantom{00}4.98 & \textcolor{red}{\bf \phantom{00}6.56} & \phantom{00}3.67 & \phantom{0}50.00 \\
 &  & 0.35 & \phantom{00}8.65 & \phantom{00}4.48 & \textcolor{darkgreen}{\bf\phantom{00}8.68} & \phantom{0}11.05 & \phantom{00}6.87 & \phantom{0}64.76 \\
 &  & 0.4 & \textcolor{darkgreen}{\bf\phantom{0}19.93} & \phantom{00}9.70 & \phantom{0}19.81 & \phantom{0}24.01 & \phantom{0}17.04 & \phantom{0}75.35 \\
 &  & 0.45 & \textcolor{darkgreen}{\bf\phantom{0}37.72} & \phantom{0}18.50 & \phantom{0}37.26 & \phantom{0}43.09 & \phantom{0}33.87 & \phantom{0}81.54 \\
 &  & 0.5 & \textcolor{darkgreen}{\bf\phantom{0}57.90} & \phantom{0}30.93 & \phantom{0}57.07 & \phantom{0}62.98 & \phantom{0}54.01 & \phantom{0}84.84 \\
 &  & 0.55 & \textcolor{darkgreen}{\bf\phantom{0}75.49} & \phantom{0}46.33 & \phantom{0}74.45 & \phantom{0}79.12 & \phantom{0}72.40 & \phantom{0}86.62 \\
\cline{2-9}
 & \multirow[t]{6}{*}{0.5} & 0.5 & \phantom{00}4.93 & \phantom{00}2.84 & \phantom{00}4.63 & \textcolor{red}{\bf \phantom{00}6.32} & \phantom{00}4.09 & \phantom{0}50.00 \\
 &  & 0.55 & \textcolor{darkgreen}{\bf\phantom{00}8.35} & \phantom{00}4.56 & \phantom{00}7.75 & \phantom{0}10.24 & \phantom{00}7.28 & \phantom{0}63.98 \\
 &  & 0.6 & \textcolor{darkgreen}{\bf\phantom{0}19.25} & \phantom{0}10.13 & \phantom{0}17.80 & \phantom{0}22.32 & \phantom{0}17.62 & \phantom{0}74.72 \\
 &  & 0.65 & \textcolor{darkgreen}{\bf\phantom{0}37.40} & \phantom{0}20.09 & \phantom{0}34.73 & \phantom{0}41.09 & \phantom{0}35.30 & \phantom{0}81.36 \\
 &  & 0.7 & \textcolor{darkgreen}{\bf\phantom{0}58.86} & \phantom{0}34.62 & \phantom{0}55.19 & \phantom{0}61.81 & \phantom{0}56.79 & \phantom{0}84.98 \\
 &  & 0.75 & \textcolor{darkgreen}{\bf\phantom{0}77.60} & \phantom{0}52.98 & \phantom{0}73.91 & \phantom{0}79.19 & \phantom{0}76.15 & \phantom{0}86.91 \\
\cline{2-9}
 & \multirow[t]{6}{*}{0.7} & 0.7 & \phantom{00}4.94 & \phantom{00}2.72 & \phantom{00}3.88 & \textcolor{red}{\bf \phantom{00}5.65} & \phantom{00}4.43 & \phantom{0}50.00 \\
 &  & 0.75 & \textcolor{darkgreen}{\bf\phantom{00}9.53} & \phantom{00}5.11 & \phantom{00}7.49 & \phantom{0}10.32 & \phantom{00}8.77 & \phantom{0}65.42 \\
 &  & 0.8 & \textcolor{darkgreen}{\bf\phantom{0}25.10} & \phantom{0}13.61 & \phantom{0}20.35 & \phantom{0}25.73 & \phantom{0}23.87 & \phantom{0}77.07 \\
 &  & 0.85 & \textcolor{darkgreen}{\bf\phantom{0}50.73} & \phantom{0}30.48 & \phantom{0}42.80 & \phantom{0}50.02 & \phantom{0}49.89 & \phantom{0}83.67 \\
 &  & 0.9 & \phantom{0}77.00 & \phantom{0}56.02 & \phantom{0}68.15 & \phantom{0}74.74 & \textcolor{darkgreen}{\bf\phantom{0}77.36} & \phantom{0}86.83 \\
 &  & 0.95 & \phantom{0}93.89 & \phantom{0}83.90 & \phantom{0}87.72 & \phantom{0}91.28 & \textcolor{darkgreen}{\bf\phantom{0}94.44} & \phantom{0}88.35 \\
\cline{2-9}
 & \multirow[t]{3}{*}{0.9} & 0.9 & \phantom{00}4.92 & \phantom{00}2.10 & \phantom{00}2.19 & \phantom{00}3.72 & \phantom{00}4.94 & \phantom{0}50.00 \\
 &  & 0.95 & \textcolor{darkgreen}{\bf\phantom{0}20.47} & \phantom{0}10.05 & \phantom{00}9.80 & \phantom{0}14.24 & \phantom{0}20.05 & \phantom{0}73.11 \\
 &  & 1.0 & \textcolor{darkgreen}{\bf\phantom{0}90.11} & \phantom{0}74.57 & \phantom{0}49.01 & \phantom{0}60.72 & \phantom{0}81.96 & \phantom{0}85.92 \\
\cline{2-9}
 & \multirow[t]{6}{*}{0.95} & 0.95 & \phantom{00}4.90 & \phantom{00}1.63 & \phantom{00}1.22 & \phantom{00}2.31 & \phantom{00}4.71 & \phantom{0}50.00 \\
 &  & 0.96 & \textcolor{darkgreen}{\bf\phantom{00}5.94} & \phantom{00}1.97 & \phantom{00}1.43 & \phantom{00}2.64 & \phantom{00}5.54 & \phantom{0}56.35 \\
 &  & 0.97 & \textcolor{darkgreen}{\bf\phantom{00}9.70} & \phantom{00}3.46 & \phantom{00}2.41 & \phantom{00}4.23 & \phantom{00}8.76 & \phantom{0}62.91 \\
 &  & 0.98 & \textcolor{darkgreen}{\bf\phantom{0}17.88} & \phantom{00}7.05 & \phantom{00}4.58 & \phantom{00}7.69 & \phantom{0}15.77 & \phantom{0}69.35 \\
 &  & 0.99 & \textcolor{darkgreen}{\bf\phantom{0}34.54} & \phantom{0}15.26 & \phantom{00}8.78 & \phantom{0}14.59 & \phantom{0}29.38 & \phantom{0}75.15 \\
 &  & 1.0 & \textcolor{darkgreen}{\bf\phantom{0}67.60} & \phantom{0}35.61 & \phantom{0}17.49 & \phantom{0}27.09 & \phantom{0}51.69 & \phantom{0}79.75 \\
\cline{2-9}
 & \multirow[t]{2}{*}{0.99} & 0.99 & \phantom{00}4.81 & \phantom{00}0.60 & \phantom{00}0.07 & \phantom{00}0.26 & \phantom{00}2.23 & \phantom{0}50.00 \\
 &  & 1.0 & \textcolor{darkgreen}{\bf\phantom{0}11.42} & \phantom{00}1.50 & \phantom{00}0.13 & \phantom{00}0.39 & \phantom{00}3.18 & \phantom{0}55.10 \\
\cline{2-9}
 & 1.0 & 1.0 & \phantom{00}4.97 & \phantom{00}0.00 & \phantom{00}0.00 & \phantom{00}0.00 & \phantom{00}0.00 & \phantom{0}50.00 \\
\cline{1-9} \cline{2-9}

\end{tabular}

	\end{table}
	
	\begin{table}[tbp]
		\centering\thisfloatpagestyle{empty}
		\vspace{-23mm}
		\caption{Rejection rates and expected
			proportion of allocations on the superior arm~(EPASA, last column), both in percentage points, for the RDP RAR procedure under several parameter configurations, tests, and trial size ~$\Iend = 960$.  Red indicates type I error rate inflation, green indicates highest power (excluding asymptotic test).}
		\label{tab:comparison_tests_RDP_960}
		 \begin{tabular}{lllrrrrrr}
\toprule
$\Iend$ & $\theta_\text{C}$ & $\theta_{\text{D}}$ &CX-S Wald & CX-SA Wald & UX Wald & Asymp. Wald & FET (corr.) & EPASA \\
\midrule
\multirow[t]{57}{*}{960} & \multirow[t]{6}{*}{0.0} & 0.0 & \phantom{00}0.14 & \phantom{00}0.00 & \phantom{00}0.00 & \phantom{00}0.00 & \phantom{00}0.00 & \phantom{0}50.00 \\
 &  & 0.02 & \textcolor{darkgreen}{\bf\phantom{0}56.83} & \phantom{00}5.17 & \phantom{0}26.51 & \phantom{0}36.39 & \phantom{0}11.55 & \phantom{0}83.13 \\
 &  & 0.04 & \textcolor{darkgreen}{\bf\phantom{0}99.14} & \phantom{0}20.41 & \phantom{0}98.34 & \phantom{0}99.21 & \phantom{0}92.75 & \phantom{0}86.50 \\
 &  & 0.06 & \textcolor{darkgreen}{\bf100.00} & \phantom{0}62.98 & \textcolor{darkgreen}{\bf100.00} & 100.00 & \phantom{0}99.96 & \phantom{0}87.68 \\
 &  & 0.08 & \textcolor{darkgreen}{\bf100.00} & \phantom{0}95.50 & \textcolor{darkgreen}{\bf100.00} & 100.00 & \textcolor{darkgreen}{\bf100.00} & \phantom{0}88.28 \\
 &  & 0.1 & \textcolor{darkgreen}{\bf100.00} & \phantom{0}99.88 & \textcolor{darkgreen}{\bf100.00} & 100.00 & \textcolor{darkgreen}{\bf100.00} & \phantom{0}88.64 \\
\cline{2-9}
 & \multirow[t]{6}{*}{0.01} & 0.01 & \phantom{00}4.90 & \phantom{00}1.41 & \phantom{00}0.16 & \phantom{00}0.35 & \phantom{00}0.06 & \phantom{0}50.00 \\
 &  & 0.03 & \textcolor{darkgreen}{\bf\phantom{0}31.44} & \phantom{00}6.38 & \phantom{0}26.70 & \phantom{0}31.62 & \phantom{0}17.52 & \phantom{0}79.41 \\
 &  & 0.05 & \textcolor{darkgreen}{\bf\phantom{0}72.93} & \phantom{0}20.49 & \phantom{0}71.27 & \phantom{0}76.25 & \phantom{0}58.79 & \phantom{0}85.38 \\
 &  & 0.07 & \textcolor{darkgreen}{\bf\phantom{0}93.51} & \phantom{0}49.13 & \phantom{0}93.34 & \phantom{0}95.10 & \phantom{0}88.43 & \phantom{0}87.19 \\
 &  & 0.09 & \textcolor{darkgreen}{\bf\phantom{0}98.94} & \phantom{0}76.21 & \textcolor{darkgreen}{\bf\phantom{0}98.94} & \phantom{0}99.30 & \phantom{0}97.82 & \phantom{0}88.01 \\
 &  & 0.11 & \phantom{0}99.87 & \phantom{0}91.90 & \textcolor{darkgreen}{\bf\phantom{0}99.88} & \phantom{0}99.93 & \phantom{0}99.70 & \phantom{0}88.48 \\
\cline{2-9}
 & \multirow[t]{6}{*}{0.05} & 0.05 & \phantom{00}4.97 & \phantom{00}2.62 & \phantom{00}4.74 & \textcolor{red}{\bf \phantom{00}6.11} & \phantom{00}2.83 & \phantom{0}50.00 \\
 &  & 0.07 & \textcolor{darkgreen}{\bf\phantom{0}14.01} & \phantom{00}6.23 & \phantom{0}13.87 & \phantom{0}16.96 & \phantom{0}10.01 & \phantom{0}71.32 \\
 &  & 0.09 & \phantom{0}37.75 & \phantom{0}15.45 & \textcolor{darkgreen}{\bf\phantom{0}37.77} & \phantom{0}43.15 & \phantom{0}31.06 & \phantom{0}81.32 \\
 &  & 0.11 & \phantom{0}64.81 & \phantom{0}29.63 & \textcolor{darkgreen}{\bf\phantom{0}64.88} & \phantom{0}70.10 & \phantom{0}57.73 & \phantom{0}85.21 \\
 &  & 0.13 & \phantom{0}84.24 & \phantom{0}48.01 & \textcolor{darkgreen}{\bf\phantom{0}84.29} & \phantom{0}87.56 & \phantom{0}79.30 & \phantom{0}86.91 \\
 &  & 0.15 & \phantom{0}94.28 & \phantom{0}66.94 & \textcolor{darkgreen}{\bf\phantom{0}94.29} & \phantom{0}95.79 & \phantom{0}91.77 & \phantom{0}87.80 \\
\cline{2-9}
 & \multirow[t]{6}{*}{0.1} & 0.1 & \phantom{00}4.98 & \phantom{00}3.08 & \phantom{00}5.00 & \textcolor{red}{\bf \phantom{00}6.53} & \phantom{00}3.56 & \phantom{0}50.00 \\
 &  & 0.12 & \phantom{0}10.26 & \phantom{00}5.58 & \textcolor{darkgreen}{\bf\phantom{0}10.29} & \phantom{0}12.89 & \phantom{00}7.98 & \phantom{0}67.14 \\
 &  & 0.14 & \textcolor{darkgreen}{\bf\phantom{0}25.50} & \phantom{0}12.39 & \textcolor{darkgreen}{\bf\phantom{0}25.50} & \phantom{0}30.17 & \phantom{0}21.35 & \phantom{0}77.79 \\
 &  & 0.16 & \textcolor{darkgreen}{\bf\phantom{0}47.10} & \phantom{0}22.73 & \phantom{0}46.99 & \phantom{0}52.74 & \phantom{0}41.78 & \phantom{0}83.10 \\
 &  & 0.18 & \textcolor{darkgreen}{\bf\phantom{0}68.10} & \phantom{0}36.30 & \phantom{0}67.90 & \phantom{0}72.99 & \phantom{0}63.35 & \phantom{0}85.67 \\
 &  & 0.2 & \textcolor{darkgreen}{\bf\phantom{0}83.57} & \phantom{0}52.02 & \phantom{0}83.35 & \phantom{0}86.82 & \phantom{0}80.17 & \phantom{0}87.01 \\
\cline{2-9}
 & \multirow[t]{6}{*}{0.3} & 0.3 & \phantom{00}4.99 & \phantom{00}3.55 & \phantom{00}4.70 & \textcolor{red}{\bf \phantom{00}6.38} & \phantom{00}4.26 & \phantom{0}50.00 \\
 &  & 0.32 & \textcolor{darkgreen}{\bf\phantom{00}7.50} & \phantom{00}4.95 & \phantom{00}7.07 & \phantom{00}9.32 & \phantom{00}6.59 & \phantom{0}62.14 \\
 &  & 0.34 & \textcolor{darkgreen}{\bf\phantom{0}15.20} & \phantom{00}9.06 & \phantom{0}14.41 & \phantom{0}18.06 & \phantom{0}13.76 & \phantom{0}71.87 \\
 &  & 0.36 & \textcolor{darkgreen}{\bf\phantom{0}27.91} & \phantom{0}15.72 & \phantom{0}26.67 & \phantom{0}31.85 & \phantom{0}25.85 & \phantom{0}78.49 \\
 &  & 0.38 & \textcolor{darkgreen}{\bf\phantom{0}43.94} & \phantom{0}24.63 & \phantom{0}42.34 & \phantom{0}48.42 & \phantom{0}41.48 & \phantom{0}82.55 \\
 &  & 0.4 & \textcolor{darkgreen}{\bf\phantom{0}60.40} & \phantom{0}35.47 & \phantom{0}58.70 & \phantom{0}64.63 & \phantom{0}57.93 & \phantom{0}84.96 \\
\cline{2-9}
 & \multirow[t]{6}{*}{0.5} & 0.5 & \phantom{00}4.99 & \phantom{00}3.64 & \phantom{00}4.36 & \textcolor{red}{\bf \phantom{00}6.08} & \phantom{00}4.49 & \phantom{0}50.00 \\
 &  & 0.52 & \textcolor{darkgreen}{\bf\phantom{00}7.21} & \phantom{00}4.91 & \phantom{00}6.35 & \phantom{00}8.57 & \phantom{00}6.59 & \phantom{0}61.29 \\
 &  & 0.54 & \textcolor{darkgreen}{\bf\phantom{0}14.12} & \phantom{00}8.79 & \phantom{0}12.66 & \phantom{0}16.19 & \phantom{0}13.17 & \phantom{0}70.76 \\
 &  & 0.56 & \textcolor{darkgreen}{\bf\phantom{0}25.83} & \phantom{0}15.28 & \phantom{0}23.56 & \phantom{0}28.69 & \phantom{0}24.47 & \phantom{0}77.56 \\
 &  & 0.58 & \textcolor{darkgreen}{\bf\phantom{0}41.14} & \phantom{0}24.19 & \phantom{0}38.20 & \phantom{0}44.44 & \phantom{0}39.50 & \phantom{0}81.93 \\
 &  & 0.6 & \textcolor{darkgreen}{\bf\phantom{0}57.50} & \phantom{0}35.18 & \phantom{0}54.34 & \phantom{0}60.68 & \phantom{0}55.86 & \phantom{0}84.59 \\
\cline{2-9}
 & \multirow[t]{6}{*}{0.7} & 0.7 & \phantom{00}4.99 & \phantom{00}3.52 & \phantom{00}3.94 & \textcolor{red}{\bf \phantom{00}5.67} & \phantom{00}4.65 & \phantom{0}50.00 \\
 &  & 0.72 & \textcolor{darkgreen}{\bf\phantom{00}7.78} & \phantom{00}5.13 & \phantom{00}6.28 & \phantom{00}8.63 & \phantom{00}7.32 & \phantom{0}62.35 \\
 &  & 0.74 & \textcolor{darkgreen}{\bf\phantom{0}16.71} & \phantom{0}10.23 & \phantom{0}13.97 & \phantom{0}18.00 & \phantom{0}15.93 & \phantom{0}72.52 \\
 &  & 0.76 & \textcolor{darkgreen}{\bf\phantom{0}31.84} & \phantom{0}19.07 & \phantom{0}27.60 & \phantom{0}33.50 & \phantom{0}30.76 & \phantom{0}79.44 \\
 &  & 0.78 & \textcolor{darkgreen}{\bf\phantom{0}50.77} & \phantom{0}31.41 & \phantom{0}45.61 & \phantom{0}52.40 & \phantom{0}49.63 & \phantom{0}83.57 \\
 &  & 0.8 & \textcolor{darkgreen}{\bf\phantom{0}69.18} & \phantom{0}46.52 & \phantom{0}64.22 & \phantom{0}70.37 & \phantom{0}68.28 & \phantom{0}85.90 \\
\cline{2-9}
 & \multirow[t]{6}{*}{0.9} & 0.9 & \phantom{00}4.98 & \phantom{00}2.97 & \phantom{00}2.98 & \phantom{00}4.63 & \phantom{00}4.74 & \phantom{0}50.00 \\
 &  & 0.92 & \textcolor{darkgreen}{\bf\phantom{0}12.63} & \phantom{00}7.26 & \phantom{00}8.27 & \phantom{0}11.57 & \phantom{0}12.10 & \phantom{0}68.46 \\
 &  & 0.94 & \textcolor{darkgreen}{\bf\phantom{0}38.49} & \phantom{0}23.38 & \phantom{0}28.55 & \phantom{0}35.44 & \phantom{0}37.75 & \phantom{0}80.69 \\
 &  & 0.96 & \textcolor{darkgreen}{\bf\phantom{0}73.10} & \phantom{0}53.23 & \phantom{0}61.37 & \phantom{0}68.71 & \phantom{0}73.00 & \phantom{0}86.05 \\
 &  & 0.98 & \phantom{0}94.95 & \phantom{0}86.63 & \phantom{0}88.79 & \phantom{0}92.22 & \textcolor{darkgreen}{\bf\phantom{0}95.59} & \phantom{0}88.11 \\
 &  & 1.0 & \textcolor{darkgreen}{\bf\phantom{0}99.99} & \phantom{0}99.96 & \phantom{0}99.22 & \phantom{0}99.79 & \phantom{0}99.96 & \phantom{0}88.99 \\
\cline{2-9}
 & \multirow[t]{6}{*}{0.95} & 0.95 & \phantom{00}4.98 & \phantom{00}2.59 & \phantom{00}2.26 & \phantom{00}3.74 & \phantom{00}4.76 & \phantom{0}50.00 \\
 &  & 0.96 & \textcolor{darkgreen}{\bf\phantom{00}8.80} & \phantom{00}4.56 & \phantom{00}4.28 & \phantom{00}6.60 & \phantom{00}8.30 & \phantom{0}63.14 \\
 &  & 0.97 & \textcolor{darkgreen}{\bf\phantom{0}22.51} & \phantom{0}12.42 & \phantom{0}12.44 & \phantom{0}17.35 & \phantom{0}21.85 & \phantom{0}74.44 \\
 &  & 0.98 & \phantom{0}47.75 & \phantom{0}30.07 & \phantom{0}30.07 & \phantom{0}38.13 & \textcolor{darkgreen}{\bf\phantom{0}48.07} & \phantom{0}81.92 \\
 &  & 0.99 & \phantom{0}77.22 & \phantom{0}60.23 & \phantom{0}56.15 & \phantom{0}64.92 & \textcolor{darkgreen}{\bf\phantom{0}78.15} & \phantom{0}85.84 \\
 &  & 1.0 & \textcolor{darkgreen}{\bf\phantom{0}98.86} & \phantom{0}96.75 & \phantom{0}82.42 & \phantom{0}91.43 & \phantom{0}97.50 & \phantom{0}87.60 \\
\cline{2-9}
 & \multirow[t]{2}{*}{0.99} & 0.99 & \phantom{00}4.97 & \phantom{00}1.52 & \phantom{00}0.64 & \phantom{00}1.40 & \phantom{00}4.52 & \phantom{0}50.00 \\
 &  & 1.0 & \textcolor{darkgreen}{\bf\phantom{0}58.01} & \phantom{0}23.29 & \phantom{0}10.46 & \phantom{0}19.43 & \phantom{0}41.61 & \phantom{0}73.80 \\
\cline{2-9}
 & 1.0 & 1.0 & \phantom{00}4.61 & \phantom{00}0.00 & \phantom{00}0.00 & \phantom{00}0.00 & \phantom{00}0.00 & \phantom{0}50.00 \\
\cline{1-9} \cline{2-9}

\end{tabular}

	\end{table}
	\FloatBarrier

	\FloatBarrier

	\begin{table}[tbp]
		\vspace{-15mm}\thisfloatpagestyle{empty}
		\centering
		\caption{Rejection rates (\%) for the EA non-RA procedure under several parameter configurations, tests, and trial size 60. All columns except the last one consider the Wald statistic.  Red indicates type I error rate inflation, green indicates highest power (excluding asymptotic test).}
		\label{tab:comparison_tests_TB_60}
		
		\begin{tabular}{lllrrrrr}
\toprule
$\Iend$ & $\theta_\text{C}$ & $\theta_{\text{D}}$ & CX-S Wald & CX-SA Wald & UX Wald & Asymp. Wald & FET \\
\midrule
\multirow[t]{51}{*}{60} & \multirow[t]{6}{*}{0.0} & 0.0 & \phantom{00}0.00 & \phantom{00}0.00 & \phantom{00}0.00 & \phantom{00}0.00 & \phantom{00}0.00 \\
 &  & 0.1 & \phantom{00}7.32 & \phantom{00}7.32 & \textcolor{darkgreen}{\bf\phantom{0}17.55} & \phantom{0}17.55 & \phantom{00}7.32 \\
 &  & 0.2 & \phantom{0}57.25 & \phantom{0}57.25 & \textcolor{darkgreen}{\bf\phantom{0}74.48} & \phantom{0}74.48 & \phantom{0}57.25 \\
 &  & 0.3 & \phantom{0}92.34 & \phantom{0}92.34 & \textcolor{darkgreen}{\bf\phantom{0}96.98} & \phantom{0}96.98 & \phantom{0}92.34 \\
 &  & 0.4 & \phantom{0}99.43 & \phantom{0}99.43 & \textcolor{darkgreen}{\bf\phantom{0}99.85} & \phantom{0}99.85 & \phantom{0}99.43 \\
 &  & 0.5 & \phantom{0}99.98 & \phantom{0}99.98 & \textcolor{darkgreen}{\bf100.00} & 100.00 & \phantom{0}99.98 \\
\cline{2-8}
 & \multirow[t]{6}{*}{0.01} & 0.01 & \phantom{00}0.00 & \phantom{00}0.00 & \phantom{00}0.00 & \phantom{00}0.00 & \phantom{00}0.00 \\
 &  & 0.11 & \phantom{00}8.08 & \phantom{00}8.08 & \textcolor{darkgreen}{\bf\phantom{0}17.90} & \phantom{0}17.94 & \phantom{00}8.08 \\
 &  & 0.21 & \phantom{0}53.15 & \phantom{0}53.15 & \textcolor{darkgreen}{\bf\phantom{0}68.67} & \phantom{0}69.08 & \phantom{0}53.15 \\
 &  & 0.31 & \phantom{0}88.51 & \phantom{0}88.51 & \textcolor{darkgreen}{\bf\phantom{0}93.85} & \phantom{0}94.32 & \phantom{0}88.51 \\
 &  & 0.41 & \phantom{0}98.58 & \phantom{0}98.58 & \textcolor{darkgreen}{\bf\phantom{0}99.32} & \phantom{0}99.46 & \phantom{0}98.58 \\
 &  & 0.51 & \phantom{0}99.92 & \phantom{0}99.92 & \textcolor{darkgreen}{\bf\phantom{0}99.96} & \phantom{0}99.98 & \phantom{0}99.92 \\
\cline{2-8}
 & \multirow[t]{6}{*}{0.05} & 0.05 & \phantom{00}0.15 & \phantom{00}0.15 & \phantom{00}0.71 & \phantom{00}0.71 & \phantom{00}0.15 \\
 &  & 0.15 & \phantom{00}9.34 & \phantom{00}9.34 & \textcolor{darkgreen}{\bf\phantom{0}16.24} & \phantom{0}17.33 & \phantom{00}9.34 \\
 &  & 0.25 & \phantom{0}43.62 & \phantom{0}43.62 & \textcolor{darkgreen}{\bf\phantom{0}52.66} & \phantom{0}57.06 & \phantom{0}43.62 \\
 &  & 0.35 & \phantom{0}78.82 & \phantom{0}78.82 & \textcolor{darkgreen}{\bf\phantom{0}83.29} & \phantom{0}86.72 & \phantom{0}78.82 \\
 &  & 0.45 & \phantom{0}95.33 & \phantom{0}95.33 & \textcolor{darkgreen}{\bf\phantom{0}96.50} & \phantom{0}97.59 & \phantom{0}95.33 \\
 &  & 0.55 & \phantom{0}99.45 & \phantom{0}99.45 & \textcolor{darkgreen}{\bf\phantom{0}99.60} & \phantom{0}99.77 & \phantom{0}99.45 \\
\cline{2-8}
 & \multirow[t]{6}{*}{0.1} & 0.1 & \phantom{00}0.94 & \phantom{00}0.94 & \phantom{00}2.33 & \phantom{00}2.60 & \phantom{00}0.94 \\
 &  & 0.2 & \phantom{00}9.57 & \phantom{00}9.57 & \textcolor{darkgreen}{\bf\phantom{0}13.34} & \phantom{0}16.23 & \phantom{00}9.57 \\
 &  & 0.3 & \phantom{0}37.16 & \phantom{0}37.16 & \textcolor{darkgreen}{\bf\phantom{0}42.55} & \phantom{0}48.69 & \phantom{0}37.16 \\
 &  & 0.4 & \phantom{0}70.51 & \phantom{0}70.51 & \textcolor{darkgreen}{\bf\phantom{0}74.31} & \phantom{0}79.44 & \phantom{0}70.51 \\
 &  & 0.5 & \phantom{0}91.35 & \phantom{0}91.35 & \textcolor{darkgreen}{\bf\phantom{0}93.04} & \phantom{0}94.89 & \phantom{0}91.35 \\
 &  & 0.6 & \phantom{0}98.53 & \phantom{0}98.53 & \textcolor{darkgreen}{\bf\phantom{0}99.04} & \phantom{0}99.27 & \phantom{0}98.53 \\
\cline{2-8}
 & \multirow[t]{6}{*}{0.3} & 0.3 & \phantom{00}2.61 & \phantom{00}2.61 & \phantom{00}3.65 & \phantom{00}4.86 & \phantom{00}2.61 \\
 &  & 0.4 & \phantom{00}7.82 & \phantom{00}7.82 & \textcolor{darkgreen}{\bf\phantom{0}11.22} & \phantom{0}12.31 & \phantom{00}7.82 \\
 &  & 0.5 & \phantom{0}25.94 & \phantom{0}25.94 & \textcolor{darkgreen}{\bf\phantom{0}34.06} & \phantom{0}35.11 & \phantom{0}25.94 \\
 &  & 0.6 & \phantom{0}56.08 & \phantom{0}56.08 & \textcolor{darkgreen}{\bf\phantom{0}64.70} & \phantom{0}66.31 & \phantom{0}56.08 \\
 &  & 0.7 & \phantom{0}83.83 & \phantom{0}83.83 & \textcolor{darkgreen}{\bf\phantom{0}88.37} & \phantom{0}89.59 & \phantom{0}83.83 \\
 &  & 0.8 & \phantom{0}97.14 & \phantom{0}97.14 & \textcolor{darkgreen}{\bf\phantom{0}98.34} & \phantom{0}98.53 & \phantom{0}97.14 \\
\cline{2-8}
 & \multirow[t]{6}{*}{0.5} & 0.5 & \phantom{00}2.74 & \phantom{00}2.74 & \phantom{00}4.67 & \textcolor{red}{\bf \phantom{00}5.19} & \phantom{00}2.74 \\
 &  & 0.6 & \phantom{00}7.67 & \phantom{00}7.67 & \textcolor{darkgreen}{\bf\phantom{0}11.55} & \phantom{0}12.33 & \phantom{00}7.67 \\
 &  & 0.7 & \phantom{0}25.94 & \phantom{0}25.94 & \textcolor{darkgreen}{\bf\phantom{0}34.06} & \phantom{0}35.11 & \phantom{0}25.94 \\
 &  & 0.8 & \phantom{0}59.64 & \phantom{0}59.64 & \textcolor{darkgreen}{\bf\phantom{0}67.08} & \phantom{0}69.24 & \phantom{0}59.64 \\
 &  & 0.9 & \phantom{0}91.35 & \phantom{0}91.35 & \textcolor{darkgreen}{\bf\phantom{0}93.04} & \phantom{0}94.89 & \phantom{0}91.35 \\
 &  & 1.0 & \phantom{0}99.98 & \phantom{0}99.98 & \textcolor{darkgreen}{\bf100.00} & 100.00 & \phantom{0}99.98 \\
\cline{2-8}
 & \multirow[t]{4}{*}{0.7} & 0.7 & \phantom{00}2.61 & \phantom{00}2.61 & \phantom{00}3.65 & \phantom{00}4.86 & \phantom{00}2.61 \\
 &  & 0.8 & \phantom{00}8.90 & \phantom{00}8.90 & \textcolor{darkgreen}{\bf\phantom{0}10.80} & \phantom{0}14.14 & \phantom{00}8.90 \\
 &  & 0.9 & \phantom{0}37.16 & \phantom{0}37.16 & \textcolor{darkgreen}{\bf\phantom{0}42.55} & \phantom{0}48.69 & \phantom{0}37.16 \\
 &  & 1.0 & \phantom{0}92.34 & \phantom{0}92.34 & \textcolor{darkgreen}{\bf\phantom{0}96.98} & \phantom{0}96.98 & \phantom{0}92.34 \\
\cline{2-8}
 & \multirow[t]{2}{*}{0.9} & 0.9 & \phantom{00}0.94 & \phantom{00}0.94 & \phantom{00}2.33 & \phantom{00}2.60 & \phantom{00}0.94 \\
 &  & 1.0 & \phantom{00}7.32 & \phantom{00}7.32 & \textcolor{darkgreen}{\bf\phantom{0}17.55} & \phantom{0}17.55 & \phantom{00}7.32 \\
\cline{2-8}
 & \multirow[t]{6}{*}{0.95} & 0.95 & \phantom{00}0.15 & \phantom{00}0.15 & \phantom{00}0.71 & \phantom{00}0.71 & \phantom{00}0.15 \\
 &  & 0.96 & \phantom{00}0.12 & \phantom{00}0.12 & \textcolor{darkgreen}{\bf\phantom{00}0.62} & \phantom{00}0.62 & \phantom{00}0.12 \\
 &  & 0.97 & \phantom{00}0.14 & \phantom{00}0.14 & \textcolor{darkgreen}{\bf\phantom{00}0.69} & \phantom{00}0.69 & \phantom{00}0.14 \\
 &  & 0.98 & \phantom{00}0.18 & \phantom{00}0.18 & \textcolor{darkgreen}{\bf\phantom{00}0.88} & \phantom{00}0.88 & \phantom{00}0.18 \\
 &  & 0.99 & \phantom{00}0.24 & \phantom{00}0.24 & \textcolor{darkgreen}{\bf\phantom{00}1.17} & \phantom{00}1.17 & \phantom{00}0.24 \\
 &  & 1.0 & \phantom{00}0.33 & \phantom{00}0.33 & \textcolor{darkgreen}{\bf\phantom{00}1.56} & \phantom{00}1.56 & \phantom{00}0.33 \\
\cline{2-8}
 & \multirow[t]{2}{*}{0.99} & 0.99 & \phantom{00}0.00 & \phantom{00}0.00 & \phantom{00}0.00 & \phantom{00}0.00 & \phantom{00}0.00 \\
 &  & 1.0 & \textcolor{darkgreen}{\bf\phantom{00}0.00} & \textcolor{darkgreen}{\bf\phantom{00}0.00} & \textcolor{darkgreen}{\bf\phantom{00}0.00} & \phantom{00}0.00 & \textcolor{darkgreen}{\bf\phantom{00}0.00} \\
\cline{2-8}
 & 1.0 & 1.0 & \phantom{00}0.00 & \phantom{00}0.00 & \phantom{00}0.00 & \phantom{00}0.00 & \phantom{00}0.00 \\
\bottomrule
\end{tabular}

	\end{table}
	\begin{table}[tbp]
		\vspace{-20mm}\thisfloatpagestyle{empty}
		\centering
		\caption{Rejection rates (\%) for the EA non-RA procedure under several parameter configurations, tests, and trial size 240. All columns except the last one consider the Wald statistic.  Red indicates type I error rate inflation, green indicates highest power (excluding asymptotic test).}
		\label{tab:comparison_tests_TB_240}
		
		\begin{tabular}{lllrrrrr}
\toprule
$\Iend$ & $\theta_\text{C}$ & $\theta_{\text{D}}$ &  CX-S Wald & CX-SA Wald & UX Wald & Asymp. Wald & FET \\
\midrule
\multirow[t]{54}{*}{240} & \multirow[t]{6}{*}{0.0} & 0.0 & \phantom{00}0.00 & \phantom{00}0.00 & \phantom{00}0.00 & \phantom{00}0.00 & \phantom{00}0.00 \\
 &  & 0.05 & \textcolor{darkgreen}{\bf\phantom{0}55.85} & \textcolor{darkgreen}{\bf\phantom{0}55.85} & \textcolor{darkgreen}{\bf\phantom{0}55.85} & \phantom{0}55.85 & \textcolor{darkgreen}{\bf\phantom{0}55.85} \\
 &  & 0.1 & \textcolor{darkgreen}{\bf\phantom{0}98.40} & \textcolor{darkgreen}{\bf\phantom{0}98.40} & \textcolor{darkgreen}{\bf\phantom{0}98.40} & \phantom{0}98.40 & \textcolor{darkgreen}{\bf\phantom{0}98.40} \\
 &  & 0.15 & \textcolor{darkgreen}{\bf\phantom{0}99.99} & \textcolor{darkgreen}{\bf\phantom{0}99.99} & \textcolor{darkgreen}{\bf\phantom{0}99.99} & \phantom{0}99.99 & \textcolor{darkgreen}{\bf\phantom{0}99.99} \\
 &  & 0.2 & \textcolor{darkgreen}{\bf100.00} & \textcolor{darkgreen}{\bf100.00} & \textcolor{darkgreen}{\bf100.00} & 100.00 & \textcolor{darkgreen}{\bf100.00} \\
 &  & 0.25 & \textcolor{darkgreen}{\bf100.00} & \textcolor{darkgreen}{\bf100.00} & \textcolor{darkgreen}{\bf100.00} & 100.00 & \textcolor{darkgreen}{\bf100.00} \\
\cline{2-8}
 & \multirow[t]{6}{*}{0.01} & 0.01 & \phantom{00}0.09 & \phantom{00}0.09 & \phantom{00}0.09 & \phantom{00}0.09 & \phantom{00}0.09 \\
 &  & 0.06 & \phantom{0}42.19 & \phantom{0}42.19 & \textcolor{darkgreen}{\bf\phantom{0}45.09} & \phantom{0}45.09 & \phantom{0}42.19 \\
 &  & 0.11 & \phantom{0}90.94 & \phantom{0}90.94 & \textcolor{darkgreen}{\bf\phantom{0}93.48} & \phantom{0}93.48 & \phantom{0}90.94 \\
 &  & 0.16 & \phantom{0}99.48 & \phantom{0}99.48 & \textcolor{darkgreen}{\bf\phantom{0}99.72} & \phantom{0}99.72 & \phantom{0}99.48 \\
 &  & 0.21 & \textcolor{darkgreen}{\bf\phantom{0}99.99} & \textcolor{darkgreen}{\bf\phantom{0}99.99} & \textcolor{darkgreen}{\bf\phantom{0}99.99} & \phantom{0}99.99 & \textcolor{darkgreen}{\bf\phantom{0}99.99} \\
 &  & 0.26 & \textcolor{darkgreen}{\bf100.00} & \textcolor{darkgreen}{\bf100.00} & \textcolor{darkgreen}{\bf100.00} & 100.00 & \textcolor{darkgreen}{\bf100.00} \\
\cline{2-8}
 & \multirow[t]{6}{*}{0.05} & 0.05 & \phantom{00}2.11 & \phantom{00}2.11 & \phantom{00}3.41 & \phantom{00}3.41 & \phantom{00}2.11 \\
 &  & 0.1 & \phantom{0}22.22 & \phantom{0}22.22 & \textcolor{darkgreen}{\bf\phantom{0}28.80} & \phantom{0}28.90 & \phantom{0}22.22 \\
 &  & 0.15 & \phantom{0}67.43 & \phantom{0}67.43 & \textcolor{darkgreen}{\bf\phantom{0}73.40} & \phantom{0}74.06 & \phantom{0}67.43 \\
 &  & 0.2 & \phantom{0}93.47 & \phantom{0}93.47 & \textcolor{darkgreen}{\bf\phantom{0}95.14} & \phantom{0}95.49 & \phantom{0}93.47 \\
 &  & 0.25 & \phantom{0}99.33 & \phantom{0}99.33 & \textcolor{darkgreen}{\bf\phantom{0}99.56} & \phantom{0}99.59 & \phantom{0}99.33 \\
 &  & 0.3 & \phantom{0}99.96 & \phantom{0}99.96 & \textcolor{darkgreen}{\bf\phantom{0}99.98} & \phantom{0}99.98 & \phantom{0}99.96 \\
\cline{2-8}
 & \multirow[t]{6}{*}{0.1} & 0.1 & \phantom{00}2.85 & \phantom{00}2.85 & \phantom{00}4.26 & \phantom{00}4.51 & \phantom{00}2.85 \\
 &  & 0.15 & \phantom{0}16.08 & \phantom{0}16.08 & \textcolor{darkgreen}{\bf\phantom{0}19.84} & \phantom{0}20.69 & \phantom{0}16.08 \\
 &  & 0.2 & \phantom{0}51.60 & \phantom{0}51.60 & \textcolor{darkgreen}{\bf\phantom{0}57.63} & \phantom{0}58.06 & \phantom{0}51.60 \\
 &  & 0.25 & \phantom{0}83.67 & \phantom{0}83.67 & \textcolor{darkgreen}{\bf\phantom{0}87.20} & \phantom{0}87.25 & \phantom{0}83.67 \\
 &  & 0.3 & \phantom{0}96.90 & \phantom{0}96.90 & \textcolor{darkgreen}{\bf\phantom{0}97.82} & \phantom{0}97.82 & \phantom{0}96.90 \\
 &  & 0.35 & \phantom{0}99.67 & \phantom{0}99.67 & \textcolor{darkgreen}{\bf\phantom{0}99.79} & \phantom{0}99.79 & \phantom{0}99.67 \\
\cline{2-8}
 & \multirow[t]{6}{*}{0.3} & 0.3 & \phantom{00}3.62 & \phantom{00}3.62 & \phantom{00}4.77 & \phantom{00}4.96 & \phantom{00}3.62 \\
 &  & 0.35 & \phantom{0}10.34 & \phantom{0}10.34 & \textcolor{darkgreen}{\bf\phantom{0}12.71} & \phantom{0}12.97 & \phantom{0}10.34 \\
 &  & 0.4 & \phantom{0}32.14 & \phantom{0}32.14 & \textcolor{darkgreen}{\bf\phantom{0}36.69} & \phantom{0}37.03 & \phantom{0}32.14 \\
 &  & 0.45 & \phantom{0}63.19 & \phantom{0}63.19 & \textcolor{darkgreen}{\bf\phantom{0}67.31} & \phantom{0}67.90 & \phantom{0}63.19 \\
 &  & 0.5 & \phantom{0}86.90 & \phantom{0}86.90 & \textcolor{darkgreen}{\bf\phantom{0}88.74} & \phantom{0}89.24 & \phantom{0}86.90 \\
 &  & 0.55 & \phantom{0}97.05 & \phantom{0}97.05 & \textcolor{darkgreen}{\bf\phantom{0}97.59} & \phantom{0}97.72 & \phantom{0}97.05 \\
\cline{2-8}
 & \multirow[t]{6}{*}{0.5} & 0.5 & \phantom{00}3.33 & \phantom{00}3.33 & \phantom{00}4.52 & \phantom{00}4.53 & \phantom{00}3.33 \\
 &  & 0.55 & \phantom{00}9.04 & \phantom{00}9.04 & \textcolor{darkgreen}{\bf\phantom{0}11.22} & \phantom{0}11.29 & \phantom{00}9.04 \\
 &  & 0.6 & \phantom{0}29.09 & \phantom{0}29.09 & \textcolor{darkgreen}{\bf\phantom{0}32.76} & \phantom{0}33.20 & \phantom{0}29.09 \\
 &  & 0.65 & \phantom{0}60.82 & \phantom{0}60.82 & \textcolor{darkgreen}{\bf\phantom{0}64.13} & \phantom{0}65.02 & \phantom{0}60.82 \\
 &  & 0.7 & \phantom{0}86.90 & \phantom{0}86.90 & \textcolor{darkgreen}{\bf\phantom{0}88.74} & \phantom{0}89.24 & \phantom{0}86.90 \\
 &  & 0.75 & \phantom{0}97.66 & \phantom{0}97.66 & \textcolor{darkgreen}{\bf\phantom{0}98.23} & \phantom{0}98.29 & \phantom{0}97.66 \\
\cline{2-8}
 & \multirow[t]{6}{*}{0.7} & 0.7 & \phantom{00}3.62 & \phantom{00}3.62 & \phantom{00}4.77 & \phantom{00}4.96 & \phantom{00}3.62 \\
 &  & 0.75 & \phantom{0}11.11 & \phantom{0}11.11 & \textcolor{darkgreen}{\bf\phantom{0}13.55} & \phantom{0}13.80 & \phantom{0}11.11 \\
 &  & 0.8 & \phantom{0}37.91 & \phantom{0}37.91 & \textcolor{darkgreen}{\bf\phantom{0}42.56} & \phantom{0}42.70 & \phantom{0}37.91 \\
 &  & 0.85 & \phantom{0}75.89 & \phantom{0}75.89 & \textcolor{darkgreen}{\bf\phantom{0}79.76} & \phantom{0}79.77 & \phantom{0}75.89 \\
 &  & 0.9 & \phantom{0}96.90 & \phantom{0}96.90 & \textcolor{darkgreen}{\bf\phantom{0}97.82} & \phantom{0}97.82 & \phantom{0}96.90 \\
 &  & 0.95 & \phantom{0}99.96 & \phantom{0}99.96 & \textcolor{darkgreen}{\bf\phantom{0}99.98} & \phantom{0}99.98 & \phantom{0}99.96 \\
\cline{2-8}
 & \multirow[t]{3}{*}{0.9} & 0.9 & \phantom{00}2.85 & \phantom{00}2.85 & \phantom{00}4.26 & \phantom{00}4.51 & \phantom{00}2.85 \\
 &  & 0.95 & \phantom{0}22.22 & \phantom{0}22.22 & \textcolor{darkgreen}{\bf\phantom{0}28.80} & \phantom{0}28.90 & \phantom{0}22.22 \\
 &  & 1.0 & \textcolor{darkgreen}{\bf\phantom{0}98.40} & \textcolor{darkgreen}{\bf\phantom{0}98.40} & \textcolor{darkgreen}{\bf\phantom{0}98.40} & \phantom{0}98.40 & \textcolor{darkgreen}{\bf\phantom{0}98.40} \\
\cline{2-8}
 & \multirow[t]{6}{*}{0.95} & 0.95 & \phantom{00}2.11 & \phantom{00}2.11 & \phantom{00}3.41 & \phantom{00}3.41 & \phantom{00}2.11 \\
 &  & 0.96 & \phantom{00}2.86 & \phantom{00}2.86 & \textcolor{darkgreen}{\bf\phantom{00}4.25} & \phantom{00}4.25 & \phantom{00}2.86 \\
 &  & 0.97 & \phantom{00}5.87 & \phantom{00}5.87 & \textcolor{darkgreen}{\bf\phantom{00}7.81} & \phantom{00}7.81 & \phantom{00}5.87 \\
 &  & 0.98 & \phantom{0}12.96 & \phantom{0}12.96 & \textcolor{darkgreen}{\bf\phantom{0}15.36} & \phantom{0}15.36 & \phantom{0}12.96 \\
 &  & 0.99 & \phantom{0}27.76 & \phantom{0}27.76 & \textcolor{darkgreen}{\bf\phantom{0}29.47} & \phantom{0}29.47 & \phantom{0}27.76 \\
 &  & 1.0 & \textcolor{darkgreen}{\bf\phantom{0}55.85} & \textcolor{darkgreen}{\bf\phantom{0}55.85} & \textcolor{darkgreen}{\bf\phantom{0}55.85} & \phantom{0}55.85 & \textcolor{darkgreen}{\bf\phantom{0}55.85} \\
\cline{2-8}
 & \multirow[t]{2}{*}{0.99} & 0.99 & \phantom{00}0.09 & \phantom{00}0.09 & \phantom{00}0.09 & \phantom{00}0.09 & \phantom{00}0.09 \\
 &  & 1.0 & \textcolor{darkgreen}{\bf\phantom{00}0.14} & \textcolor{darkgreen}{\bf\phantom{00}0.14} & \textcolor{darkgreen}{\bf\phantom{00}0.14} & \phantom{00}0.14 & \textcolor{darkgreen}{\bf\phantom{00}0.14} \\
\cline{2-8}
 & 1.0 & 1.0 & \phantom{00}0.00 & \phantom{00}0.00 & \phantom{00}0.00 & \phantom{00}0.00 & \phantom{00}0.00 \\
\bottomrule
\end{tabular}

	\end{table}
	\begin{table}[tbp]
		\vspace{-20mm}\thisfloatpagestyle{empty}
		\centering
		\caption{Rejection rates (\%) for the EA non-RA procedure under several parameter configurations, tests, and trial size 960. All columns except the last one consider the Wald statistic.  Red indicates type I error rate inflation, green indicates highest power (excluding asymptotic test).}
		\label{tab:comparison_tests_TB_960}
		
		\begin{tabular}{lllrrrrr}
\toprule
$\Iend$ & $\theta_\text{C}$ & $\theta_{\text{D}}$ & CX-S Wald & CX-SA Wald & UX Wald & Asymp. Wald & FET \\
\midrule
\multirow[t]{57}{*}{960} & \multirow[t]{6}{*}{0.0} & 0.0 & \phantom{00}0.00 & \phantom{00}0.00 & \phantom{00}0.00 & \phantom{00}0.00 & \phantom{00}0.00 \\
 &  & 0.02 & \textcolor{darkgreen}{\bf\phantom{0}91.83} & \textcolor{darkgreen}{\bf\phantom{0}91.83} & \textcolor{darkgreen}{\bf\phantom{0}91.83} & \phantom{0}91.83 & \textcolor{darkgreen}{\bf\phantom{0}91.83} \\
 &  & 0.04 & \textcolor{darkgreen}{\bf\phantom{0}99.99} & \textcolor{darkgreen}{\bf\phantom{0}99.99} & \textcolor{darkgreen}{\bf\phantom{0}99.99} & \phantom{0}99.99 & \textcolor{darkgreen}{\bf\phantom{0}99.99} \\
 &  & 0.06 & \textcolor{darkgreen}{\bf100.00} & \textcolor{darkgreen}{\bf100.00} & \textcolor{darkgreen}{\bf100.00} & 100.00 & \textcolor{darkgreen}{\bf100.00} \\
 &  & 0.08 & \textcolor{darkgreen}{\bf100.00} & \textcolor{darkgreen}{\bf100.00} & \textcolor{darkgreen}{\bf100.00} & 100.00 & \textcolor{darkgreen}{\bf100.00} \\
 &  & 0.1 & \textcolor{darkgreen}{\bf100.00} & \textcolor{darkgreen}{\bf100.00} & \textcolor{darkgreen}{\bf100.00} & 100.00 & \textcolor{darkgreen}{\bf100.00} \\
\cline{2-8}
 & \multirow[t]{6}{*}{0.01} & 0.01 & \phantom{00}2.07 & \phantom{00}2.07 & \phantom{00}2.27 & \phantom{00}2.27 & \phantom{00}2.07 \\
 &  & 0.03 & \phantom{0}53.88 & \phantom{0}53.88 & \textcolor{darkgreen}{\bf\phantom{0}57.78} & \phantom{0}57.78 & \phantom{0}53.88 \\
 &  & 0.05 & \phantom{0}95.96 & \phantom{0}95.96 & \textcolor{darkgreen}{\bf\phantom{0}96.68} & \phantom{0}96.68 & \phantom{0}95.96 \\
 &  & 0.07 & \phantom{0}99.90 & \phantom{0}99.90 & \textcolor{darkgreen}{\bf\phantom{0}99.92} & \phantom{0}99.92 & \phantom{0}99.90 \\
 &  & 0.09 & \textcolor{darkgreen}{\bf100.00} & \textcolor{darkgreen}{\bf100.00} & \textcolor{darkgreen}{\bf100.00} & 100.00 & \textcolor{darkgreen}{\bf100.00} \\
 &  & 0.11 & \textcolor{darkgreen}{\bf100.00} & \textcolor{darkgreen}{\bf100.00} & \textcolor{darkgreen}{\bf100.00} & 100.00 & \textcolor{darkgreen}{\bf100.00} \\
\cline{2-8}
 & \multirow[t]{6}{*}{0.05} & 0.05 & \phantom{00}3.56 & \phantom{00}3.56 & \phantom{00}4.42 & \phantom{00}4.57 & \phantom{00}3.56 \\
 &  & 0.07 & \phantom{0}21.44 & \phantom{0}21.44 & \textcolor{darkgreen}{\bf\phantom{0}24.16} & \phantom{0}24.64 & \phantom{0}21.44 \\
 &  & 0.09 & \phantom{0}64.05 & \phantom{0}64.05 & \textcolor{darkgreen}{\bf\phantom{0}67.30} & \phantom{0}67.80 & \phantom{0}64.05 \\
 &  & 0.11 & \phantom{0}91.99 & \phantom{0}91.99 & \textcolor{darkgreen}{\bf\phantom{0}93.21} & \phantom{0}93.36 & \phantom{0}91.99 \\
 &  & 0.13 & \phantom{0}99.11 & \phantom{0}99.11 & \textcolor{darkgreen}{\bf\phantom{0}99.30} & \phantom{0}99.31 & \phantom{0}99.11 \\
 &  & 0.15 & \phantom{0}99.95 & \phantom{0}99.95 & \textcolor{darkgreen}{\bf\phantom{0}99.96} & \phantom{0}99.96 & \phantom{0}99.95 \\
\cline{2-8}
 & \multirow[t]{6}{*}{0.1} & 0.1 & \phantom{00}3.93 & \phantom{00}3.93 & \phantom{00}4.69 & \phantom{00}4.78 & \phantom{00}3.93 \\
 &  & 0.12 & \phantom{0}14.44 & \phantom{0}14.44 & \textcolor{darkgreen}{\bf\phantom{0}16.03} & \phantom{0}16.36 & \phantom{0}14.44 \\
 &  & 0.14 & \phantom{0}44.37 & \phantom{0}44.37 & \textcolor{darkgreen}{\bf\phantom{0}46.96} & \phantom{0}47.52 & \phantom{0}44.37 \\
 &  & 0.16 & \phantom{0}76.63 & \phantom{0}76.63 & \textcolor{darkgreen}{\bf\phantom{0}78.69} & \phantom{0}79.06 & \phantom{0}76.63 \\
 &  & 0.18 & \phantom{0}93.99 & \phantom{0}93.99 & \textcolor{darkgreen}{\bf\phantom{0}94.81} & \phantom{0}94.91 & \phantom{0}93.99 \\
 &  & 0.2 & \phantom{0}99.06 & \phantom{0}99.06 & \textcolor{darkgreen}{\bf\phantom{0}99.23} & \phantom{0}99.24 & \phantom{0}99.06 \\
\cline{2-8}
 & \multirow[t]{6}{*}{0.3} & 0.3 & \phantom{00}4.24 & \phantom{00}4.24 & \phantom{00}4.89 & \phantom{00}4.99 & \phantom{00}4.24 \\
 &  & 0.32 & \phantom{00}9.02 & \phantom{00}9.02 & \textcolor{darkgreen}{\bf\phantom{0}10.07} & \phantom{0}10.22 & \phantom{00}9.02 \\
 &  & 0.34 & \phantom{0}24.18 & \phantom{0}24.18 & \textcolor{darkgreen}{\bf\phantom{0}26.13} & \phantom{0}26.36 & \phantom{0}24.18 \\
 &  & 0.36 & \phantom{0}48.09 & \phantom{0}48.09 & \textcolor{darkgreen}{\bf\phantom{0}50.43} & \phantom{0}50.79 & \phantom{0}48.09 \\
 &  & 0.38 & \phantom{0}72.48 & \phantom{0}72.48 & \textcolor{darkgreen}{\bf\phantom{0}74.25} & \phantom{0}74.67 & \phantom{0}72.48 \\
 &  & 0.4 & \phantom{0}89.09 & \phantom{0}89.09 & \textcolor{darkgreen}{\bf\phantom{0}90.03} & \phantom{0}90.28 & \phantom{0}89.09 \\
\cline{2-8}
 & \multirow[t]{6}{*}{0.5} & 0.5 & \phantom{00}4.20 & \phantom{00}4.20 & \phantom{00}4.89 & \phantom{00}4.89 & \phantom{00}4.20 \\
 &  & 0.52 & \phantom{00}8.26 & \phantom{00}8.26 & \textcolor{darkgreen}{\bf\phantom{00}9.33} & \phantom{00}9.33 & \phantom{00}8.26 \\
 &  & 0.54 & \phantom{0}21.38 & \phantom{0}21.38 & \textcolor{darkgreen}{\bf\phantom{0}23.32} & \phantom{0}23.32 & \phantom{0}21.38 \\
 &  & 0.56 & \phantom{0}43.09 & \phantom{0}43.09 & \textcolor{darkgreen}{\bf\phantom{0}45.65} & \phantom{0}45.65 & \phantom{0}43.09 \\
 &  & 0.58 & \phantom{0}67.35 & \phantom{0}67.35 & \textcolor{darkgreen}{\bf\phantom{0}69.63} & \phantom{0}69.65 & \phantom{0}67.35 \\
 &  & 0.6 & \phantom{0}85.96 & \phantom{0}85.96 & \textcolor{darkgreen}{\bf\phantom{0}87.31} & \phantom{0}87.36 & \phantom{0}85.96 \\
\cline{2-8}
 & \multirow[t]{6}{*}{0.7} & 0.7 & \phantom{00}4.24 & \phantom{00}4.24 & \phantom{00}4.89 & \phantom{00}4.99 & \phantom{00}4.24 \\
 &  & 0.72 & \phantom{00}9.23 & \phantom{00}9.23 & \textcolor{darkgreen}{\bf\phantom{0}10.32} & \phantom{0}10.48 & \phantom{00}9.23 \\
 &  & 0.74 & \phantom{0}25.79 & \phantom{0}25.79 & \textcolor{darkgreen}{\bf\phantom{0}27.81} & \phantom{0}28.11 & \phantom{0}25.79 \\
 &  & 0.76 & \phantom{0}52.57 & \phantom{0}52.57 & \textcolor{darkgreen}{\bf\phantom{0}54.98} & \phantom{0}55.37 & \phantom{0}52.57 \\
 &  & 0.78 & \phantom{0}78.82 & \phantom{0}78.82 & \textcolor{darkgreen}{\bf\phantom{0}80.56} & \phantom{0}80.83 & \phantom{0}78.82 \\
 &  & 0.8 & \phantom{0}94.06 & \phantom{0}94.06 & \textcolor{darkgreen}{\bf\phantom{0}94.76} & \phantom{0}94.87 & \phantom{0}94.06 \\
\cline{2-8}
 & \multirow[t]{6}{*}{0.9} & 0.9 & \phantom{00}3.93 & \phantom{00}3.93 & \phantom{00}4.69 & \phantom{00}4.78 & \phantom{00}3.93 \\
 &  & 0.92 & \phantom{0}16.26 & \phantom{0}16.26 & \textcolor{darkgreen}{\bf\phantom{0}18.38} & \phantom{0}18.48 & \phantom{0}16.26 \\
 &  & 0.94 & \phantom{0}58.66 & \phantom{0}58.66 & \textcolor{darkgreen}{\bf\phantom{0}62.00} & \phantom{0}62.35 & \phantom{0}58.66 \\
 &  & 0.96 & \phantom{0}94.93 & \phantom{0}94.93 & \textcolor{darkgreen}{\bf\phantom{0}95.79} & \phantom{0}95.91 & \phantom{0}94.93 \\
 &  & 0.98 & \textcolor{darkgreen}{\bf\phantom{0}99.98} & \textcolor{darkgreen}{\bf\phantom{0}99.98} & \textcolor{darkgreen}{\bf\phantom{0}99.98} & \phantom{0}99.98 & \textcolor{darkgreen}{\bf\phantom{0}99.98} \\
 &  & 1.0 & \textcolor{darkgreen}{\bf100.00} & \textcolor{darkgreen}{\bf100.00} & \textcolor{darkgreen}{\bf100.00} & 100.00 & \textcolor{darkgreen}{\bf100.00} \\
\cline{2-8}
 & \multirow[t]{6}{*}{0.95} & 0.95 & \phantom{00}3.56 & \phantom{00}3.56 & \phantom{00}4.42 & \phantom{00}4.57 & \phantom{00}3.56 \\
 &  & 0.96 & \phantom{00}8.88 & \phantom{00}8.88 & \textcolor{darkgreen}{\bf\phantom{0}10.49} & \phantom{0}10.63 & \phantom{00}8.88 \\
 &  & 0.97 & \phantom{0}29.84 & \phantom{0}29.84 & \textcolor{darkgreen}{\bf\phantom{0}33.19} & \phantom{0}33.26 & \phantom{0}29.84 \\
 &  & 0.98 & \phantom{0}67.55 & \phantom{0}67.55 & \textcolor{darkgreen}{\bf\phantom{0}70.81} & \phantom{0}70.81 & \phantom{0}67.55 \\
 &  & 0.99 & \phantom{0}95.96 & \phantom{0}95.96 & \textcolor{darkgreen}{\bf\phantom{0}96.68} & \phantom{0}96.68 & \phantom{0}95.96 \\
 &  & 1.0 & \textcolor{darkgreen}{\bf100.00} & \textcolor{darkgreen}{\bf100.00} & \textcolor{darkgreen}{\bf100.00} & 100.00 & \textcolor{darkgreen}{\bf100.00} \\
\cline{2-8}
 & \multirow[t]{2}{*}{0.99} & 0.99 & \phantom{00}2.07 & \phantom{00}2.07 & \phantom{00}2.27 & \phantom{00}2.27 & \phantom{00}2.07 \\
 &  & 1.0 & \textcolor{darkgreen}{\bf\phantom{0}34.88} & \textcolor{darkgreen}{\bf\phantom{0}34.88} & \textcolor{darkgreen}{\bf\phantom{0}34.88} & \phantom{0}34.88 & \textcolor{darkgreen}{\bf\phantom{0}34.88} \\
\cline{2-8}
 & 1.0 & 1.0 & \phantom{00}0.00 & \phantom{00}0.00 & \phantom{00}0.00 & \phantom{00}0.00 & \phantom{00}0.00 \\
\bottomrule
\end{tabular}

	\end{table}
	
	\FloatBarrier

	\begin{table}[h!]
		\caption{Upper critical values for the unconditional exact test for different trial sizes, statistics, and RA procedures. The one-sided significance level is set to~$2.5\%$ in each setting. Note: critical values for $\Iend = 960$ found by restricting the null set $[0,1]$ to $0.00:0.01:1.00$.}\label{tab:critical}
\centering
		
		\begin{tabular}{cllll}\hline
			\multicolumn{1}{c}{\rule{0pt}{13pt}$\Iend$}  & UX FET RDP & Wald RDP & UX FET EA & Wald EA\\\hline
			10 & 0.1 & 1.7272595098492414 & 0.13333333333333333 & 1.959965156484713\\
			20 & 0.08668730650154799 & 1.9215378456610455 & 0.14035087719298245 & 1.853047161780774\\
			30 & 0.06666666666666667 & 2.0359633906020442 & 0.10811407483071651 & 1.9329712334408418\\
			40 & 0.056160926432242395 & 2.082478408222474 & 0.09480249480249481 & 1.9625514979929637\\
			50 & 0.05384897107006595 & 2.077347117755887 & 0.08450068146047536 & 1.9970918228698624\\
			60 & 0.050815732062114774 & 2.0953869552983915 & 0.06977998727279189 & 2.06568306450296\\
			70 & 0.05111272283085849 & 2.0980279891303883 & 0.08221013616964479 & 1.967736176247307\\
			80 & 0.04984354552541328 & 2.096976978695872 & 0.07211385187049967 & 2.0160645150967422\\
			90 & 0.049230363163264006 & 2.0944721097402654 & 0.07937738341030817 & 1.9592844056604022\\
			100 & 0.04629884007947289 & 2.1014769412666388 & 0.07040879622570355 & 2.0046321754009315\\
			120 & 0.045915383653536694 & 2.1027665720396795 & 0.06709588404416586 & 2.0099557916716355\\
			150 & 0.04508205644330902 & 2.10273332962542 & 0.06898224231605116 & 1.9787001382948695\\
			180 & 0.04403334678418491 & 2.1061977615142933 & 0.06816766165704137 & 1.979735020132453\\
			210 & 0.04423613894350655 & 2.103763815417522 & 0.066778869568174 & 1.9696196403913138\\
			240 & 0.043880073891202416 & 2.103157864866603 & 0.06528413110170263 & 1.9711384650967143\\
			960 & 0.0425576537814003 & 2.0995182034464555 & 0.05681751737279753 & 1.9694556533734013\\\hline
		\end{tabular}
	\end{table}
	\FloatBarrier

	\section{Further computational details and results for the modified play-the-winner trial application}
	
	\label{sect:reiertsentrial_extra}
	
	\subsection{Justification of log-rank test}
	This section justifies the use of the log-rank test in the design of~\citet{reiertsen1993}.
	Let~$(\tilde{L}_k)_{k=1}^\infty$,  be an independent sequence of positive real-valued event times (random variables) and~$W_k\in\{\C,\D\}$ be a set of treatment indicators such that each~$\tilde{L}_k\mid W_k = a$ has the same cumulative distribution function~$F_a$ 
	for \mbox{all~$k\in\{1,2,\dots\}
		$}, \mbox{$a\in\{\C,\D\}$.} Let~$\delta_k$ be an (observed) set of positive real-valued censoring times \mbox{and~$L^\prime_k=\min(\delta_k,\tilde{L}_k)$} be the observed times for~$k\in\{1,\dots,\kappa\}$. It is assumed that the censoring times are independent of the event times and treatment indicators. 
	Following~\citet{lee_survival}, under these assumptions, the log-rank test, taking as input a realisation~$\Bell^\prime,\bw$ of~$\bL^\prime,\bW$, as well as the censoring times~$\bdelta$, is an asymptotic test for testing~$F_{C}(\ell)=F_D(\ell)~$ for all~$\ell\leq \max_k \ell^\prime_k$ versus~$\exists \ell\leq \max_k \ell^\prime_k$ such that $F_C(\ell)\neq F_D(\ell)$. 
	
	Under the \mbox{M-PTW} design and the i.i.d. Bernoulli outcomes model, the treatment sequence lengths~$L^\prime_k$ can be considered as observed times corresponding to geometric event times, i.e., we \mbox{have~$F_a(\ell) = 1-\theta_a^{\floor{\ell}}$.} When the censoring times are independent of the event times,  the theoretical guarantees of the log-rank test hold.
	Censoring the multiple \mbox{M-PTW} sequences at the end of the M-PTW sequences may however induce length time bias, as there is a higher probability that extraordinarily long trial sequences are occurring at the end of the M-PTW sequence. When increasing the trial size to infinity, the effect of this will vanish. 
	
	\subsection{Markov chain formulation}\label{MPTW_MCform}
	We now describe the Markov chain used to model the design in~\citet{reiertsen1993}. The Markov chain modeling the M-PTW DRA procedure is the Markov chain described in Example~1, augmented with~$W_i$~and~$L_i$.
	The length of the \mbox{M-PTW} sequences were not reported in~\citet{reiertsen1993} and hence an assumption needs to be made on this part of the data. In order to analyze the power of the trial, it is assumed that the length of the concurrent sequences are roughly the same as the lengths reported in \citet[Figure 2]{reiertsen1993}, scaled such that it matches the actual sample size. In practice, the lengths of the \mbox{M-PTW} sequences could be derived from a surgery schedule.~\autoref{triallengths} presents the assumed \mbox{M-PTW} sequence lengths. 
	\FloatBarrier
	\begin{table}[tbp]
		\centering
		\caption{Assumed \mbox{M-PTW} sequence lengths.}\label{triallengths}
		\begin{tabular}{lcccccl}
			\hline
			M-PTW sequence               & 1                      & 2                      & 3                      & 4                     & 5                     & 6 \\ \hline
			Monday                     & 18                     & 15                     & 15                     & 15                    & 10                    & - \\
			Tuesday                    & 16                     & 16                     & 10                     & 8                     & -                     & - \\
			Wednesday                  & 19                     & 16                     & 16                     & 13                    & 10                    & 8 \\
			Thursday                   & 18                     & 15                     & 15                     & 12                    & -                     & - \\
			\multicolumn{1}{l}{Friday} & \multicolumn{1}{l}{19} & \multicolumn{1}{l}{16} & \multicolumn{1}{l}{13} & \multicolumn{1}{l}{9} & \multicolumn{1}{l}{5} & - \\ \hline
		\end{tabular}
	\end{table}
	\FloatBarrier

	The multiple independent \mbox{M-PTW} sequences started in the trial are modeled by restarting the \mbox{M-PTW} procedure (keeping the successes and allocations per arm) whenever the length of the current \mbox{M-PTW} sequence reaches a value in~\autoref{triallengths}, i.e., letting~$\ell_1,\ell_2,\dots, \ell_m$ be the length of the trial sequences~(e.g.,~$18,15,15,\dots,16,16,10,\dots,9,5$), we define~$\mathcal{L}=\{\ell_1,\ell_1+\ell_2,\dots, \textstyle\sum_{k=1}^{|\mathcal{L}|-1} \ell_k\}$ to be the decision epochs where \mbox{M-PTW} restarts.

	We introduce a Markov chain describing the evolution of the collected data under the \mbox{M-PTW} design.
	To accurately describe the behaviour under~\mbox{M-PTW}, we take~$i_t=t$ (hence we can use~$i$ and~$t$ interchangeably) and augment the state for the Markov chain described in Example~1  with~$W_t$ and~$L_t$, \mbox{i.e.,~$\bX_t=(\bS_t,\bN_t,W_t,L_t)$.}
	The process~$\bX$ is a Markov chain with  initial state~$$\bX_0 = \begin{cases}
		((0,0),(0,0), \C,1),\quad \text{with probability~$1/2$,}\\
		((0,0),(0,0), \D,1),\quad \text{with probability~$1/2$,}
	\end{cases}$$
	state space~$\mathcal{X}=\cup_t\mathcal{X}_t$ with~$$\mathcal{X}_t=\{((s'_\C,s'_\D),(n'_\C,n'_\D), w, \ell):((s'_\C,s'_\D),(n'_\C,n'_\D))\in\mathcal{X}^\text{SS}_{t},\, w\in\{\C,\D\}, \ell\in\calI_{\leq 15}\},$$
	and transition dynamics according to~(1) where, letting~$w,\ell$ be functions such that~$w(\bX_t)=W_t,$ \mbox{and~$\ell(\bX_t)=L_t$,} we have that~$q^\pi(\bx_{t},\bx_{t+1}) =1$ when we have~$t+1\notin\mathcal{L}$, $ n_{w(\bx_{t})}(\bx_{t+1}) = n_{w(\bx_{t})}(\bx_t)+1$
	$\sum_{a\in\{\C,\D\}}|\partial n_a(\bx_t,\bx_{t+1})|=1$, and either 
	\begin{enumerate}[label={(\alph*)}]
		\item~$\ell(\bx_{t+1})=\ell(\bx_t)+1$,~$\partial s_{w(\bx_t)}(\bx_t,\bx_{t+1}) = 1$,~$w(\bx_{t+1}) = w(\bx_t)$,\label{case1}
		\item~$\ell(\bx_{t+1})=1$,~$\ell(\bx_t) = 15$,~$w(\bx_{t+1}) \neq w(\bx_t)$,
		\item~$\ell(\bx_{t+1})=1$,~$\partial s_{w(\bx_t)}(\bx_t,\bx_{t+1}) = 0$,~$w(\bx_{t+1}) \neq w(\bx_t)$.
	\end{enumerate}
	Furthermore,~$q^\pi(\bx_{t},\bx_{t+1}) =1/2$ when~$t+1\in\mathcal{L}$,~$\sum_{a\in\{\C,\D\}}|\partial n_a(\bx_t,\bx_{t+1})|=1$, \\$\ell(\bx_{t+1})=1$,~$\partial n_{w(\bx_{t})}(\bx_{t},\bx_{t+1}) = 1$, and~$\partial s_{w(\bx_{t})}(\bx_t,\bx_{t+1}) = y$ for~$y\in\{0,1\}$.

	Misclassification of the outcomes as a zero occurred rarely~\citep{reiertsen1993}, out of~327 outcomes only~2 successes were   \mbox{misclassified ($p_\text{miscl}\approx0.612\%$).}  Hence, the choice is made not to take misclassification into account in the Markov chain, although (assuming independence and that the probability of misclassification is constant over time)  this could be done by, e.g., the introduction of a randomization probability for the allocations, i.e., by changing case~\ref{case1} above to 
	\begin{equation*}
		q^\pi(\bx_{t},\bx_{t+1}) = \begin{cases}
			p_\text{miscl}, \quad &\text{if~$\ell(\bx_{t+1})=1$},\\&\partial s_{w(\bx_t)}(\bx_t,\bx_{t+1}) = 1,\\&\text{and}~w(\bx_{t+1})\neq w(\bx_t),\\
			1-p_\text{miscl},  &\text{if~$\ell(\bx_{t+1})=\ell(\bx_{t})+1$},\\&\partial s_{w(\bx_t)}(\bx_t,\bx_{t+1}) = 1,\\&\text{and}~w(\bx_{t+1})= w(\bx_t).
		\end{cases}
	\end{equation*}

	The absolute rejection rates for M-PTW under the considered tests are given in \autoref{tab:comparison_tests_ex1}, where it is shown that the power of the considered tests is actually around~$40\%$ for the reported success rates.
	
	\begin{table}[tbp]
		\subsection{Additional results}
		\centering
		\caption{Rejection rates for the    CX-S Wald, UX Wald, and log-rank test (upper and lower significance level equal to $2.5\%$) and EPASA (in percentage points)  for~\hbox{$\theta_\C =0.748$}, $\theta_\D\geq \theta_\C$ under the \mbox{M-PTW} design. The 95\% confidence radius shown for the log-rank test is based on a normal approximation. The critical value for the UX Wald test was~1.9626231638655138.  }
		\label{tab:comparison_tests_ex1}
		\begin{tabular}{llllll}
\toprule
$\theta_\text{C}$ & $\theta_{\text{D}}$ & CX-S Wald & UX Wald & log-rank test & EPASA \\
\midrule
\multirow[t]{7}{*}{0.748} & 0.748 & \phantom{00}4.93 & \phantom{00}4.85 & \phantom{00}5.15 +/- 0.14 & \phantom{0}50.00 \\
 & 0.800 & \phantom{0}19.82 & \phantom{0}19.53 & \phantom{0}18.65 +/- 0.24 & \phantom{0}54.77 \\
 & 0.830 & \phantom{0}43.53 & \phantom{0}43.00 & \phantom{0}40.45 +/- 0.30 & \phantom{0}57.91 \\
 & 0.850 & \phantom{0}62.61 & \phantom{0}61.99 & \phantom{0}58.06 +/- 0.31 & \phantom{0}60.19 \\
 & 0.900 & \phantom{0}95.17 & \phantom{0}94.90 & \phantom{0}92.36 +/- 0.16 & \phantom{0}66.63 \\
 & 0.950 & \phantom{0}99.96 & \phantom{0}99.95 & \phantom{0}99.83 +/- 0.03 & \phantom{0}74.25 \\
 & 1.000 & 100.00 & 100.00 & 100.00 +/- 0.00 & \phantom{0}83.18 \\
\cline{1-6}
\bottomrule
\end{tabular}

	\end{table}
	
	\section{Further computational details and results for the ARREST trial application}\label{extra_results_arrest}
	\subsection{Markov chain formulation for determining UX stopping threshold}\label{ARREST_MC}
	In this section, the Markov chain used to derive the UX OST for the ARREST trial is described.
	Let~$\bX_t = (\bS_{t}, \bN_{t}, M_t)$ where~$M_t = \ubar{\text{Proj}}_\mathcal{M}(\max_{t'\leq t}\tilde{\pi}_\C(\bX_{t'}))$, and we define~$\ubar{\text{Proj}}_E(x)=\max\{y\in E:y\leq x\}$ for~$x\in [0,1],\,E\subset [0,1]$, and~$\{0.5\}\subseteq\mathcal{M}\subset [0.5,1]$ with~$|\mathcal{M}|<\infty$. The state variable~$M_t$ denotes the highest value in~$\mathcal{M}$ that was crossed by the posterior probability~$\tilde{\pi}_\C(\bX_{t'})$ that treatment~$\C$ is optimal based on~$\bX_{t'}$, for time~$t'$ up to and including~$t$. Assume that the assignment in blocks is done using~PBD.  
	Then, letting~$i_0=0$ and~$i_t = \sum_{u=1}^tb_t$ where~$(b_t)_t$ is the sequence of block sizes and~$m$ be a function such that~$m(\bX_t)=M_t$, the process~$\bX$  is a Markov chain with 
	state space~$\mathcal{X}=\cup_t\mathcal{X}_t$
	with~$\mathcal{X}_t=\{((s'_\C,s'_\D),(n'_\C,n'_\D), m):((s'_\C,s'_\D),(n'_\C,n'_\D))\in\mathcal{X}^{\text{SS}}_{i_t},\, m\in\mathcal{M} \},$
	initial state $\bX_0=((0,0),(0,0), 0.5)$ and  transition structure~(1) with~$q^\pi(\bx_{t},\bx_{t+1})~$ equal to
	\begin{equation}
		\begin{cases}
			\pi_\C(\bx_t)b_{t} - \floor{
				\pi_\C(\bx_t)b_{t}}, \; &\text{if }\partial \bn(\bx_{t},\bx_{t+1})=( \ceil{\pi_\C(\bx_t)b_{t}},b_t - \ceil{\pi_\C(\bx_t)b_{t}})\\&\text{ and } \max(m(\bx_t), \ubar{\text{Proj}}_\mathcal{M}(\tilde{\pi}_\C(\bx_{t+1}))) = m(\bx_{t+1})\\
			\ceil{\pi_\C(\bx_t)b_{t}}-\pi_\C(\bx_t)b_{t},  &\text{if }\partial \bn(\bx_{t},\bx_{t+1})=( \floor{\pi_\C(\bx_t)b_{t}},b_t - \floor{\pi_\C(\bx_t)b_{t}})\\&\text{ and } \max(m(\bx_t), \ubar{\text{Proj}}_\mathcal{M}(\tilde{\pi}_\C(\bx_{t+1}))) = m(\bx_{t+1})\\
			0,&\text{else,}
		\end{cases} \label{defn:allocprob_blocked}
	\end{equation} for all ~$\bx_t\in\mathcal{X}_{t},\bx_{t+1}\in\mathcal{X}_{t+1}$,   where~$\floor{\delta}=\ceil{\delta}-1$ for all~$\delta\in\mathcal{D}.$  
	When calculating the UX OST, symmetry was used to calculate the two-sided critical value as, given~$\theta_C=\theta_D$,
	\begin{align*}
		& \arg\min\{p^*\in \mathcal{M}:\mathbb{P}^\pi_\btheta(\max_{t\leq {\Tend}}\max\{\Pi(\theta_\D\geq\theta_\C\mid\bX_t),\,\Pi(\theta_\C\geq\theta_\D\mid\bX_t)\}\geq p^*)\leq \alpha\} \\&= \arg\min\{p^*\in \mathcal{M}:\mathbb{P}^\pi_\btheta(m(\bX_t)\geq p^*)\leq \alpha/2\}.
	\end{align*}
	Note that the critical value calculated in this manner really controls type I errors, but it can be too conservative if~$\mathcal{M}$ is too discrete.

	\subsection{Additional results}
\begin{figure}[h!]
		\centering
		\includegraphics[width=.95\linewidth]{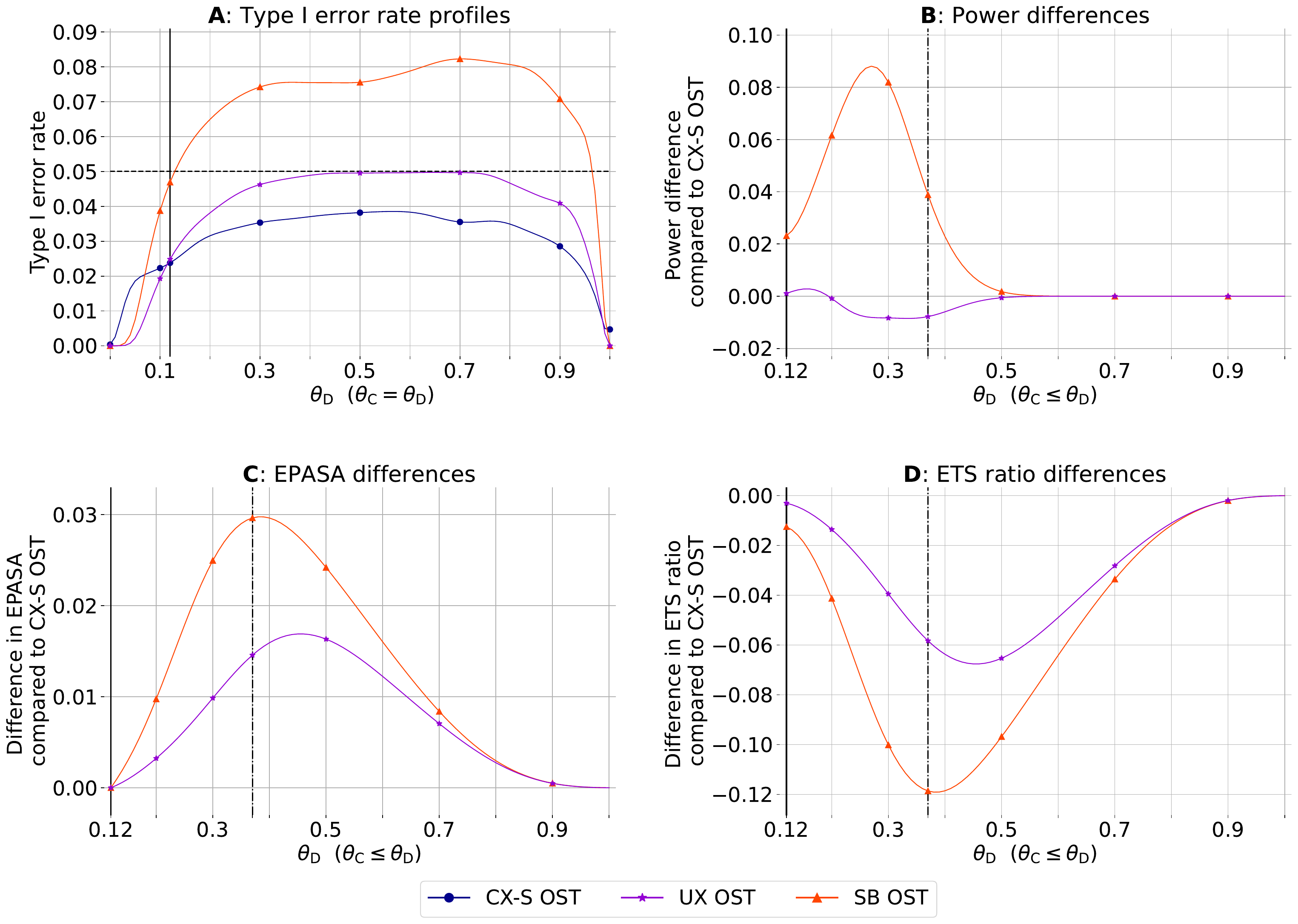}
		\caption{ ARREST trial. Type~I~error rate under the SB, CX-S, and UX OST for~$\theta_\C=\theta_\D$ (Subfigure A), the vertical solid line denotes~$\theta_\C=\theta_\D=0.12$. Power difference (Subfigure B), expected proportion of 
			participants allocated to the superior arm (EPASA, Subfigure C), and expected trial size ratio~$\mathbb{E}_\btheta^\pi[i_{u(\bX_{\Tend})}]/\Iend$ (ETS ratio, Subfigure~D) 
			for the SB and UX OST  compared to CX-S OST,~$\theta_\C = 0.12$ and~$\theta_\D\geq \theta_\C$, where the vertical dash-dotted line \hbox{denotes~$\theta_\D=0.37.$} The upper and lower significance levels both equal 2.5\%.}\label{fig:comparison_ARREST}
	\end{figure}

	\autoref{fig:comparison_ARREST} shows, in addition to the results in~\autoref{fig:comparison_ARREST_top} of the paper, the expected participant outcomes, and expected trial size ratio~$\mathbb{E}_\btheta^\pi[i_{u(\bX_{\Tend})}]/\Iend$ for the \emph{simulation-based}~(SB) and UX~OSTs versus the \mbox{CX-S OST.} 
	The EPASA, assuming that after early stopping due to superiority of treatment~$\D$, all remaining participants are allocated the developmental treatment, is defined as $$(N_{\D,i_{u(\bX_{\Tend})}} + (\Iend-i_{u(\bX_{\Tend})})\mathbb{I}(\text{Optional stopping in favor of~$\D$}))/\Iend.$$ 
	In terms of expected trial size and EPASA, the SB OST is best (i.e., shows higher expected treatment outcomes and smaller expected trial sizes), while the UX OST performs best out of the exact tests. The UX OST results in a decrease of about~$1.5\%$ in EPASA and an increase in expected trial size ratio  of around  6\% in comparison to the SB OST.
	
		\autoref{tab:comparison_tests_ex2}  shows the operating characteristics, together with EPASA up to optional stopping, where this OC is highest for CX-S as optional stopping occurs at later time points.

	\FloatBarrier	
	
	\begin{table}[h!]
	\footnotesize
	\centering
	\vspace{-7mm}
	\caption{ARREST trial: Operating characteristics (in percentage points) for the  SB, CX-S, and UX optional stopping thresholds for~$\theta_\C =0.12$,~$\theta_\D\geq \theta_\C$. RR denotes rejection rate~(type I error rate or power). The expected proportion of participants allocated to the superior arm excluding optional stopping (EPASA, excl. OS) is defined as~$N_{\D,i_{u(\bX_{\Tend})}}/{\Iend}$. The UX OST was~0.9918742236024845. Both the upper and lower significance level was set to~2.5\%.}
	\label{tab:comparison_tests_ex2}
	\begin{tabular}{llllllllll}
\toprule
 & $\theta_\text{C}$ & \multicolumn{8}{l}{0.12} \\
 & $\theta_{\text{D}}$ & 0.12 & 0.2 & 0.3 & 0.37 & 0.5 & 0.7 & 0.9 & 1.0 \\
OST & Measure &  &  &  &  &  &  &  &  \\
\midrule
\multirow[t]{4}{*}{SB} & RR & \phantom{00}4.69 & \phantom{0}20.54 & \phantom{0}67.62 & \phantom{0}90.46 & \phantom{0}99.78 & 100.00 & 100.00 & 100.00 \\
 & EPASA & \phantom{0}50.00 & \phantom{0}63.60 & \phantom{0}74.87 & \phantom{0}80.51 & \phantom{0}86.50 & \phantom{0}89.38 & \phantom{0}89.98 & \phantom{0}90.00 \\
 & EPASA (excl. OS) & \phantom{0}50.00 & \phantom{0}60.63 & \phantom{0}63.29 & \phantom{0}61.37 & \phantom{0}56.46 & \phantom{0}51.50 & \phantom{0}50.04 & \phantom{0}50.00 \\
 & Trial size ratio & \phantom{0}98.18 & \phantom{0}92.38 & \phantom{0}71.74 & \phantom{0}54.59 & \phantom{0}33.44 & \phantom{0}22.46 & \phantom{0}20.07 & \phantom{0}20.00 \\
\cline{1-10}
\multirow[t]{4}{*}{CX-S} & RR & \phantom{00}2.38 & \phantom{0}14.38 & \phantom{0}59.43 & \phantom{0}86.58 & \phantom{0}99.60 & 100.00 & 100.00 & 100.00 \\
 & EPASA & \phantom{0}50.00 & \phantom{0}62.62 & \phantom{0}72.38 & \phantom{0}77.54 & \phantom{0}84.08 & \phantom{0}88.54 & \phantom{0}89.93 & \phantom{0}90.00 \\
 & EPASA (excl. OS) & \phantom{0}50.00 & \phantom{0}61.37 & \phantom{0}65.63 & \phantom{0}64.79 & \phantom{0}60.63 & \phantom{0}53.51 & \phantom{0}50.17 & \phantom{0}50.00 \\
 & Trial size ratio & \phantom{0}99.43 & \phantom{0}96.51 & \phantom{0}81.76 & \phantom{0}66.45 & \phantom{0}43.12 & \phantom{0}25.82 & \phantom{0}20.27 & \phantom{0}20.00 \\
\cline{1-10}
\multirow[t]{4}{*}{UX} & RR & \phantom{00}2.49 & \phantom{0}14.29 & \phantom{0}58.60 & \phantom{0}85.80 & \phantom{0}99.55 & 100.00 & 100.00 & 100.00 \\
 & EPASA & \phantom{0}50.00 & \phantom{0}62.94 & \phantom{0}73.36 & \phantom{0}79.00 & \phantom{0}85.72 & \phantom{0}89.25 & \phantom{0}89.98 & \phantom{0}90.00 \\
 & EPASA (excl. OS) & \phantom{0}50.00 & \phantom{0}61.09 & \phantom{0}64.47 & \phantom{0}62.77 & \phantom{0}57.64 & \phantom{0}51.81 & \phantom{0}50.05 & \phantom{0}50.00 \\
 & Trial size ratio & \phantom{0}99.13 & \phantom{0}95.15 & \phantom{0}77.81 & \phantom{0}60.62 & \phantom{0}36.59 & \phantom{0}23.00 & \phantom{0}20.08 & \phantom{0}20.00 \\
\cline{1-10}
\bottomrule
\end{tabular}

\end{table}
\FloatBarrier

	\end{appendix}

\end{document}